\documentclass[%
 reprint,
 amsmath,amssymb,
 aps,
pra,
]{revtex4-2}

\usepackage{graphicx}% Include figure files
\usepackage{dcolumn}% Align table columns on decimal point
\usepackage{bm}% bold math
\usepackage{tikz}
\usepackage{tikz-feynman}

\begin{document}

\preprint{APS/123-QED}

\title{Strange Metal to Insulator Transitions in the Lowest Landau Level} % Force line breaks with \\

\author{Ben Currie}
\author{Evgeny Kozik}%
\affiliation{%
 Department of Physics, King’s College London, Strand, London WC2R 2LS, United Kingdom
}%

\date{\today}% It is always \today, today,

\begin{abstract}
We study the microscopic model of electrons in the partially-filled lowest Landau level interacting via the Coulomb potential by the diagrammatic theory within the GW approximation. In a wide range of filling fractions and temperatures, we find a homogeneous non-Fermi liquid (nFL) state similar to that found in the Sachdev-Ye-Kitaev (SYK) model, with logarithmic corrections to the anomalous dimension. In addition, the phase diagram is qualitatively similiar to that of SYK: a first-order transition terminating at a critical end-point separates the nFL phase from a band insulator that corresponds to the fully-filled Landau level. This critical point, as well as that of the SYK model---whose critical exponents we determine more precisely---are shown to both belong to the Van der Waals universality class. The possibility of a charge density wave (CDW) instability is also investigated, and we find the homogeneous nFL state to extend down to the ground state for fillings $0.2 \lesssim \nu \lesssim 0.8$, while a CDW appears outside this range of fillings at sufficiently low temperatures. Our results suggest that the SYK-like nFL state should be a generic feature of the partially-filled lowest Landau level at intermediate temperatures.
\end{abstract}

\maketitle

%\tableofcontents

\section{\label{sec:Introduction}Introduction}

Two-dimensional electrons subject to strong magnetic fields exhibit an extreme variety of highly correlated quantum phases. Key examples are the fractional quantum Hall (FQH) states  \cite{Tsui1982,Stormer1983},
whose strong correlations are manifest in the emergence of low-energy quasi-particles possessing fractional charge and statistics \cite{Arovas1984,Halperin1984}, which arise due to strong Coulomb interactions within a flat band.  Being an inherently non-perturbative phenomenon, the development of a microscopic theory of the FQH states -- based solely on the electronic degrees of freedom and their Coulomb interactions -- remains an important open problem. An understanding has been achieved using trial wavefunctions \cite{Laughlin1983,Jain1989} and effective field theories of emergent fermions -- such as composite fermions (CFs) \cite{Jain1989,Jain1990,Lopez1991,Zhang1989,HLR1993,Shankar1997}. At mean-field level, the CFs, which are bound states of electrons and an even number of flux quanta, move in a partially cancelled magnetic field, resulting in insulating behaviour at odd-denominator filling fractions and metallic properties at even-denominators.
While the existence of emergent fermions that experience a weaker magnetic field has been confirmed spectacularly by experiments at low temperatures \cite{Jiang1989,Willett1990,Willett1993,Kang1993,Willett1993_2,Du1993,Willett1997}, the nature of the state of the system at higher temperatures remains unknown.

Owing to the strong coupling nature of the problem, controlled solutions of the electronic Hamiltonian have been limited to exact diagonalisation \cite{YHL1983,Laughlin1983b,Haldane1983,Haldane1985,Fano1986,d'Ambrunmenil1989,Morf2002} and DMRG calculations \cite{Naokazu2001,Feiguin2008,Kovrizhin2010,Zaletel2012,Hu2012,Estienne2013,Zaletel2013} at finite system size. These methods allow exact calculation of the ground and excited state properties, but are limited to small system sizes. In the thermodynamic limit, no method has yielded results consistent with experiment. Hartree-Fock calculations were performed early-on \cite{FPA1979,Yoshioka1979,Yoshioka1983}, and yielded charge density wave (CDW) states for all filling fractions. In addition, using the (partially self-consistent) $GW$ approximation, Tao and Thouless considered a periodic occupation of the lowest Landau level single-particle wavefunctions, and found the existence of a gap to all excitations at all integer-denominator filling fractions $\nu = \frac{1}{m}$ \cite{TaoThouless1983,Tao1984}.  
A fully self-consistent homogeneous solution of the $GW$ equations was obtained by Haussmann \cite{Haussmann1996} in an effort to understand the nature of the pseudo-gap in the single-electron density of states observed in experiments \cite{Ashoori1990,Ashoori1993,Eisenstein1992,Brown1994,Eisenstein1995,Eisenstein2016}. Since a metallic state was observed instead, its physical implications and relevance to the lowest Landau level physics as well as related models was not realised. 
 
In this paper, we revisit the $GW$ theory for this problem and analyse it in full detail without further approximations. We obtain the phase diagram in the full space of parameters, which reveals two distinct types of metal-to-insulator transition. We pay particular attention to the metallic non-Fermi liquid (nFL) state found in a wide range of parameters at intermediate temperatures, as well as the critical point of its transition to a fully-filled Landau level band insulator. As a byproduct, we reconcile the different diagrammatic results, and discuss the shortcomings and successes of each.

We start from the microscopic model of electrons in the lowest Landau level (LLL) interacting via the long-range Coulomb potential in the thermodynamic limit. At partial-filling of the LLL, the non-interacting limit yields a macroscopically degenerate ground state, thus rendering an approach based on an expansion around it very challenging. The lowest-order terms, corresponding to the mean-field (Hartree-Fock) theory, have been understood for some time \cite{FPA1979,Yoshioka1979,Yoshioka1983}, and yield CDW band insulator ground states for all filling fractions. This, however, contradicts the experimental observation that FQH states possess no long-range order \cite{Chang1983}. 
Moreover, the expansion in the bare interaction cannot be properly justified in the thermodynamic limit since terms beyond first-order are divergent due to the long range of the Coulomb interaction. For this reason, an expansion in the screened interaction is necessary.
We access non-perturbative physics by formulating a diagrammatic series that is renormalised self-consistently to infinite order in both the one-electron ($G$) and screened Coulomb interaction ($W$) channels. In this study, the diagrammatic expansion is truncated at the level of the $GW$ approximation, in which the electronic self-energy $\Sigma$ is given by the $GW$ diagram and the polarisation $\Pi$ renormalises the interaction by $GG$ \cite{Hedin1965}. We note that our approach is the same as that used by Haussmann \cite{Haussmann1996}, but differs from the approach of Tao and Thouless, which uses only partial self-consistency. Importantly, we find that the Tao Thouless states are in fact not (fully) self-consistent solutions of the $GW$ equations (this point will be further discussed in Section \ref{sec: GW general}). We envisage that this theory will be systematically extended to sufficiently higher orders in the future by means of advanced diagrammatic Monte Carlo techniques~\cite{VanHoucke2010, Kozik2010, VanHoucke2019, Chen2019, Kozik2023combinatorial}, which could enable \textit{a priori} control of the potential systematic error. Our $GW$ results, however, are a necessary starting point for such extensions. Most importantly, being controlled in the high temperature limit, they already suggest a non-trivial and rich physical picture of the LLL at finite temperatures.

In particular, we demonstrate that the CDW physics obtained at the mean-field level is altered fundamentally: there is no continuous transition to density-wave ordering down to (and including, at the $GW$ level) $T=0$ for $ 0.2 \lesssim \nu \lesssim 0.8$, and the resulting metallic state is an nFL of the type described by the prototypical Sachdev-Ye-Kitaev (SYK) model~\cite{Sachdev1993,Kitaev2015,SYKReview2022}---a model with random all-to-all $q$-fermion coupling, denoted $\text{SYK}_{q}$ (with $q=4$ in the original case). Specifically, this nFL phase displays quantum criticality, characterised by a power-law low-frequency scaling of the fermion self-energy, 
\begin{equation} \label{nFL self-energy}
    \Sigma (i\omega) = \lambda e^{-i\mathrm{sign}(\omega)(\pi/2+\theta)}|\omega|^{1-2\Delta}, ~~\text{for} ~\omega \to 0,
\end{equation}
with the so-called anomalous dimension $0<\Delta<1/2$, constant $\lambda > 0$, and particle-hole asymmetry parameter $\theta$, which controls the filling fraction of the LLL and which must satisfy $-\pi\Delta < \theta < \pi\Delta$. A solution of this form is relevant to the normal state of high temperature superconductors \cite{EsterlisSchmalian2019,Wang2020,Hauck2020} and is closely related to extremal black-holes due to the emergent low-energy conformal symmetry and the saturation of the chaos bound on the Lyapunov exponent \cite{Sachdev2015,Maldacena2016,Jensen2016}.
Physically, it describes a compressible metal without coherent electronic quasi-particle excitations. In certain SYK-inspired models in non-zero spatial dimension, it is known that Eq.~\eqref{nFL self-energy} leads to linear-in-temperature resistivity \cite{Song2017,Chowdhury2018,Patel2018,Cha2020}, which is the main signature of the ubiquitous strange metal phase \cite{Martin1990,Takagi1992,Valla2000}.

\begin{figure}[b]
\includegraphics[trim={10 2 25 23},clip,width = 0.5\textwidth]{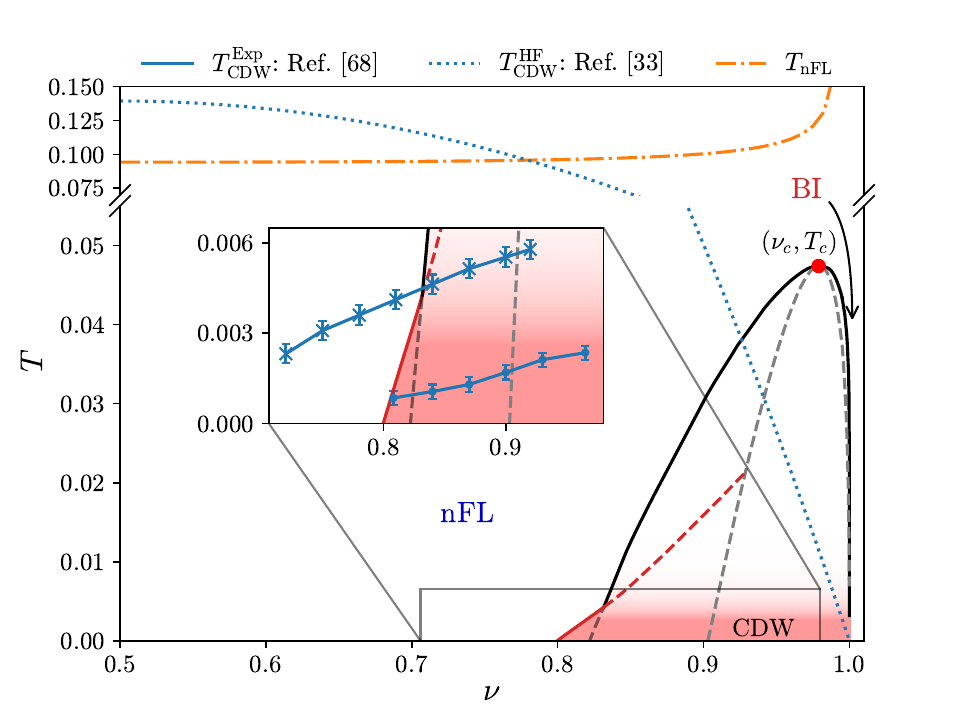}
\caption{\label{fig:T density phase diagram} 
Phase diagram of the model~\eqref{Hamiltonian} within the $GW$ approximation in the temperature ($T$, in units of $e^2/\ell_B$ which at typical magnetic fields $B = 13\,\text{T}$ corresponds to $\sim 200 \, \textrm{K}$) against filling fraction ($\nu$) plane. The non-Fermi liquid (nFL) phase of quantum-critical fermions exists down to low $T$ and for a broad range of $\nu$, and below the free fermion thermal state to nFL crossover temperature $T_\text{nFL}$ (orange dash-dot line). At some transition filling $\nu_\text{nFL-BI}(T)$, the nFL undergoes a first-order transition to a band-insulating (BI) phase corresponding to the fully-filled LLL (solid black line). Phase separation occurs inside this region, in which the the nFL and BI phases coexist, while metastable phases exist between the solid black and grey dashed lines. The second-order critical point of this transition is labelled by $(\nu_c,T_c)$ (red point). In addition, a second-order charge-density wave (CDW) transition is observed from the nFL phase at low temperatures $T_{\mathrm{CDW}}(\nu)$ (red line). Inside the phase-separation region, the location of the CDW transition is unknown, and the boundary of the CDW phase is indicated by the blurring. The metastable nFL phase undergoes a CDW transition shown by the red dashed line. Experimental CDW transition temperatures reported in Ref. \cite{Chen2006} are shown for both heterojunction (dots) and quantum well (crosses) samples. Also shown is the Hartree-Fock charge-density-wave temperature $T_{\text{CDW}}^{\text{HF}}$ obtained in Ref. \cite{FPA1979}.}
\end{figure}

Our nFL solution is described by Eq.~\eqref{nFL self-energy} with $\Delta = \frac{1}{4}$ 
in the theoretical limit $\omega \to 0$; however, at any practically relevant frequency, the long-range nature of the Coulomb interaction gives rise to appreciable logarithmic corrections to $\Delta$. This state is found in a broad range of filling fractions centred around one-half, as summarised in the phase diagram in Fig. \ref{fig:T density phase diagram}, 
and emerges from the high-temperature (free fermion) thermal state at the crossover temperature $T_\mathrm{nFL} \sim 0.1 e^2/\ell_B$.
Similarly to the SYK model \cite{SYKPhaseDiagram2018,PhasesofMelonicQM2019}, as the filling fraction is increased at low temperatures, the nFL state undergoes a first-order transition to a fully-filled-LLL band insulator, with the transition line ending at a critical point $(\nu_c, T_c)$ (see Fig.~\ref{fig:T density phase diagram}). 

We study the critical point specifically, calculating the critical exponents and comparing them to those in the SYK model. To this end, we reconsider the critical point of the $\text{SYK}_4$ model, and demonstrate that the corresponding critical exponents actually differ from those reported previously~\cite{SYKPhaseDiagram2018,PhasesofMelonicQM2019}. Instead, the values we obtain for $\text{SYK}_4$ agree with those found recently in the $q\to \infty$ limit in Ref.~\cite{Louw2023}. Remarkably, we find precisely the same exponents for the LLL model, indicating that the critical points of the LLL and the $\text{SYK}_q$ models for all $q$ belong to the same universality class: the Van der Waals universality class, also describing an extremal black hole. This answers in the affirmative the question raised in Ref.~\cite{Louw2023} of whether the critical points of the $\text{SYK}_4$ and $\text{SYK}_{q=\infty}$ models are the same.

Compared to the SYK model, the phase diagram is enriched by the presence of the CDW insulating state. Experimentally, CDW phases are known to exist only for extremal Landau level filling fractions $\nu \lesssim 0.25$ and $\nu \gtrsim 0.85$ \cite{Willett1988,Jiang1990,Goldman1990,Jiang1991,Williams1991,Paalanen1992_2}, and our results predict a CDW ground state in qualitative agreement with these observations (the $\nu \lesssim 0.25$ range is a mirror image of $\nu \gtrsim 0.85$ due to the particle-hole symmetry). 
At larger filling fractions close to the nFL-to-band insulator transition and at low temperature, we find that the transition to the CDW phase is continuous. In addition, the corresponding critical temperatures $T_{\mathrm{CDW}}(\nu)$ are the same order of magnitude as those observed in experiments performed on GaAs/AlGaAs heterojunctions and AlGaAs/GaAs/AlGaAs quantum wells \cite{Paalanen1992_2,Kukushkin1993,Chen2006} (data from Ref. \cite{Chen2006} are shown in Fig. \ref{fig:T density phase diagram} for both types of sample).

Although the $GW$ approximation is controlled in the high temperature limit, the observed nFL state exists only at low-temperatures, for which the higher-order terms cannot be neglected \textit{a priori}.
Nonetheless, additional observations allow us to suggest that the nFL physics predicted by $GW$ should indeed describe the normal phase from which the fractional quantum Hall states emerge as the temperature is lowered. Specifically, the $GW$ results are in qualitative agreement with the exact results in the thin-torus (TT) limit, in which the ground states are the Tao-Thouless states \cite{Bergholtz2005,Bergholtz_2006,Bergholtz2007,Bergholtz2008}: We find that the low-temperature $GW$ solution in this limit is indeed a TT state, while at intermediate temperatures it transforms to a nFL phase that is qualitatively similar to that observed in the thermodynamic limit. Thus, since $GW$ gives correct physics both in the limit of high temperatures and in the TT limit at low temperatures, by continuity we expect that it captures the physics in the thermodynamic limit at some intermediate temperatures as well, where the nFL state is predicted.

Our finding of the nFL at intermediate temperatures within the LLL is also consistent with empirical evidence from other systems where interactions completely dominate the kinetic energy.
One example is the twisted bilayer graphene near the magic angle, in which nFL quantum critical states have been found experimentally
\cite{Cao2020,Lyu2021,Jaoui2022}.
There, the normal state preceding the low-temperature fractionally filled insulator is also an nFL state~\cite{Cao2018,Cao2020,Zhang2022}.
In addition, it has been proposed~\cite{Chen2018,Brzezinska2023} that a graphene flake subject to a strong perpendicular magnetic field could serve as an experimental realisation of SYK physics, due to the absence of any dispersion and the spatial randomness of the interaction stemming from the irregular shape of the boundary.
More generally, it is known that the generic class of Hamiltonians which conserve both centre of mass momentum \emph{and} centre of mass position cannot possess gapless ground states that are ordinary metals (here defined as phases which possess a non-zero Drude weight) \cite{Seidel2005}. Since the LLL Hamiltonian belongs to this class, any metallic state must be an nFL in nature.
Experimental work on the high temperature properties of the partially filled LLL, which is currently lacking, would be crucial to verify our predictions experimentally.

This paper is organised as follows. In Sec.~\ref{Microscopic model}, we derive the microscopic Hamiltonian and discuss the single-particle physics. In Sec.~\ref{sec: GW general}, we present the diagrammatic formulation of this problem and the $GW$ approximation.  In Sec.~\ref{sec: GW homogeneous}, we consider translationally symmetric solutions of the $GW$ equations. The nFL solution is discussed in detail, as well as the phase diagram displaying the nFL to band insulator transition. The corresponding critical point is also analysed, with particular focus on calculation of the critical exponents. In Sec.~\ref{sec: GW CDW}, continuous phase transitions to density wave ordered solutions are discussed. Finally, we put the $GW$ results in a broader context and discuss their relevance to the LLL physics in Sec.~\ref{sec: conclusion}.

\section{\label{Microscopic model} Model}
The two-dimensional electron system is subject to a large homogeneous and perpendicular magnetic field $B = (0,0,-B)$. The strong magnetic field implies that the electrons are spin polarised and thus only one spin component is considered.  
We work in Landau gauge, in which the vector potential is $A = (0,-Bx,0)$. This choice of gauge explicitly breaks the translation invariance in the $x$-direction. Since, however, the $y$-translation symmetry is preserved, the $y$ component of momentum $k$ is a good quantum number and can be used to label the single-particle states.

The lowest Landau level (LLL) eigenfunctions are 
\begin{equation}
    \psi_k = \frac{1}{\sqrt{\ell_B\sqrt{\pi}}}e^{iky}e^{-\frac{1}{2\ell_B^2}(x-k\ell_B^2)^2}, \label{eq:wavefunction}
\end{equation}
where $\ell_B = \sqrt{\frac{\hbar}{eB}}$ is the magnetic length, $k = 2\pi n/L_y$, and $L_x$ and $L_y$ are the system sizes in the $x$ and $y$ direction. Notably, the wavefunctions are localised in the $x$-direction at the position $x = k\ell_B^2$, with localisation width $\ell_B$. This is a manifestation of the  explicitly broken $x$-translation symmetry in Landau gauge. The corresponding eigenenergies are independent of $k$ and are equal to $E = \hbar\omega_B/2$ with $\omega_B = eB/m$ the cyclotron frequency. Therefore the LLL contains a macroscopic number of states $D = \frac{L_xL_y}{2\pi\ell_B^2}$, all of which are degenerate. The higher Landau levels are separated from the LLL by a gap $\propto B$, which allows us to ignore them entirely. In the following we set $E=0$, and consider the thermodynamic limit $L_x,L_y \to \infty$, in which the momentum $k$ is continuous.

In the basis of single-electron states (\ref{eq:wavefunction}), the Hamiltonian in terms of the corresponding creation ($c^\dagger_k$) and annihilation ($c_k$) operators is  
\begin{equation}\label{Hamiltonian}
    H_1 = \frac{1}{2} \sum_{m}\sum_{kq}V_{kqm}c^\dagger_{m+q}c^\dagger_{m+k}c_{m+q+k}c_{m} - \sum_{m}\mu c^\dagger_mc_m,
\end{equation}
where  the momentum summations are a shorthand for $\sum_k = \int \frac{dk}{2\pi}$ (we use this notation throughout), and
\begin{equation} \label{V matrix elements}
    V_{kqm} \equiv V_{kq} = \int \frac{dp}{2\pi} \frac{2\pi}{\sqrt{p^2+q^2}}e^{-\frac{1}{2}(p^2+q^2)\ell_B^2}e^{ipk\ell_B^2}
\end{equation}
is the Coulomb potential $V(r)=\frac{1}{r}$ projected onto the LLL and represented diagrammatically as 
\begin{equation}
V_{kqm} 
~~= 
\begin{gathered}
\begin{tikzpicture}
\begin{feynman}
\vertex (c);
\vertex [above = 0.5cm of c](d){$m+q$};
\vertex [below = 0.5cm of c] (e) {$m$};
\vertex [right = 0.6cm of c](a);
\vertex [right = 2cm of a] (b);
\vertex [right = 0.6cm of b] (f);
\vertex [above = 0.5cm of f] (h){{$m+k$}};
\vertex [below = 0.5cm of f] (g) {$m+k+q$};
\diagram*{
(a) -- [photon,edge label = $q$] (b),
(a) -- [fermion] (d),
(e) -- [fermion] (a),
(b) -- [fermion] (h),
(g) -- [fermion] (b),
};
\end{feynman}
\end{tikzpicture}
\end{gathered}. \label{diagram:interaction_vertex}
\end{equation}
In the following, we use units in which $\ell_B  =1$, and measure all energies in units of $e^2/\ell_B$. 

In realistic systems, a uniform positive background charge density is required to ensure charge-neutrality and hence thermodynamic stability. In two-dimensions and in a non-zero external magnetic field, the stability of interacting electrons is very fragile because of the large exchange energy, which can lead to negative compressibility $\kappa$ \cite{Tao1989,Kravchenko1990,Eisenstein1992a,Eisenstein1994}. In the inversion-layer materials used in experiments, the two-dimensional plane in which the electrons reside is located at the interface between two semiconductors with differing band-gaps, and the plane of positive background charge (here the donor ions) is separated from this electron plane by a small distance $d$ \cite{PrangeGirvin1990}. While the $d=0$ case leads to $\kappa <0$ for generic densities, non-zero $d$ introduces a positive capacitance energy, which can offset and nullify the (negative) exchange energy for large enough separation $d$. This ensures $\kappa >0$ for all electron densities. In this paper, for simplicity, we consider a separation such that the (homogeneous) exchange term is cancelled exactly, which allows us to simply set the spatially uniform components of this term to zero.
We note that while dynamical properties of the nFL phase are unchanged by the value of $d$, the thermodynamic properties -- for example the compressibility -- are altered. Therefore, the parameters at which phase transitions occur at non-zero $d$ will in general differ from those reported here at $d=0$.

\section{\label{sec: GW general} Diagrammatic formulation and $GW$ approximation}

Our main observable, containing the information about the state of the system and its excitation properties, is the finite-temperature (Matsubara) Green's function of the electrons in the LLL, which is defined as~\cite{AGD}
\begin{equation} \label{Defn of one particle GF}
    G_k(\tau) = -\left\langle \mathcal {T} c_k(\tau)c_k(0)^\dagger \right\rangle,
\end{equation}
where $\mathcal{T}$ is the time-ordering operator for the imaginary time $\tau$. 
It can be obtained through a calculation of the electronic self-energy $\Sigma$, which contains the combined effect of electron interactions on one-particle excitations, via the Dyson equation~\cite{AGD}
\begin{equation} \label{Dyson}
    G_{k}(i\omega) = \frac{1}{i\omega+\mu - \Sigma_{k}(i\omega)},
\end{equation}
where $\omega$ is the Matsubara frequency. In the diagrammatic theory, $\Sigma$ is formulated as an expansion in the powers of the coupling vertex, with each term represented graphically by the corresponding number of vertices (\ref{diagram:interaction_vertex}) connected by the Green's functions and integrated over all internal variables. In particular, the first-order self-energy diagrams yield 
\begin{equation} \label{First order full self energy}
    \Sigma_{k}(i\omega) = \frac{1}{\beta}\sum_{\nu}\int \frac{dq}{2\pi}\left(V_{q,0}-V_{0,q}\right)G_{k-q}(i\omega-i\nu)
\end{equation}
where the first and second terms correspond to the Hartree and Fock contributions of the mean-field theory, respectively.
In the homogeneous high-temperature phase, the Hartree-Fock self-energy is a trivial (i.e. momentum and frequency independent) energy shift. However, as discussed in the introduction, it develops an instability towards CDW order at intermediate temperatures $T_\text{CDW}^\text{HF}\sim 0.1 e^2/\ell_B$ for all LLL fillings (blue dotted curve in Fig. \ref{fig:T density phase diagram}: see Appendix \ref{HF appendix} for more details). Since the first order gives no new physics in the homogeneous phase, which is of primary interest to us here, we now move on to higher-order.

Systematically going beyond the mean-field by including higher order in $V$ terms is plagued by Dyson's collapse~\cite{Dyson1952} in the LLL model as at $B=0$, which renders the expansion in the bare $V$ divergent with zero convergence radius. There are currently a number of ways to address the problem, such as, e.g., by an arbitrary dressing of the interaction with the explicit inclusion of the corresponding counter-terms in the expansion~\cite{Wu2017, Rossi2016, Chen2019}, or by a more general homotopy of the microscopic model~\cite{homotopic_action}. Both methods, however, are only supposed to yield physically meaningful results at sufficiently high orders of the expansion. We thus adopt the standard approach of reformulating the expansion in terms of the interaction $W$ that is self-consistently screened by the electron polarisation $\Pi$~\cite{Hedin1965}:
\begin{equation} \label{momentum dependent W}
\begin{split}
    &W_{k_1,k_2;q}(i\nu) = V_{(k_1-k_2)q} \\
    &+ \sum_{k_3} V_{(k_1-k_3)q} \, \Pi_{k_3;q}(i\nu) \, W_{k_3,k_2;q}(i\nu),
\end{split}
\end{equation}
where $\omega$ and $\nu$ are fermionic and bosonic Matsubara frequencies, respectively. Here, the self-energy $\Sigma$ and polarisation $\Pi$ are irreducible (skeleton) expansions in terms of $G$ and $W$ which themselves are to be computed self-consistently~\cite{Hedin1965,AGD}. This formulation is formally exact and could in principle be evaluated to high expansion orders by bold-line DiagMC techniques~\cite{Prokofev1998,VanHoucke2010, Kozik2010, VanHoucke2012, VanHoucke2019} to yield results with controlled error bars. However, the DiagMC approach with self-consistency in two channels poses substantial technical challenges and typically builds upon an accurate deterministic solution of the problem at the lowest order of the diagrammatic expansion~\cite{VanHoucke2019}, which we derive here. This is the so-called self-consistent $GW$ approximation, where
\begin{equation} \label{momentum-dependent self energy}
    \Sigma_k(i\omega) = -\frac{1}{\beta}\sum_{i\nu}\int \frac{dq}{2\pi} W_{k-q,k-q;q}(i\nu)G_{k-q}(i\omega-i\nu)
\end{equation}
and
\begin{equation} \label{momentum-dependent polarisation}
    \Pi_{k_3;q}(i\nu) = \frac{1}{\beta}\sum_{i\omega} G_{k_3+q}(i\omega+i\nu)G_{k_3}(i\omega).
\end{equation}
Equations \eqref{Dyson}, \eqref{momentum dependent W} \eqref{momentum-dependent self energy} and \eqref{momentum-dependent polarisation} form a closed set, which we solve here by iterations without any additional approximations. For this we utilise the discrete Lehmann representation \cite{DLR2022}, which enables compact representation of the frequency dependence of all quantities.

While the truncation to $GW$ is uncontrolled in general, it can be shown that at high temperature, the $GW$ diagram is leading order in $T$. This is a consequence of the long-range nature of the Coulomb interaction, and is similar to the dominance of the same diagrams in the zero magnetic field electron gas \cite{Bruus2004}.
The same is true for large $\mu$, and hence also for large (and small) filling fractions. 

Tao and Thouless solved the $GW$ equations at zero temperature, by approximating the self-energy to be frequency independent \cite{TaoThouless1983,Tao1984}. This is similar to the quasi-particle self-consistent $GW$ approximation \cite{Schilfgaarde2006,Kotani2007,Huser2013}. The resulting states, named TT states, were found to have microscopic (in the sense that their wavelength $\lambda \ll \ell_B$) periodic modulations of the charge density, and a charge gap for all fractional fillings, including those with even denominators. Details of the TT approach, and the related thin-torus limit in which the TT states are the exact ground states, are given in Appendix \ref{TT appendix}. We find, however, that including the full dynamics due to the frequency dependence destroys the TT states in the thermodynamic limit, as shown below.

\section{\label{sec: GW homogeneous} Homogeneous solution}

In this section we confine ourselves to the generic case of transitionally symmetric solutions and consider the possibility of breaking this symmetry separately in the next section. Due to our choice of the gauge, the $y$-component of the momentum $k$ plays an unusual role. The lack of the translational invariance in the $x$-direction implies that the system becomes inhomogeneous whenever $G$ features a non-trivial dependence on $k$.
This can be seen by considering the density operator in the momentum space
\begin{equation} \label{density definition}
    \begin{split}
        \rho\left(\bf q \right) = \int d^2 r~ e^{-i\bf q\cdot \bf r} \phi^\dagger(\bf r) \phi(\bf  r),
    \end{split}
\end{equation}
where $\phi({\mathbf r}) = \sum_{k} \psi_k({\mathbf r}) c_k$, which yields
\begin{equation}
    \rho\left( \bf q \right) = e^{-\frac{1}{4}{\bf q}^2\ell_B^2} \int\frac{dk}{2\pi} ~e^{-iq_xk\ell_B^2}c_{k-q_y/2}^\dagger c_{k+q_y/2}.
\end{equation}
If the $y$-translational symmetry is unbroken, then $\langle c^\dagger_{k+q_y/2}c_{k-q_y/2} \rangle = n_k 2\pi\delta(q_y)$,  where $n_k=\langle c^\dagger_{k}c_{k} \rangle$ is the momentum distribution. This leads, e.g., for a single mode along the $x$-axis $n_k = e^{iQk}n$ to 
\begin{equation}
    \langle \rho\left(\bf q\right)\rangle = 2\pi ne^{-\frac{1}{4}q_x^2\ell_B^2}\delta(q_y)\delta(q_x\ell_B^2-Q)
\end{equation}
which corresponds to a state which is uniform in the $y$-direction but has CDW order with wavevector $Q$ in the $x$-direction.

If the solution is translationally invariant, the resulting momentum independence of $G$ substantially simplifies Eqs.~\eqref{momentum dependent W}, \eqref{momentum-dependent self energy} and \eqref{momentum-dependent polarisation}. Since in this case $W$ depends only on the difference between the leg momenta $k = k_1-k_2$, 
\begin{equation}
     W_{kq}(i\nu) = \int dp~ e^{-ipk}\frac{U(p,q)}{1-\Pi(i\nu)U(p,q)}
\end{equation}
with $U(p,q) = \frac{1}{\sqrt{p^2+(q\ell_B^2)^2}}e^{-\frac{1}{2\ell_B^2}\left(p^2+(q\ell_B^2)^2\right)}$, the polarisation and self-energy become 
\begin{equation} \label{bubble}
    \Pi(i\nu) = \frac{1}{\beta}\sum_{i\omega}G(i\omega+i\nu)G(i\omega)
\end{equation} and 
\begin{equation} \label{momentum independent self energy}   \Sigma(i\omega) =-\frac{1}{\beta}\sum_{i\nu}G(i\omega-i\nu)\mathcal{W}(i\nu)\end{equation}
where 
\begin{equation}\label{momentum independent W}
    \mathcal{W}(i\nu) = \ell_B^2\int_0^\infty d\rho~ \rho  \frac{U(\rho)}{1-\Pi(i\nu)U(\rho)}
\end{equation}
is a totally momentum independent effective interaction with $U(\rho) = e^{-\frac{1}{2}\rho^2\ell_B^2}/\rho\ell_B^2$.

Let us first show that all solutions of Eqs.~\eqref{Dyson}, \eqref{bubble}, \eqref{momentum independent self energy}, \eqref{momentum independent W} are necessarily gapless in the $T=0$ limit. Closely following the argument of Ref.~\cite{Sachdev1993} developed for the SYK model, let us assume that the Green's function has a gap $\Delta$, i.e. that the corresponding spectral function $\rho(\omega) = -\frac{1}{\pi}\operatorname{Im} G(\omega+i\eta)$ satisfies $\rho(\omega)=0$ in some range of frequencies $|\omega| < \Delta/2$. By performing a spectral decomposition of Eqs. \eqref{bubble} and \eqref{momentum independent self energy}, it follows that $\mathcal{W}$ has a gap $2\Delta$ and that $\Sigma$ has a gap $3\Delta$, but then the Dyson equation \eqref{Dyson} implies that the Green's function must have a gap $3\Delta$. Hence the only consistent value is $\Delta = 0$ and the solution is gapless. 

Tunnelling experiments indicate that at low temperature the electron spectral function exhibits a pseudo-gap (i.e. a strong suppression) at low frequencies \cite{Ashoori1990,Ashoori1993,Eisenstein1992,Brown1994,Eisenstein1995,Eisenstein2016}. This behaviour has been found theoretically within the composite Fermion picture, which yields an exponential suppression of the spectral function, $\rho(\omega) \propto e^{-\omega_0/|\omega|}$, with some energy scale $\omega_0>0$~\cite{HePlatzmanHalperin1993,Kim1994}. Following a similar argument to that given above, it is possible to show that this spectral function is also inconsistent with the $GW$ equations.

\subsection{Non-Fermi liquid solution}

The set of equations \eqref{Dyson}, \eqref{bubble}, \eqref{momentum independent self energy} and \eqref{momentum independent W} can be solved analytically for low frequencies and $T \to 0$, resulting in the nFL solution of the form \eqref{nFL self-energy}.
It corresponds to a broad singularity in the spectral function $\rho(\omega) = A|\omega|^{2\Delta - 1}$ at low frequencies, which indicates an absence of well-defined electron-like quasi-particles, or, equivalently, an infinitesimally short quasi-particle lifetime. This self-energy, and thus the corresponding single-electron excitation spectrum, is the same as that of the nFL solution realised in the SYK model. 

For $\omega \to 0$ the exponent $\Delta$ solves the equation (the calculation is outlined in Appendix \ref{Anomalous dimension appendix})
\begin{equation}\label{Delta equation}
    f(\Delta) + g(\Delta)\log(\omega) = 1 
\end{equation}
with $f(\Delta)$ and $g(\Delta)$ being functions of $\Delta$ only, and 
\begin{equation} \label{g function}
g(\Delta) = \frac{(4\Delta-1)^2}{(2\Delta-1)\left[\sec(2\pi\Delta)-1 \right]}\frac{\sin(\pi\Delta)^2}{\sin(\pi\Delta-\theta)\sin(\pi\Delta+\theta)}.
\end{equation}
Thus $g(\Delta) \to  0$ for $\omega \to 0$, which yields $\Delta = 1/4$ (and $f(\Delta) = 0$) for all $\theta$. The independence of $\Delta$ on the filling fraction is a non-trivial result, and in this case, is a consequence of the logarithmic corrections. To see this, note that the log in the second term on the RHS of Eq.\eqref{Delta equation} implies that the solutions to Eq.\eqref{Delta equation} are given by the roots of the function $g(\Delta)$, which are robust with respect to variations of $\theta$. Without the log-term in Eq.\eqref{Delta equation}, the solution $\Delta$ is no longer a root of the function $g(\Delta)$, and therefore it acquires some weak-dependence on $\theta$. 

In practice, the asymptotic form Eq.~\eqref{Delta equation} is applicable only for frequencies as small as $\omega \ll 10^{-6}$.
In more practical frequency regimes we must solve the $GW$ equations numerically. To this end, we examine the functions 
\begin{equation} \label{log derivative function}
    \Delta_{\Sigma}(i\omega) = \frac{\partial \log\left(|\Sigma(i\omega)|\right)}{\partial \log(\omega)}, ~ \Delta_{G}(i\omega) = \frac{\partial \log\left(|G(i\omega)|\right)}{\partial \log(\omega)}
\end{equation}
against frequency at the lowest accessible temperatures (in practice, $T=2 \times 10^{-5}$) and at half-filling ($\mu =\theta = 0$). Since $G$ and $\Sigma$ are related by the Dyson equation \eqref{Dyson}, $\Delta_G$ and $\Delta_\Sigma$ will lead to the same estimate of $\Delta$ for sufficiently small $\omega$. 

\begin{figure}[h]
\includegraphics[trim={2 2 25 37},clip,width = 0.5\textwidth]{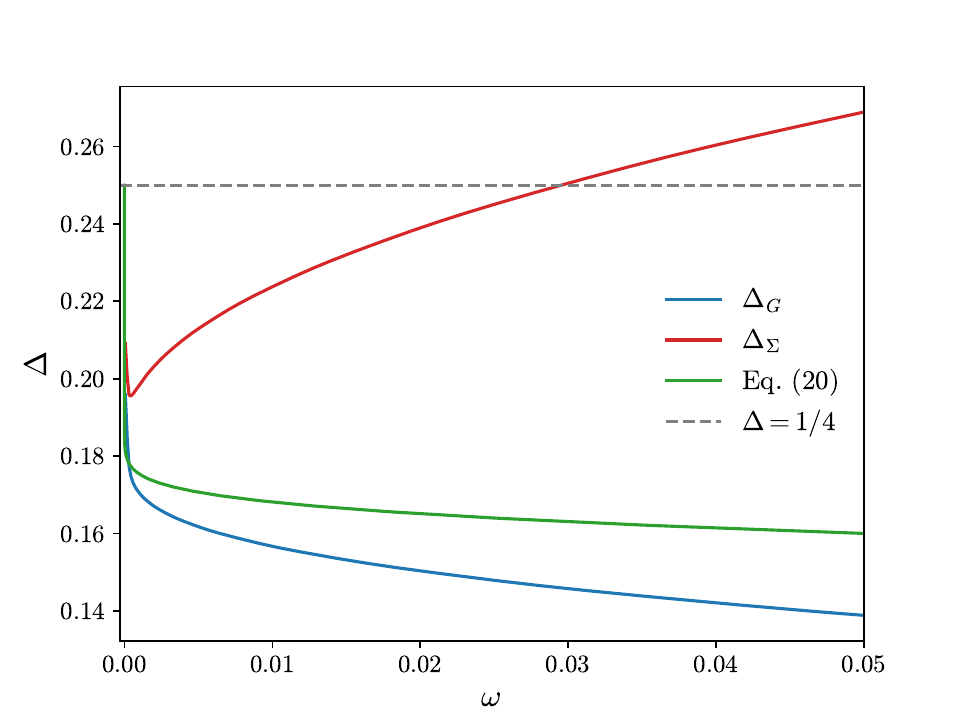}% Here is how to import EPS art
\caption{\label{fig:omega dependence of exponent} 
The exponent $\Delta$ estimated from 
the functions $\Delta_G$ (blue curve) and $\Delta_\Sigma$ (red curve) defined in Eq.~\eqref{log derivative function} at half-filling $\nu = 1/2$ ($\mu,\theta = 0$). This is shown for temperature $T = 2 \times 10^{-5}$. Finite temperature effects are seen at small frequencies $T \lesssim\omega\lesssim  10T$ as rapid up-turns.
As $\omega \to 0$, the two curves approach asymptotically. Also shown (green curve) is the value of $\Delta$ that solves Eq.\eqref{Delta equation}. At arbitrarily low-frequency, all three curves will coincide, and approach the value $\Delta = 1/4$ as $\omega \to 0$. }
\end{figure}

The result is shown in Fig.~\ref{fig:omega dependence of exponent}. It is clear that, for realistic frequencies, the deviation of the apparent $\Delta(\omega)$ from $1/4$ due to the logarithmic in frequency corrections is substantial, meaning that the asymptotic value $\Delta=1/4$ is not observable in practice. 
There are also notable differences between the estimates of $\Delta$ from the asymptotics of $G(i\omega)$ and $\Sigma(i\omega)$, stemming from $G_0$ in the Dyson equation. As the frequency is lowered, however, this difference is seen to eventually become less important compared to the remaining logarithmic corrections. 

The nFL character (i.e. the anamolous scaling of $\Sigma(i\omega)$) of the solution will only be present at sufficiently low temperatures. At sufficiently high temperature, the system is in a free-fermion thermal state.
To estimate the crossover temperature to the nFL regime, we consider the function $\delta\Sigma = \operatorname{Im}\Sigma(i\omega_0)-\operatorname{Im}\Sigma(i\omega_1)$ where $\omega_0 = \pi T$ and $\omega_1 = 3\pi T$ are the lowest and first Matsubara frequencies respectively, and adopt the convention in which a metallic state corresponds to $\delta \Sigma <0$, while the thermal state has $\delta \Sigma >0$. In these terms, the crossover temperature $T_\mathrm{nFL}$ is defined by the equation $\delta \Sigma = 0$, which yields $T_\mathrm{nFL} \sim 0.1 e^2/\ell_B$ roughly independent of the filling fraction, as indicated in Fig. \ref{fig:T density phase diagram}.

The appearance of SYK physics in the microscopic description of the LLL could be seen as surprising. Qualitatively, a connection can be made in terms of the constraints placed on the Green's function by the requirement of a homogeneous charge-density -- which imply that all single particle properties are momentum independent -- and the flat band nature of the non-interacting physics. Mathematically, the form of the SYK self-energy
\begin{equation}
\Sigma_{\text{SYK}}(i\omega) = - \frac{U^2}{\beta}\sum_{\nu}G(i\omega-i\nu)\Pi(i\nu), \label{Sigma_SYK} 
\end{equation}
where $U$ is the SYK interaction strength and $\Pi$ is the polarisation defined in Eq.~\eqref{bubble}, corresponds to that for the LLL, Eq.~\eqref{momentum independent self energy}, with the single particle-hole bubble $\Pi$ in place of the full screened Coulomb interaction $W$. Therefore the SYK character of the $GW$ solution can be understood in terms of the dominance of the effect of the single-bubble contribution in $\mathcal{W}$, up to the logarithmic corrections due to the long-range nature of the bare Coulomb potential.
Our preliminary calculations suggest that a similar nFL state is also present for the Trugman-Kivelson potential, for which the Laughlin wavefunction is the exact ground state \cite{Trugman1985}. A more focused study for this potential is left for future work.

While we have demonstrated that the single-particle exciation spectrum is (nearly) identical to that of the SYK model, it is possible that the multi-particle excitation spectrum, which is not investigated here, is distinct from that of the SYK model. For example the diagrammatic expansion for the density-density correlator -- which determines the spectrum of two-particle excitations -- that is thermodynamically consistent (in the Baym-Kadanoff sense \cite{Baym1961}) with the $GW$ self-energy contains vertex corrections, while for the SYK model the vertex corrections vanish due to the large-$N$ limit \cite{Sachdev1993,Parcollet1999}. Investigation of the two-particle spectrum for the current model is however beyond the scope of the present work.

\subsection{Phase diagram}
\begin{figure}
    \centering
    \includegraphics[trim={2 2 25 37},clip,width = 0.5\textwidth]{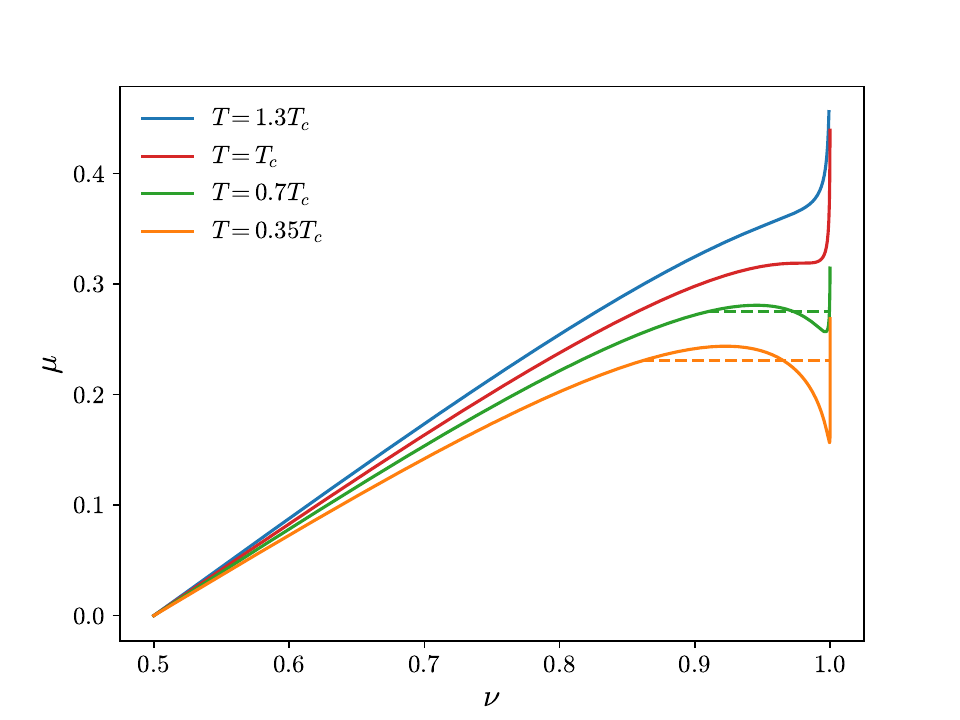}
    \caption{Chemical potential $\mu$ against filling $\nu$ for various temperatures. The dashed lines, obtained using Maxwell's construction, show the value of $\mu$ at which the density jumps discontinuously as $\mu$ is varied in equilibrium. 
    }
    \label{fig:density against mu}
\end{figure}

Similarly to the SYK model \cite{SYKPhaseDiagram2018,PhasesofMelonicQM2019}, we find that a first-order transition to a band insulating phase, corresponding to a fully-filled LLL, is triggered at some finite  value $\mu_\text{nFL-BI}$.  
In Fig. \ref{fig:density against mu}, we show the chemical potential against density for different temperatures. The curves are obtained by fixing the charge density $\nu$ and calculating the corresponding value of $\mu$. We observe the typical behaviour of a first-order transition:  below some critical temperature, we find a range of $\mu$ values for which three different solutions $\nu(\mu)$ exist. The solution with the intermediate charge density has a negative compressibility $K \propto \frac{d\nu}{d\mu}<0$ and is therefore thermodynamically unstable.  
The chemical potential $\mu_\text{nFL-BI}$, at which $\nu$ jumps discontinuously, is found via Maxwell's construction (see appendix \ref{Maxwell appendix} for details), depicted in Fig. \ref{fig:density against mu} as the horizontal dashed line. 

\begin{table*}[t]
\begin{tabular}{c| c c c | c c c c | c c c c } 
 \hline
 & $T_c$ & $\mu_c$ & $\nu_c$ & $\tilde{\gamma}_{+}$ & $\tilde{\gamma}_{-}$ & $q_{+}$ & $q_{-}$ & $\tilde{\gamma}^{\mathrm{c}}_{+}$ & $\tilde{\gamma}^{\mathrm{c}}_{-}$ & $q^{\mathrm{c}}_{+}$ & $q^{\mathrm{c}}_{-}$ \\ [0.1cm] 
 \hline
 $\text{SYK}_4$ & 0.068372(2)& 0.344713(3) & 0.9498(2) & 0.66(1) & 0.66(1) & 0.33(1) & 0.33(1) & -0.30(2) & -0.35(4) & -0.6(1) & -0.65(5) \\ 
 LLL & 0.04735(2) & 0.3190(1) & 0.3190(1) & 0.66(1) & 0.66(1) & 0.33(1) & 0.33(1) & -0.29(2) & -0.35(5) & -0.65(5) & -0.6(2)
\end{tabular}
\caption{Critical parameters $(T_c,\mu_c,\nu_c)$, critical exponents $(\tilde{\gamma}_{+},\tilde{\gamma}_{-},q_{+},q_{-})$ and the corresponding sub-leading correction exponents $(\tilde{\gamma}^{\mathrm{c}}_{+},\tilde{\gamma}^{\mathrm{c}}_{-},q^{\mathrm{c}}_{+},q^{\mathrm{c}}_{-})$ for both $\text{SYK}_4$ and LLL models. The subscript on the exponents specifies the direction in which the crtical point is approached. }
\label{critical point table}
\end{table*}

\begin{figure}[b]
\includegraphics[trim={10 2 25 40},clip,width = 0.5\textwidth]{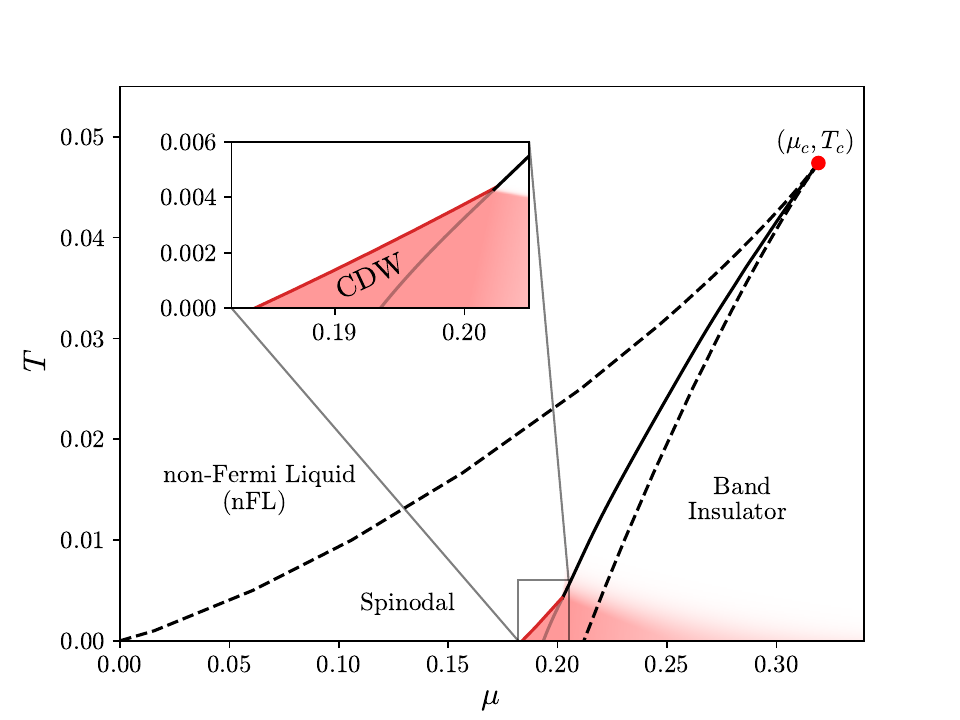}
\caption{\label{fig:T mu phase diagram} Phase diagram in the $T-\mu$ plane. The solid black line marks the first order transition between the non-Fermi liquid and  band-insulator  phases at chemical potential $\mu_\text{nFL-BI}(T)$, which terminates at a critical end-point $(\mu_c,T_c)$ (red point). In the spinodal region, bounded by the dotted lines, both phases can be realised, however one is metastable. The second-order transition from the nFL to a charge-density-wave is seen at low temperatures. On the band-insulator side, we expect the CDW to also be present, and for this to emerge from the band-insulator phase via a first-order phase transition. We have not determined the location of this transition, but indicate its presence by the blurred region. }
\end{figure}

The phase diagram in the $T-\mu$ plane is shown in Fig. \ref{fig:T mu phase diagram} (black solid and dotted curves). We focus solely on positive $\mu$, since the negative $\mu$ behaviour can be inferred from the particle-hole symmetry. In the small-$\mu$ region we find the nFL phase discussed in the previous section. In the large $\mu$-region, the Green's function is exponential in imaginary time, $G(\tau>0) \sim -e^{\mu\tau}$, for $\mu/T \gg 1$, which corresponds to a band-insulating phase with unity filling of the LLL. The two phases are separated by a first-order transition line, which ends at the second-order critical end point $(\mu_c, T_c)$. At all temperatures below $T_c$, there is some finite range of $\mu$, bounded by the dotted curves and labelled the spinodal region, in which both the nFL and BI solutions are found, but one is metastable.
The critical parameters are $T_c = 0.04735(2)$, $\mu_c = 0.3190(1)$, which correspond to $\nu_c = 0.979(1)$. 

The phase diagram in the $T-\nu$ plane is shown in Fig. \ref{fig:T density phase diagram}. The solid black curve delimits the phase-separation region in which the nFL and band-insulating phases coexist. The critical point lies at the top of the phase-separation `dome'.

\subsection{Critical point of the LLL and SYK models}
We now focus on the critical point of the continuous transition, determine its universality class and compare the results to those for the SYK model. To this end, we calculate the critical exponents $\tilde{\gamma}_{\pm}$ of the compressibility $\chi(T,\mu) = \frac{\partial n }{\partial \mu}$ as a function of temperature, defined as 
\begin{equation} \label{gamma exponent definition}
    \chi(t,\mu_c) \sim |t|^{-\tilde{\gamma}_{\pm}}~\text{for } t \to 0^{\pm}
\end{equation}
where $t = (T-T_c)/T_c$ is the reduced temperature, as well as the exponents $q_{\pm}$ for the density 
\begin{equation}
    |\nu(T,\mu_c)-\nu(T_c,\mu_c)| \sim |t|^{q_\pm}~\text{for } t \to 0^{\pm}.
\end{equation}

\begin{figure}
\includegraphics[trim={11 6 25 37},clip,width = 0.5\textwidth]{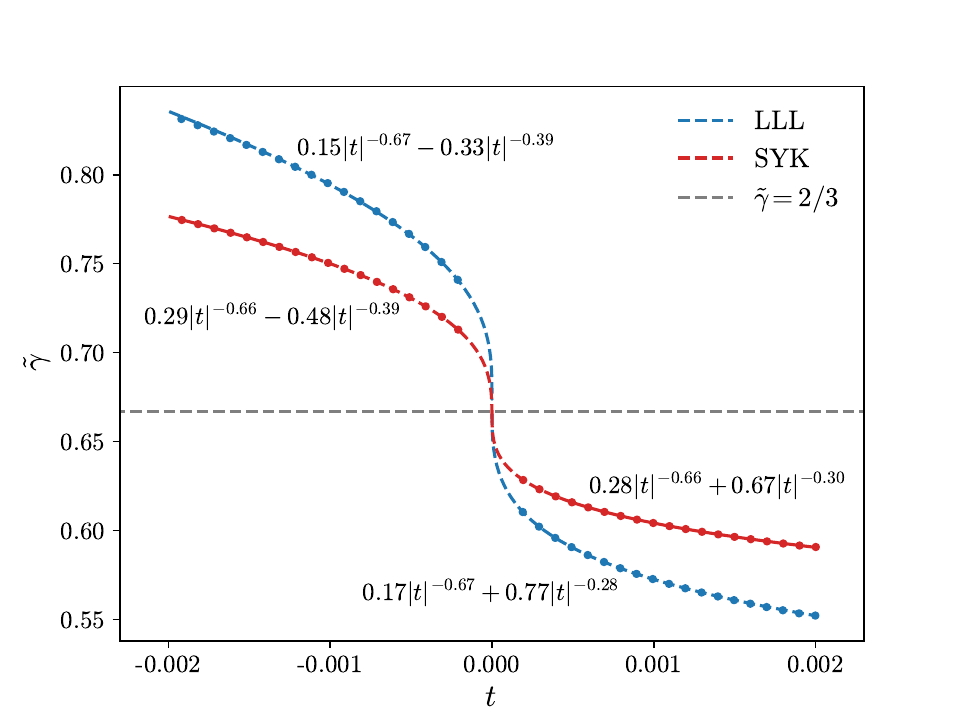}
\caption{\label{fig: critical exponents} Estimate of the susceptibility critical exponents $\tilde{\gamma}_{\pm}$. The function $\tilde{\gamma}(t)$, defined in Eq.~\eqref{exponent via log derivative}, is plotted against the reduced temperature $t=(T-T_c)/T_c$ at the critical chemical potential $\mu_c$ for the lowest Landau level model (blue) and SYK model (red). The dashed lines are the fits by Eq.~\eqref{fitting function}, revealing that the critical exponents for both LLL and SYK models and positive and negative $t$ are identical, and are in perfect agreement with the value $\tilde{\gamma}=2/3$ obtained analytically for the $\text{SYK}_q$ model in the $q \to \infty$ limit in Ref.~\cite{Louw2023}. The fitting parameters for $f_\mathrm{fit}(t)$, defined in Eq.~\eqref{fitting function}, are shown next to their respective curves. 
The corrections to the critical scaling are asymmetric around $t=0$, and $\tilde{\gamma}(t)$ approaches the limiting value $\tilde{\gamma}$ with infinite slope.}
\end{figure}

The critical exponents can be evaluated by considering their temperature-dependent estimates
\begin{equation} \label{exponent via log derivative}
\tilde{\gamma}(t) = \frac{\partial \log \chi(t,\mu_c)}{\partial \log t},
\end{equation}
and similarly for $q(t)$. In previous studies of the critical point in the SYK model \cite{SYKPhaseDiagram2018,PhasesofMelonicQM2019}, the exponents $\tilde{\gamma}_{\pm}$, $q_{\pm}$ were obtained by a linear extrapolation of $\tilde{\gamma}(t)$, $q(t)$ to $t \to 0^{\pm}$. Since we have very small error bars on the critical values $\mu_c, T_c,\nu_c$, we are able to approach the critical point much more closely, observing that the limiting values of the critical exponents are actually approached with infinite slope (see Fig.~\ref{fig: critical exponents}). This results in significant error in the estimate if a linear extrapolation is performed. We use instead the fitting ansatz
\begin{equation} \label{fitting function}
    f_\text{fit}(t) = a_{\pm}|t|^{\alpha_\pm} + b_\pm|t|^{\beta_\pm} ~~\text{for}~~ t \to 0^{\pm} 
\end{equation}
which includes the leading-order corrections to the critical behaviour.  
To further reduce systematic errors from the use of a finite range of $t$ in the fit, we extract the fit parameters (including the value of $T_c$) for a set of different ranges of $t$ (with order of magnitude roughly $t \sim 10^{-4}$) and choices of $\mu_c$ within its error bar, and extrapolate the resulting values to the $t\to 0$ limit; the uncertainty of this extrapolation together with the spread of the data with $\mu_c$ is the source of the claimed error bars. An example of such a fit in the range $t \in [2\times 10^{-4},2\times 10^{-3}]$ is shown in Fig.~\ref{fig: critical exponents}. In this way, we obtain the critical exponents with high accuracy, as well as the exponents characterising the leading-order corrections. Since in Refs.~\cite{SYKPhaseDiagram2018,PhasesofMelonicQM2019} the exponents for the SYK model have been found to differ on the two sides of the transition, we calculate both independently.

The location of the critical point, as well as the critical exponents and subleading exponents for the LLL and $\text{SYK}_4$ models are summarised in Table~\ref{critical point table}. For the LLL model we find the values $\tilde{\gamma}_{+} = 0.66(1)$, $\tilde{\gamma}_{-} = 0.66(1)$ and $q_{+} = 0.33(1)$, $q_{-} = 0.33(1)$. These exponents are different from those reported previously for the $\text{SYK}_4$ model.
However, by evaluating the $\text{SYK}_4$ critical parameters with error bars at least two orders of magnitudes smaller than the previous estimates [$T_c^{\text{SYK}} = 0.068372(2)$, $\mu_c^{\text{SYK}} = 0.344713(3)$ and $\nu_c^{\text{SYK}} = 0.9498(2)$], and using the fit \eqref{fitting function}, we find the $\text{SYK}_4$ exponents $\tilde{\gamma}^{\text{SYK}}_{+} = 0.66(1)$, $\tilde{\gamma}^{\text{SYK}}_{-} = 0.66(1)$ and $q^{\text{SYK}}_{+} = 0.33(1)$, $q^{\text{SYK}}_{-} = 0.33(1)$, which are identical to our LLL exponents.
Interestingly, while the exponents (both critical and correction) as well as the leading coefficients are symmetric around the critical point, we find that the coefficients of the sub-leading terms are asymmetric.
This is likely the cause of the apparent asymmetry of the critical exponents obtained previously by linear fitting.

Very recently, an analytical study of the generalised $\text{SYK}_q$ model 
has been performed in the $q \to \infty$ limit \cite{Louw2023}. The resulting exponents are identical to those obtained here for the LLL and SYK ($\text{SYK}_{4}$ in these notations) models, and were found to belong to the Van der Waals/Ising mean-field universality class. It is emphasised by the authors of Ref. \cite{Louw2023} that the critical exponents depend upon the direction in the $T,\mu$ plane along which the critical point is approached. This is the reason for the difference between the exponents given here and those usually quoted for this universality class (for more details on this topic, the reader is referred to Ref. \cite{Louw2023}). Therefore we also conclude that both the LLL and the $\text{SYK}_{4}$ models belong to this universality class, and it is very likely that this is also true for all possible $q$ by continuity. A proper investigation of all $\text{SYK}_q$ critical points is beyond the present scope.

\section{GW CDW instability} \label{sec: GW CDW}
Having established the nature of the homogeneous phase, we now analyse the possibility of CDW instabilities within the $GW$ approximation. 
We detect the occurrence of a continuous phase transition to a charge-ordered phase by the divergence of the charge susceptibility $\chi_p$ at the relevant wavevector $p$. 
We obtain $\chi_p$ via the Bethe-Salpeter equation, which is expressed in terms of the two-particle irreducible vertex $\Gamma$ (see appendix \ref{HF appendix} for the definition of the susceptibility). Since the $GW$ approximation is $\Phi$-derivable, calculating $\Gamma$ via the functional derivative
\begin{equation}
    \Gamma = \frac{1}{T}\frac{\delta \Sigma}{\delta G}
\end{equation}
ensures thermodynamic consistency \cite{Baym1961}. Determining the vertex in this way also allows a different interpretation of the divergence of the susceptibility: it signals the switchover from iterative stability to instability of a self-consistent solution of the Dyson equation. This is easily seen by linearising the Dyson equation around a solution. 
This viewpoint has also been adopted within dynamical mean field theory, wherein the variation with respect to the hybridisation is considered \cite{vanLoon2020,vanLoon2022}.
Therefore our expression for the irreducible vertex within the $GW$ approximation is represented diagrammatically as 
\begin{widetext}
\begin{equation} \label{ring}
\begin{split}
\Gamma_{\omega\omega',kk'}~~=
\begin{gathered}
    \begin{tikzpicture}
    \begin{feynman}
    \vertex (c);
    \vertex [above = 0.5cm of c] (d) {\(k'\)};
    \vertex [below = 0.5cm of c] (e) {\(k\)};
    \vertex [right = 0.5cm of c] (a);
    \vertex [right = 1.5cm of a] (t);
    \vertex [right = 0.5cm of t] (y);
    \vertex [above = 0.5cm of y] (h) {\(k'\)};
    \vertex [below = 0.5cm of y] (g) {\(k\)};
    \diagram*{
    (a) -- [scalar,edge label = $k-k'$] (t),
    (a) -- [fermion] (d),
    (e) -- [fermion] (a),
    (h) -- [fermion] (t),
    (t) -- [fermion] (g),
    };
    \end{feynman}
    \end{tikzpicture}
    \end{gathered} 
    ~~+~~
    \begin{gathered}
    \begin{tikzpicture}
    \begin{feynman}
    \vertex (c);
    \vertex [below = 0.5cm of c] (e){\(k\)};
    \vertex [right = 0.5cm of c](a);
    \vertex [right = 1.5cm of a] (t);
    \vertex [right = 0.5cm of t] (y);
    \vertex [below = 0.5cm of y] (g){\(k\)};
    \vertex [above = 1.5cm of a] (z);
    \vertex [right = 1.5cm of z] (w);
    \vertex [above = 2cm of c] (b){\(k'\)};
    \vertex [above = 2cm of y] (d){\(k'\)};
    \diagram*{
    (e) --[fermion] (a),
    (t) -- [fermion] (g),
    (a) -- [fermion,edge label = $k-q$] (t),
    (a) -- [scalar,edge label = $q$] (z),
    (w) -- [scalar,edge label = $q$] (t),
    (w) -- [fermion,edge label' = $k'-q$] (z),
    (z) -- [fermion] (b),
    (d) -- [fermion] (w)
    };
    \end{feynman}
    \end{tikzpicture}
    \end{gathered}
    ~~+~~
    \begin{gathered}
    \begin{tikzpicture}
    \begin{feynman}
    \vertex (c);
    \vertex [below = 0.5cm of c] (e){\(k\)};
    \vertex [right = 0.5cm of c](a);
    \vertex [right = 1.5cm of a] (t);
    \vertex [right = 0.5cm of t] (y);
    \vertex [below = 0.5cm of y] (g){\(k\)};
    \vertex [above = 1.5cm of a] (z);
    \vertex [right = 1.5cm of z] (w);
    \vertex [above = 2cm of c] (b){\(k'\)};
    \vertex [above = 2cm of y] (d){\(k'\)};
    \diagram*{
    (e) --[fermion] (a),
    (t) -- [fermion] (g),
    (a) -- [fermion,edge label' = $k-q$] (t),
    (a) -- [scalar] (w),
    (t) -- [scalar] (z),
    (w) -- [fermion,edge label' = $k'+q$] (z),
    (z) -- [fermion] (b),
    (d) -- [fermion] (w)
    };
    \end{feynman}
    \end{tikzpicture}
    \end{gathered}
\end{split}
\end{equation}
\end{widetext}
where the dotted line represents $W$. In this expression, we set the transfer frequencies and momenta to zero since only these components are required to study the charge-density ordering instability. 
The index $k$ corresponds to position in the $x$-direction, and therefore we study the susceptibility as a function of the wavenumber $p$, which is the Fourier conjugate to $k$. Since we have a translation-invariant normal phase, implying dependence on only the difference $k - k'$, the Bethe-Salpeter equation for $\chi$ is diagonal in the wavenumber basis.

\begin{figure}
\includegraphics[width = 0.5\textwidth]{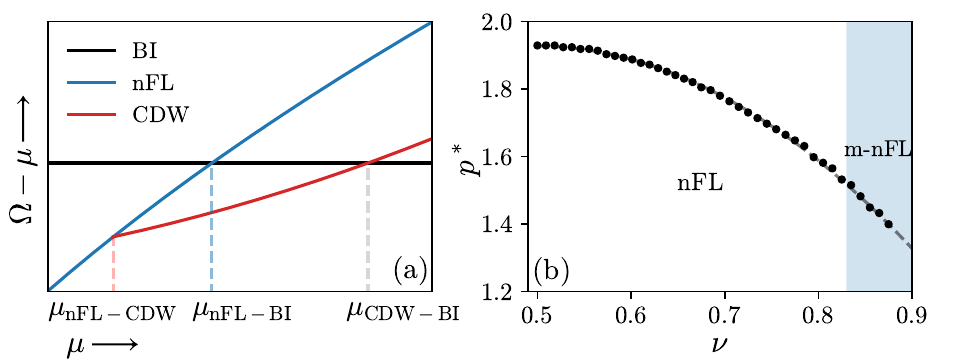}
\caption{\label{fig:ordering momentum} (a) Schematic illustration of the variation of the grand potential $\Omega$ with the chemical potential at constant (arbitrary) temperature $T$. The CDW curve must intersect the band-insulator (BI) curve at a chemical potential $\mu_\text{CDW-BI}\geq\mu_\text{nFL-BI}$, thus justifying the existence of the blurred region in Fig.\ref{fig:T mu phase diagram} wherein the CDW exists. (b) The momentum $p^*$ at which the susceptibility is a maximum as a function of filling fraction $\nu$ at fixed temperature $T = 0.005$. In the shaded region, there are CDW transitions from the metastable nFL phase. In the pure nFL phase, there are no CDW transitions at this temperature. This shows no special features of the density fluctuations as the filling is varied.}
\end{figure}

The obtained critical temperature for the nFL-to-CDW transition as a function of the chemical potential is shown in Fig. \ref{fig:T mu phase diagram} (red curve). We observe a small region at low temperature adjacent to the nFL-to-BI transition in which CDW order emerges. 
For $\mu \lesssim 0.18$ (corresponding to $0.5 \leq \nu \lesssim 0.8$), we find that the susceptibility roughly behaves as $\chi^{-1} \sim aT + C$, where $a,C$ are positive constants (with $C \to 0$ as $\mu \to 0.18$ or $\nu \to 0.8$), at the momentum at which $\chi^{-1}$ is the smallest. This suggests that no CDW transition takes place down to zero temperature below $\mu \approx 0.18$ ($\nu \approx 0.8$). Within the BI phase at $\mu \gtrsim 0.2$ ($\nu \approx 1$, almost independently of $\mu$) and at low temperatures, charge fluctuations are strongly suppressed, which eliminates the possibility of a continuous transition to a CDW in this region of the phase diagram. However, generically, the CDW should be present here too, separated from the BI phase by a first-order transition. This can be seen by considering the grand potentials of all three phases (nFL, CDW and BI, see an illustration in Fig.~\ref{fig:ordering momentum}a). At the nFL-to-CDW transition (red dashed line in  Fig.~\ref{fig:ordering momentum}a and red line in Fig.~\ref{fig:T mu phase diagram}), the grand potential $\Omega_\text{CDW}$ of the CDW phase emerges below that of the nFL phase $\Omega_\text{nFL}$. This means $\Omega_\text{CDW}$ must be smaller than the BI grand potential, $\Omega_\text{BI}$, at the location of the first-order nFL-BI transition, where $\Omega_\text{nFL}= \Omega_\text{BI}$ (blue dashed line in Fig.~\ref{fig:ordering momentum}a and black line in the inset of Fig.~\ref{fig:T mu phase diagram}). This implies that in general the CDW will persist into the BI region, undergoing a first-order transition to the BI phase at some chemical potential $\mu_\text{CDW-BI}$ (gray dashed line in Fig.~\ref{fig:ordering momentum}a). Detecting first-order transitions is beyond the present study, and thus we indicate the corresponding boundary of the CDW region only roughly by blurring in Figs.~\ref{fig:T density phase diagram} and \ref{fig:T mu phase diagram}.

The CDW transition line is also shown in the temperature-density plane in Fig.~\ref{fig:T density phase diagram}. Between roughly $\nu = 0.8$ and $0.82$, the CDW emerges from the pure nFL phase via a second order phase transition. In the phase separation region $\nu \approx 0.82$ to $\nu \approx 1$, our calculated Green's functions, which correspond to a pure phase, do not capture the equilibrium properties of the system. Within the metastable nFL phase, the transition to the CDW is shown in Fig.~\ref{fig:T density phase diagram} by the dashed red line. Generically, from the analysis of the grand potentials of the corresponding phases (Fig. \ref{fig:ordering momentum}a), we expect that the ground state in the phase-separation region is a CDW, with a (first-order) transition temperature that depends on the filling fraction. 

We have investigated the wavenumber dependence of $\chi$ at fixed temperature, by plotting the wavenumber $p_*$ at which $\chi$ is maximum as a function of $\nu$, as shown in Fig. \ref{fig:ordering momentum}b at temperature $T = 0.005$. Note that at this temperature, there are no CDW transitions from the pure nFL phase. However, transitions do occur from the metastable nFL phase at certain filling fractions, as marked in Fig.~\ref{fig:ordering momentum}. This shows no special features as $\nu$ is varied, but we note that in contrast to Hartree-Fock, $p_*$ depends on $\nu$. This is because the vertex is self-consistently determined and hence depends on $\nu$. 

The phase diagram obtained in the Hartree-Fock approximation~\cite{FPA1979} is drastically different from our self-consistent $GW$ theory: The Hartree-Fock ground state is a CDW at all filling fractions, with a typical transition temperature an order of magnitude higher than that of the CDW state in $GW$. Because of its dynamically fluctuating nature, the nFL state, which dominates the phase diagram and suppresses the possibility of ordering, is entirely missed by the Hartree-Fock theory. 

We also compare our results to the experimental critical temperatures \cite{Chen2006}. We find that the values of filling and temperature for the onset of the CDW phase are in qualitative agreement with experiment.
Therefore, while the $GW$ approximation does not capture the low-temperature FQH states at odd-denominator fillings in the range $0.2 \lesssim \nu \lesssim 0.8$, qualitative agreement with experiment is achieved outside this range of $\nu$. An explanation for this agreement could lie in the fact that $GW$, being controlled in the large-$\mu$ limit, is intrinsically more accurate at larger filling fractions.

\section{\label{sec: conclusion} Discussion} 
We have analysed the microscopic model of electrons in the lowest Landau level interacting via the Coulomb potential by the self-consistent (skeleton) diagrammatic theory truncated at the level of the $GW$ approximation. Our central observation is a peculiar finite-temperature nFL state, realised in a wide range of LLL filling fractions, $0.2 \lesssim \nu \lesssim 0.8$. The state is qualitatively similar to that found in the paradigmatic SYK model and features a power-law---rather than linear as in a Fermi liquid---frequency dependence of the self-energy at low frequencies. We demonstrate that the anomalous dimension, which characterises this power-law, deviates from that in the SYK model by sizeable logarithmic corrections due to the Coulomb interaction. The SYK-like nFL state exhibits transitions to different insulating states at low temperatures. At large (or small, by particle-hole symmetry) fillings, we observe a first-order transition from the nFL state to a band insulator phase corresponding to a full (or empty) LLL, with a critical end point. The critical point is found to belong to the same universality class as its counterpart in the $\text{SYK}_q$ models for all $q$, which itself is determined to be in the Van der Waals universality class. Near the nFL-BI transition, the nFL state also exhibits a second-order instability towards CDW order in a narrow range of filling fractions $0.8 \lesssim \nu \lesssim 0.82$ and at low temperatures, which are qualitatively consistent with those observed for CDW ordering experimentally.

Notably, in the wide range $0.2 \lesssim \nu \lesssim 0.8$, the nFL phase remains stable down to the ground state and the fractional quantum Hall (FQH) states at odd denominator fillings are not observed within the $GW$ theory. In fact, we demonstrate that the fully self-consistent $GW$ approximation necessarily gives a gapless phase if the solution is assumed translation-symmetric, as is the case for a FQH state. One implication of this fact is that the Tao-Thouless states \cite{TaoThouless1983,Tao1984}, found earlier by a $GW$ theory similar to the quasi-particle self-consistent $GW$ approximation \cite{Schilfgaarde2006,Kotani2007,Huser2013}, are in fact not valid in the fully-self consistent case.

At even-denominator filling fractions, experiments indicate a metallic state at low temperatures, which at first glance seems consistent with our observations at these fillings. However, the nFL state we observe is of a fundamentally different character to the experimentally observed Fermi liquid states of composite Fermions. This is clear from the studies of tunnelling between two parallel electron planes in the high magnetic field limit~\cite{Ashoori1990,Ashoori1993,Eisenstein1992,Brown1994,Eisenstein1995,Eisenstein2016}, which directly probe the spectral function $\rho(\omega)$ of the electrons in the partially filled LLL. It was found that, for the even-denominator compressible states, $\rho(\omega)$ exhibits an `orthogonality catastrophe' at low temperature, i.e. a strong (exponential) suppression of $\rho(\omega)$ at $\omega \to 0$. Similar features are obtained theoretically using the composite Fermion approach \cite{HePlatzmanHalperin1993,Kim1994}, by a classical treatment of the electron liquid \cite{Efros1993} and by approximating the system as a Wigner crystal (which is thought to accurately capture the short-range correlations) \cite{Johansson1993}, as well as numerically for small numbers of electrons \cite{Hatsugai1993}. Despite the suppression of $\rho(0)$, the system exhibits metallic transport properties because the conducting quasi-particles are not the microscopic electrons but rather the composite Fermions, which have little overlap with the electrons \cite{Halperin1994}. From these considerations, it is evident that extending the microscopic theory beyond the $GW$ approximation is necessary to capture the experimentally observed behaviour of the partially-filled LLL at very low temperatures.

Calculations that would systematically include diagrams beyond $GW$, e.g., by means of diagrammatic Monte Carlo techniques~\cite{VanHoucke2010, Kozik2010, VanHoucke2019, Chen2019, Kozik2023combinatorial}, are also needed to evaluate the systematic errors of our results. Nonetheless, since the $GW$ approximation is controlled in the limit of high temperatures and correctly captures the thin-torus limit at all temperatures, we expect that the nFL phase found here at intermediate temperatures should be observable in the LLL system. It would exist within an intermediate range of temperatures below the free-fermion thermal state to non-Fermi liquid crossover temperature $T_{\mathrm{nFL}} \sim 0.1 e^2/\ell_B\sim 20\,\, \mathrm{K}$, but above the temperature at which the correlations present in the FQH state begin to develop, which roughly could be taken as at most the gap scale $T_{\mathrm{gap}} \sim 0.05 e^2/\ell_B\sim  10\,\, \mathrm{K}$ \cite{VillegasRosales2021}, where the temperatures in Kelvin are shown for the typical magnetic field strength $B = 13\,\,\mathrm{T}$.
If this is the case, then the FQH states, and indeed the metallic states at even denominators, could emerge from this state of incoherent electrons as the temperature is lowered. While direct calculation of the resistivity for the LLL model is left for future studies, we conjecture that, experimentally, a signature of this nFL phase could be an anomalous scaling of the resistivity with temperature of the form $\rho_\text{dc}(T) \sim T^{2-4\Delta}$ with $\Delta$ the anomalous dimension defined in Eq.~\eqref{nFL self-energy}, as is found for models of itinerant electrons exhibiting an SYK-like phase \cite{Parcollet1999}.
In the LLL model, the value $\Delta= 1/4$ is only valid in the strict $\omega \to 0$ limit, meaning that at finite temperatures the substantial logarithmic corrections to $\Delta$ would likely lead to observable deviations from the linear-in-$T$ behaviour expected for $\Delta = 1/4$. While experimental results exist for the variation of the resistivity with the temperature \cite{Chang1983,Maryenko2018}, the temperatures ranges are low and the system is in the FQH regime. As far as we are aware, experimental data at higher temperatures does not yet exist in the literature. Our results suggest that the physics at these intermediate temperatures is potentially very rich and its further investigation can shed new light on the nature of the FQH states. 

This work was supported by EPSRC through Grant No. EP/R513064/1.

\appendix

\section{\label{HF appendix} Hartree-Fock CDW Instability}

\subsection{Detecting CDW instability}

The susceptibility of the system can be defined by considering the following extra term in the Hamiltonian
\begin{equation}
    H'(\tau) = -\sum_k\zeta_k(\tau) c_k^\dagger(\tau)c_k(\tau),
\end{equation}
where $\zeta_k(\tau)$ is an momentum dependent external field. Since this is an imaginary-time dependent perturbation, its action is to bring the system out of equilibrium. However we are only interested in the limit $\zeta \to 0 $, which restores thermodynamic equilibrium.
The relevant susceptibility is the density-density correlator
\begin{equation}
\begin{split}
    \chi_{k-k'}(\tau-\tau') = \frac{1}{\beta}\frac{\delta \langle n_k(\tau)\rangle }{\delta \zeta_{k'}(\tau')}\bigg|_{\zeta_{k'}\to 0} =  \langle n_k(\tau) n_{k'}(\tau') \rangle& \\- \langle n_k\rangle \langle n_{k'}\rangle &
\end{split}
\end{equation}
(the numerator $\langle n_k(\tau)\rangle$ is time-dependent because we have not yet taken the limit $\zeta \to 0$). The static part of the Fourier transform (with respect to $k-k'$) $\chi_p(i\nu = 0)$ with wavenumber $p$ diverges at the (second-order) phase transition between the homogeneous and CDW phases.

We calculate the static response $\chi_{k-k'}(i\nu = 0)$ by considering the generalised susceptibility, which in particle-hole notation for the frequency dependence can be expressed in terms of the (particle-hole) irreducible vertex $\Gamma_{ph}$ via the following Bethe-Salpeter equation
\begin{equation} \label{full BSE}
        \chi^{\omega \omega^\prime\nu}=\chi_{0}^{\omega \omega^\prime\nu}-\frac{1}{\beta^2} \sum_{\omega_1\omega_2} \chi_{0}^{\omega \omega_1\nu} \Gamma_{ph}^{\omega_1 \omega_2\nu} \chi^{\omega_2 \omega^\prime\nu},
\end{equation}
where $\chi_{0}^{\omega \omega^\prime\nu} = - \beta G(i\omega)G(i\omega^\prime+i\nu)\delta_{\omega\omega^\prime}$ is the bare susceptibility and we have suppressed momentum indices for now (we use similar notations and definitions to those found in \cite{Rohringer2018}).
Since the CDW response function is static we will only consider the $\nu = 0$ components of $\chi$. As a matrix equation, Eq.\eqref{full BSE} becomes
\begin{equation}
    \chi = \frac{\chi_{0}}{1+T^2\chi_0\Gamma}.
\end{equation}
The vertex $\Gamma_{kk'}$ can be written in terms of $G$ and $\Sigma$ as (considering only the $\nu = 0,q=0$ components)
\begin{equation}
    \Gamma_{\omega\omega',kk'} = \frac{1}{T}\frac{\delta \Sigma_{k,\omega}}{\delta G_{k',\omega'}}.
\end{equation}
Since the normal phase is translation invariant and in the LLL model the momentum $k$ corresponds to position (in the $x$-direction), we have $\Gamma_{kk'} = \Gamma_{k-k'}$, which can be seen by explicit calculation. Therefore, the inverse susceptibility can be written in terms of the wavenumber $p$ as 
\begin{equation} \label{Inverse susceptibility matrix}
    (\chi^{-1})_{\omega\omega',p} = -T\frac{1}{G(i\omega)^2}\left[\delta_{\omega,\omega'} - TG(i\omega)^2\Gamma_{\omega\omega',p}\right].
\end{equation}
If $(\chi^{-1})_{\omega\omega',p}$ has a zero eigenvalue, then this implies that the response function 
\begin{equation} \label{response function definition}
    \chi_p = \frac{1}{\beta^2}\sum_{\omega, \omega'}\chi_{\omega\omega',p}
\end{equation} 
is divergent. We can therefore calculate the eigenvalues of $(\chi^{-1})_{\omega\omega',p}$ and determine the critical point as the point at which an eigenvalue vanishes.

\subsection{CDW in the Hartree-Fock approximation}

At the level of the Hartree-Fock (mean-field) approximation, it is possible to find the critical point analytically without considering the eigenvalues. Since the self-energy is frequency-independent, the self-energy \eqref{First order full self energy} only depends on momentum:
\begin{equation} 
    \xi_{k} = \int \frac{dq}{2\pi} \left(V_{k-q,0}-V_{0,k-q}\right)n_F(\xi_{q}-\mu)
\end{equation}
where $n_F(x) = (e^{\beta x}+1)^{-1}$ is the Fermi function.
Introducing a wavenumber label $p$ and Fourier transforming this equation gives
\begin{equation} \label{HF energies in Fourier space}
 \xi_p = U_p n_p,
\end{equation}
where 
\begin{equation} \label{FT of density}
    n_p = \int \frac{dq}{2\pi} e^{ipq}n_F(\xi_q-\mu)
\end{equation}
and the effective potential is
\begin{equation}
    U_p =  \frac{1}{|p|}e^{-\frac{p^2}{2\ell_B^2}}-\sqrt{\frac{\pi}{2\ell_B^2}}e^{-\frac{p^2}{4\ell_B^2}}I_0\left(\frac{p^2}{4\ell_B^2}\right)
\end{equation}
where $I_0$ is the modified Bessel function of the first kind. We therefore have a set of decoupled modes, which correspond to different wavelength charge density waves, as discussed in Section \ref{Microscopic model}. Specifically, $\Gamma_{\omega\omega',p} = U_p$ in the Hartree-Fock approximation, and we can write the Bethe-Salpeter equation for $\chi$ in terms of the wavenumber $p$ as
\begin{equation}
    \chi_{\omega\omega',p} = -\beta G^2(i\omega)\delta_{\omega\omega'} + \frac{1}{\beta}\sum_{\omega_1}G(i\omega)^2 U_p \chi_{\omega_1\omega',p}
\end{equation}
from which it follows that the physical response $\chi_p = \frac{1}{\beta^2}\sum_{\omega\omega'} \chi_{\omega\omega',p}$ is
\begin{equation} \label{analytic HF susceptibility}
    \chi_p = \frac{\beta \nu (1-\nu)}{1+\beta U_p\nu(1-\nu)}
\end{equation}
which is the usual expression for the RPA density-density response. 
Since $U_p$ has a negative minimum at $p^* =  1.568$ with value $U_{p^*} = -0.557$, we have a critical temperature 
\begin{equation} \label{HF Tc}
    T_c = 0.557\nu(1-\nu)
\end{equation}
below which the system is unstable to the formation of a CDW with wavevector $p^*$.

Using the fact that the critical temperature is $T_c = -U_p\nu(1-\nu)$, we can rewrite Eq.\eqref{analytic HF susceptibility} as 
\begin{equation}
    \chi_p = \frac{\nu(1-\nu)}{T-T_c},
\end{equation}
which gives the critical exponent $\gamma_+ = 1$ where $\chi_p \sim |T - T_c|^{-\gamma_+}$ for $T\to T_c^+$.

In addition to this second order phase transition, it is known that Hartree-Fock also gives first order transitions with a slightly higher transition temperature away from half-filling \cite{FPA1979}. In addition, the ordering momentum is lowered as one moves away from half-filling. 

\section{Tao-Thouless theory and TT limit} \label{TT appendix}

The equations \eqref{Dyson}, \eqref{momentum dependent W} \eqref{momentum-dependent self energy} and \eqref{momentum-dependent polarisation} were solved approximately by Tao and Thouless \cite{TaoThouless1983,Tao1984}. For filling fraction $\nu = \frac{1}{q}$, they approximate the self-energy to be frequency independent and periodic in momentum with period $q$ in units of the momentum spacing (it is not clear how to interpret this in the thermodynamic limit, since then the spacing vanishes). Under these assumptions, it is shown that the system is gapped for all $q$, including for $q$ even.  This shortcoming of the theory as well as some others \cite{Giuliani1985,Thouless1985} meant that this picture was quickly abandoned as a correct description. In addition, their solution is not fully self-consistent, but instead relies on a quasi-particle type self-consistency, which approximates the full self-energy by its lowest-frequency value. By extending their calculation to include full self-consistency, we have found that the frequency dependence of $\Sigma$ destroys the gap and leads to solutions that are momentum independent. Therefore we conclude that the TT-type ansatz is not a valid solution at the level of fully self-consistent $GW$. 

The Tao-Thouless theory is now known to be relevant for the quantum Hall system in the so-called thin torus (TT) limit \cite{Bergholtz2005,Bergholtz_2006,Bergholtz2007,Bergholtz2008}. On a torus with small circumference $L_y \to 0$, the Hamiltonian can be solved exactly and one obtains gapped CDW states for the odd-denominator filling $\nu = 1/q$. These states have the same qualitative properties as the Laughlin states, such as fractionally charged excitations and ground state degeneracy $q$ on the torus. These CDW states are also observed for even denominators in the strict $L_y \to 0$, but as $L_y$ is increased (to roughly $L_y=5$) they are seen to undergo a quantum phase transition to a metallic state of neutral fermions that suggests a one-dimensional luttinger liquid description \cite{Bergholtz2005}.
The CDW states for odd $q$ and the Luttinger liquid states for even $q$ are then conjectured to be adiabatically connected to the states in the experimentally relevant limit $L_y \to \infty$, which is also supported numerically \cite{Seidel2005}.

\section{Non-Fermi liquid solution}\label{Anomalous dimension appendix}
To solve Eqs. \eqref{Dyson}, \eqref{bubble}, \eqref{momentum independent self energy} and \eqref{momentum independent W} we assume a power-law form of the self-energy 
\begin{equation} \label{sigma ansatz}
    \Sigma(i\omega) = \lambda e^{-i\mathrm{sign}(\omega)(\pi/2+\theta)} |\omega|^{1-2\Delta}, ~~\omega \to 0
\end{equation}
with $0<\Delta< 1/2$, $\lambda >0$ and particle-hole asymmetry parameter $-\pi\Delta < \theta < \pi \Delta$. For this range of $\Delta$, $\Sigma$ dominates the bare Green's function in the Dyson equaton, and hence the corresponding Green's function is 
\begin{equation}
    G(i\omega) = -\frac{1}{\lambda}e^{i\mathrm{sign}(\omega)(\pi/2+\theta)} |\omega|^{-(1-2\Delta)},~~\omega\to 0.
\end{equation}
In the time-domain, this corresponds to 
\begin{equation}\label{G time}
    G(\tau) = -\frac{\Gamma(2\Delta)\sin(\pi\Delta-\operatorname{sign}(\tau)\theta)\operatorname{sign}(\tau)}{\pi\lambda |\tau|^{2\Delta}}
\end{equation}
and hence 
\begin{equation}
    \Pi(\tau) = - \frac{\Gamma(2\Delta)^2\sin(\pi\Delta-\theta)\sin(\pi\Delta+\theta)}{\pi^2\lambda^2}\frac{1}{|\tau|^{4\Delta}}.
\end{equation}
Noting the identity
\begin{equation}
    \int_{-\infty}^\infty \frac{1}{|x|^\alpha}e^{i\omega x} dx = \frac{2\sin\left(\frac{\pi\alpha}{2}\right)\Gamma(1-\alpha)}{|\omega|^{1-\alpha}}~~ \text{for}~0<\alpha<1, 
\end{equation}
we see that if $0<\Delta<\frac{1}{4}$, then 
\begin{align}
    \Pi(i\nu) =\int_{-\infty}^\infty d\tau~ e^{i\nu \tau}\Pi(\tau)= -\frac{1}{\lambda^2}C_\Delta |\nu|^{4\Delta-1},~~\nu\to 0
\end{align}
with the constant
\begin{equation}
\begin{split}
    &C_\Delta = \frac{4}{\pi^2}\sin(\pi\Delta)\sin(\pi\Delta-\theta)\sin(\pi\Delta+\theta)\cos(\pi\Delta)\\
    &\hspace{2cm}\times \Gamma(1-4\Delta)\Gamma(2\Delta)^2.
\end{split}
\end{equation}
Note that in order for $\operatorname{Re}\Pi(i\nu)$ to be strictly negative (as required by the spectral representation), $C_\Delta$ must be positive which is consistent with the restriction $0<\Delta<1/4$. Note that this also implies $\Pi(i\nu \to 0) \to -\infty$.

Equation \eqref{momentum dependent W} can be rewritten as 
\begin{equation}
    \mathcal{W}(i\nu) = \frac{\lambda^2}{C_\Delta|\nu|^{4\Delta-1}}\int_0^\infty d\rho ~\rho \frac{\lambda^{-2}C_\Delta|\nu|^{4\Delta-1}}{e^{\frac{1}{2}\rho^2}\rho + \lambda^{-2}C_\Delta|\nu|^{4\Delta-1}}.
\end{equation}
At small $\nu$, the term $\lambda^{-2}C_\Delta|\nu|^{4\Delta-1}$ in the denominator dominates the other term for values of $\rho \ll \Lambda$ where $\Lambda e^{\frac{1}{2}{\Lambda^2}} = \lambda^{-2}C_\Delta|\nu|^{4\Delta-1}$. The integrand for $\rho \gg \Lambda $ is vanishingly small. Therefore, we utilise $\Lambda$ as a hard UV-cutoff as follows
\begin{equation}
    \mathcal{W}(i\nu)  = \frac{\lambda^2}{C_\Delta|\nu|^{4\Delta-1}}\int_0^\Lambda d\rho ~\rho  = \frac{\lambda^2}{C_\Delta|\nu|^{4\Delta-1}}\frac{\Lambda^2}{2}.
\end{equation}
In the limit $\nu\to 0$, the equation $\Lambda e^{\frac{1}{2}{\Lambda^2}} = \lambda^{-2}C_\Delta|\nu|^{4\Delta-1}$ has the asymptotic solution 
\begin{equation}
    \Lambda^2 = \log\left(\frac{\left(C_\Delta|\nu|^{4\Delta - 1}\right)^2}{2\lambda^4\log\left(\frac{C_\Delta|\nu|^{4\Delta-1}}{\lambda^2}\right)}\right).
\end{equation}
Ignoring the double-log term, which is sub-leading as $\nu \to 0$, this yields
\begin{equation}
    \mathcal{W}(i\nu) = \frac{\lambda^2}{C_\Delta|\nu|^{4\Delta-1}}\log\left(\frac{C_\Delta|\nu|^{4\Delta - 1}}{\lambda^2}\right),~~\nu\to 0.
\end{equation}
Moving to the real-frequency axis, we find
\begin{equation}
    \operatorname{Im} \mathcal{W} (\nu + i\eta) = -\lambda^2 B_{\Delta}\nu^{1-4\Delta} - \lambda^2 D_\Delta (4\Delta - 1)\nu^{1-4\Delta}\log (\nu)
\end{equation}
where
\begin{align}
    B_\Delta &= \frac{\sin\left(\frac{\pi}{2}(1-4\Delta)\right)\log\left(\frac{C_\Delta}{\lambda^2}\right)}{C_\Delta} \\
    &\hspace{0.5cm} + \frac{\pi}{2C_\Delta}(1-4\Delta)\cos\left(\frac{\pi}{2}(1-4\Delta)\right) \nonumber, \\
    D_\Delta &= \frac{\sin\left(\frac{\pi}{2}(1-4\Delta)\right)}{C_\Delta}.
\end{align}
On the real-frequency axis, the imaginary-part of the self-energy is given by 
\begin{equation} \label{sigma}
    \operatorname{Im} \Sigma(\omega) =  \int_0^\omega d\Omega~ \operatorname{Im} \mathcal{W}(\omega-\Omega)\rho(\Omega) 
\end{equation}
where $\omega>0$.
Performing the integration yields
\begin{multline}
     \operatorname{Im} \Sigma(\omega) = -\lambda\sin(\pi\Delta-\theta)\omega^{1-2\Delta}\frac{\Gamma(2-4\Delta)\Gamma(2\Delta)}{\pi\Gamma(2-2\Delta)}\Bigg[B_{\Delta} \\
     + D_\Delta(4\Delta-1)\left(\frac{\Gamma'(2-4\Delta)}{\Gamma(2-4\Delta)} - \frac{\Gamma'(2-2\Delta)}{\Gamma(2-2\Delta)}+\log(\omega)\right)\Bigg]
\end{multline}
for $\omega>0$. Comparing this to Eq.\eqref{sigma ansatz} on the real frequency axis then  gives the equation
\begin{equation} \label{delta equation appendix}
    f(\Delta) + g(\Delta)\log(\omega) = 1 
\end{equation}
with $g(\Delta)$ defined in Eq.\eqref{g function} and 
\begin{equation}
\begin{split}
    f(\Delta) &= \frac{\Gamma(2-4\Delta)\Gamma(2\Delta)}{\pi\Gamma(2-2\Delta)}\Bigg[B_{\Delta} \\
     &+ D_\Delta(4\Delta-1)\left(\frac{\Gamma'(2-4\Delta)}{\Gamma(2-4\Delta)} - \frac{\Gamma'(2-2\Delta)}{\Gamma(2-2\Delta)}\right)\Bigg].
\end{split}
\end{equation}
In the limit $\omega \to 0$, Eq.\eqref{delta equation appendix} is solved by $g(\Delta) = 0$, yielding $\Delta = \frac{1}{4}$.
Note, for this to be a good approximation, the cutoff needs to satisfy $\Lambda \gg 2$. This corresponds to a frequency of around $\nu \ll 10^{-6}$ for the expected values of $\Delta$.

\section{Maxwell Construction} \label{Maxwell appendix}
We find the chemical potential $\mu_\text{nFL-BI}$ of the coexistence of the nFL and BI phases across the first-order transition as follows. Equilibrium between point $A$ in one phase and point $B$ in the other phase occurs when the grand potentials of the two phases are equal: $\Omega_A=\Omega_B$. Considering the points $A$ and $B$ to be at the same temperature $T$ and volume $V$, the equilibrium condition can be expressed via an integral of $(d\Omega)_{T,V}=-Nd\mu$ from $A$ to $B$. However, since the function $N(\mu)$ is multivalued, we first Legendre transform $\Omega$ to the free energy $F = \Omega +\mu N$, and compute its (isothermal and isochoric) change as
\begin{equation}
    F_{B}-F_{A} = \int_{N_A}^{N_B}\mu(N)dN,
\end{equation}
where the equilibrium between the two phase requires $\mu(N_A) = \mu(N_B) \equiv \mu_\text{nFL-BI}$. Thus, going back to the grand potential, we obtain
\begin{multline}
    \Omega_B - \Omega_A = F_{B} - F_{A} - \mu(N_B)N_B+\mu(N_A)N_A \\
    = \int_{N_A}^{N_B}(\mu(N)-\mu_\text{nFL-BI})dN = 0,
\end{multline}
the solution to which determines $\mu_\text{nFL-BI}$.

\bibliography{biblio}% Produces the bibliography via BibTeX.

\providecommand{\noopsort}[1]{}\providecommand{\singleletter}[1]{#1}%
\begin{thebibliography}{123}%
\makeatletter
\providecommand \@ifxundefined [1]{%
 \@ifx{#1\undefined}
}%
\providecommand \@ifnum [1]{%
 \ifnum #1\expandafter \@firstoftwo
 \else \expandafter \@secondoftwo
 \fi
}%
\providecommand \@ifx [1]{%
 \ifx #1\expandafter \@firstoftwo
 \else \expandafter \@secondoftwo
 \fi
}%
\providecommand \natexlab [1]{#1}%
\providecommand \enquote  [1]{``#1''}%
\providecommand \bibnamefont  [1]{#1}%
\providecommand \bibfnamefont [1]{#1}%
\providecommand \citenamefont [1]{#1}%
\providecommand \href@noop [0]{\@secondoftwo}%
\providecommand \href [0]{\begingroup \@sanitize@url \@href}%
\providecommand \@href[1]{\@@startlink{#1}\@@href}%
\providecommand \@@href[1]{\endgroup#1\@@endlink}%
\providecommand \@sanitize@url [0]{\catcode `\\12\catcode `\$12\catcode `\&12\catcode `\#12\catcode `\^12\catcode `\_12\catcode `\%12\relax}%
\providecommand \@@startlink[1]{}%
\providecommand \@@endlink[0]{}%
\providecommand \url  [0]{\begingroup\@sanitize@url \@url }%
\providecommand \@url [1]{\endgroup\@href {#1}{\urlprefix }}%
\providecommand \urlprefix  [0]{URL }%
\providecommand \Eprint [0]{\href }%
\providecommand \doibase [0]{https://doi.org/}%
\providecommand \selectlanguage [0]{\@gobble}%
\providecommand \bibinfo  [0]{\@secondoftwo}%
\providecommand \bibfield  [0]{\@secondoftwo}%
\providecommand \translation [1]{[#1]}%
\providecommand \BibitemOpen [0]{}%
\providecommand \bibitemStop [0]{}%
\providecommand \bibitemNoStop [0]{.\EOS\space}%
\providecommand \EOS [0]{\spacefactor3000\relax}%
\providecommand \BibitemShut  [1]{\csname bibitem#1\endcsname}%
\let\auto@bib@innerbib\@empty
%</preamble>
\bibitem [{\citenamefont {Tsui}\ \emph {et~al.}(1982)\citenamefont {Tsui}, \citenamefont {Stormer},\ and\ \citenamefont {Gossard}}]{Tsui1982}%
  \BibitemOpen
  \bibfield  {author} {\bibinfo {author} {\bibfnamefont {D.~C.}\ \bibnamefont {Tsui}}, \bibinfo {author} {\bibfnamefont {H.~L.}\ \bibnamefont {Stormer}},\ and\ \bibinfo {author} {\bibfnamefont {A.~C.}\ \bibnamefont {Gossard}},\ }\bibfield  {title} {\bibinfo {title} {{Two-Dimensional Magnetotransport in the Extreme Quantum Limit}},\ }\href {https://doi.org/10.1103/PhysRevLett.48.1559} {\bibfield  {journal} {\bibinfo  {journal} {Phys. Rev. Lett.}\ }\textbf {\bibinfo {volume} {48}},\ \bibinfo {pages} {1559} (\bibinfo {year} {1982})}\BibitemShut {NoStop}%
\bibitem [{\citenamefont {Stormer}\ \emph {et~al.}(1983)\citenamefont {Stormer}, \citenamefont {Chang}, \citenamefont {Tsui}, \citenamefont {Hwang}, \citenamefont {Gossard},\ and\ \citenamefont {Wiegmann}}]{Stormer1983}%
  \BibitemOpen
  \bibfield  {author} {\bibinfo {author} {\bibfnamefont {H.~L.}\ \bibnamefont {Stormer}}, \bibinfo {author} {\bibfnamefont {A.}~\bibnamefont {Chang}}, \bibinfo {author} {\bibfnamefont {D.~C.}\ \bibnamefont {Tsui}}, \bibinfo {author} {\bibfnamefont {J.~C.~M.}\ \bibnamefont {Hwang}}, \bibinfo {author} {\bibfnamefont {A.~C.}\ \bibnamefont {Gossard}},\ and\ \bibinfo {author} {\bibfnamefont {W.}~\bibnamefont {Wiegmann}},\ }\bibfield  {title} {\bibinfo {title} {{Fractional Quantization of the {H}all Effect}},\ }\href {https://doi.org/10.1103/PhysRevLett.50.1953} {\bibfield  {journal} {\bibinfo  {journal} {Phys. Rev. Lett.}\ }\textbf {\bibinfo {volume} {50}},\ \bibinfo {pages} {1953} (\bibinfo {year} {1983})}\BibitemShut {NoStop}%
\bibitem [{\citenamefont {Arovas}\ \emph {et~al.}(1984)\citenamefont {Arovas}, \citenamefont {Schrieffer},\ and\ \citenamefont {Wilczek}}]{Arovas1984}%
  \BibitemOpen
  \bibfield  {author} {\bibinfo {author} {\bibfnamefont {D.}~\bibnamefont {Arovas}}, \bibinfo {author} {\bibfnamefont {J.~R.}\ \bibnamefont {Schrieffer}},\ and\ \bibinfo {author} {\bibfnamefont {F.}~\bibnamefont {Wilczek}},\ }\bibfield  {title} {\bibinfo {title} {{Fractional Statistics and the Quantum Hall Effect}},\ }\href {https://doi.org/10.1103/PhysRevLett.53.722} {\bibfield  {journal} {\bibinfo  {journal} {Phys. Rev. Lett.}\ }\textbf {\bibinfo {volume} {53}},\ \bibinfo {pages} {722} (\bibinfo {year} {1984})}\BibitemShut {NoStop}%
\bibitem [{\citenamefont {Halperin}(1984)}]{Halperin1984}%
  \BibitemOpen
  \bibfield  {author} {\bibinfo {author} {\bibfnamefont {B.~I.}\ \bibnamefont {Halperin}},\ }\bibfield  {title} {\bibinfo {title} {{Statistics of Quasiparticles and the Hierarchy of Fractional Quantized {H}all States}},\ }\href {https://doi.org/10.1103/PhysRevLett.52.1583} {\bibfield  {journal} {\bibinfo  {journal} {Phys. Rev. Lett.}\ }\textbf {\bibinfo {volume} {52}},\ \bibinfo {pages} {1583} (\bibinfo {year} {1984})}\BibitemShut {NoStop}%
\bibitem [{\citenamefont {Laughlin}(1983{\natexlab{a}})}]{Laughlin1983}%
  \BibitemOpen
  \bibfield  {author} {\bibinfo {author} {\bibfnamefont {R.~B.}\ \bibnamefont {Laughlin}},\ }\bibfield  {title} {\bibinfo {title} {{Anomalous Quantum {H}all Effect: An Incompressible Quantum Fluid with Fractionally Charged Excitations}},\ }\href {https://doi.org/10.1103/PhysRevLett.50.1395} {\bibfield  {journal} {\bibinfo  {journal} {Phys. Rev. Lett.}\ }\textbf {\bibinfo {volume} {50}},\ \bibinfo {pages} {1395} (\bibinfo {year} {1983}{\natexlab{a}})}\BibitemShut {NoStop}%
\bibitem [{\citenamefont {Jain}(1989)}]{Jain1989}%
  \BibitemOpen
  \bibfield  {author} {\bibinfo {author} {\bibfnamefont {J.~K.}\ \bibnamefont {Jain}},\ }\bibfield  {title} {\bibinfo {title} {{Composite-fermion approach for the fractional quantum {H}all effect}},\ }\href {https://doi.org/10.1103/PhysRevLett.63.199} {\bibfield  {journal} {\bibinfo  {journal} {Phys. Rev. Lett.}\ }\textbf {\bibinfo {volume} {63}},\ \bibinfo {pages} {199} (\bibinfo {year} {1989})}\BibitemShut {NoStop}%
\bibitem [{\citenamefont {Jain}(1990)}]{Jain1990}%
  \BibitemOpen
  \bibfield  {author} {\bibinfo {author} {\bibfnamefont {J.~K.}\ \bibnamefont {Jain}},\ }\bibfield  {title} {\bibinfo {title} {{Theory of the fractional quantum {H}all effect}},\ }\href {https://doi.org/10.1103/PhysRevB.41.7653} {\bibfield  {journal} {\bibinfo  {journal} {Phys. Rev. B}\ }\textbf {\bibinfo {volume} {41}},\ \bibinfo {pages} {7653} (\bibinfo {year} {1990})}\BibitemShut {NoStop}%
\bibitem [{\citenamefont {Lopez}\ and\ \citenamefont {Fradkin}(1991)}]{Lopez1991}%
  \BibitemOpen
  \bibfield  {author} {\bibinfo {author} {\bibfnamefont {A.}~\bibnamefont {Lopez}}\ and\ \bibinfo {author} {\bibfnamefont {E.}~\bibnamefont {Fradkin}},\ }\bibfield  {title} {\bibinfo {title} {{Fractional quantum {H}all effect and {C}hern-{S}imons gauge theories}},\ }\href {https://doi.org/10.1103/PhysRevB.44.5246} {\bibfield  {journal} {\bibinfo  {journal} {Phys. Rev. B}\ }\textbf {\bibinfo {volume} {44}},\ \bibinfo {pages} {5246} (\bibinfo {year} {1991})}\BibitemShut {NoStop}%
\bibitem [{\citenamefont {Zhang}\ \emph {et~al.}(1989)\citenamefont {Zhang}, \citenamefont {Hansson},\ and\ \citenamefont {Kivelson}}]{Zhang1989}%
  \BibitemOpen
  \bibfield  {author} {\bibinfo {author} {\bibfnamefont {S.~C.}\ \bibnamefont {Zhang}}, \bibinfo {author} {\bibfnamefont {T.~H.}\ \bibnamefont {Hansson}},\ and\ \bibinfo {author} {\bibfnamefont {S.}~\bibnamefont {Kivelson}},\ }\bibfield  {title} {\bibinfo {title} {{Effective-Field-Theory Model for the Fractional Quantum {H}all Effect}},\ }\href {https://doi.org/10.1103/PhysRevLett.62.82} {\bibfield  {journal} {\bibinfo  {journal} {Phys. Rev. Lett.}\ }\textbf {\bibinfo {volume} {62}},\ \bibinfo {pages} {82} (\bibinfo {year} {1989})}\BibitemShut {NoStop}%
\bibitem [{\citenamefont {Halperin}\ \emph {et~al.}(1993)\citenamefont {Halperin}, \citenamefont {Lee},\ and\ \citenamefont {Read}}]{HLR1993}%
  \BibitemOpen
  \bibfield  {author} {\bibinfo {author} {\bibfnamefont {B.~I.}\ \bibnamefont {Halperin}}, \bibinfo {author} {\bibfnamefont {P.~A.}\ \bibnamefont {Lee}},\ and\ \bibinfo {author} {\bibfnamefont {N.}~\bibnamefont {Read}},\ }\bibfield  {title} {\bibinfo {title} {{Theory of the half-filled {L}andau level}},\ }\href {https://doi.org/10.1103/PhysRevB.47.7312} {\bibfield  {journal} {\bibinfo  {journal} {Phys. Rev. B}\ }\textbf {\bibinfo {volume} {47}},\ \bibinfo {pages} {7312} (\bibinfo {year} {1993})}\BibitemShut {NoStop}%
\bibitem [{\citenamefont {Shankar}\ and\ \citenamefont {Murthy}(1997)}]{Shankar1997}%
  \BibitemOpen
  \bibfield  {author} {\bibinfo {author} {\bibfnamefont {R.}~\bibnamefont {Shankar}}\ and\ \bibinfo {author} {\bibfnamefont {G.}~\bibnamefont {Murthy}},\ }\bibfield  {title} {\bibinfo {title} {{Towards a Field Theory of Fractional Quantum {H}all States}},\ }\href {https://doi.org/10.1103/PhysRevLett.79.4437} {\bibfield  {journal} {\bibinfo  {journal} {Phys. Rev. Lett.}\ }\textbf {\bibinfo {volume} {79}},\ \bibinfo {pages} {4437} (\bibinfo {year} {1997})}\BibitemShut {NoStop}%
\bibitem [{\citenamefont {Jiang}\ \emph {et~al.}(1989)\citenamefont {Jiang}, \citenamefont {Stormer}, \citenamefont {Tsui}, \citenamefont {Pfeiffer},\ and\ \citenamefont {West}}]{Jiang1989}%
  \BibitemOpen
  \bibfield  {author} {\bibinfo {author} {\bibfnamefont {H.~W.}\ \bibnamefont {Jiang}}, \bibinfo {author} {\bibfnamefont {H.~L.}\ \bibnamefont {Stormer}}, \bibinfo {author} {\bibfnamefont {D.~C.}\ \bibnamefont {Tsui}}, \bibinfo {author} {\bibfnamefont {L.~N.}\ \bibnamefont {Pfeiffer}},\ and\ \bibinfo {author} {\bibfnamefont {K.~W.}\ \bibnamefont {West}},\ }\bibfield  {title} {\bibinfo {title} {{Transport anomalies in the lowest {L}andau level of two-dimensional electrons at half-filling}},\ }\href {https://doi.org/10.1103/PhysRevB.40.12013} {\bibfield  {journal} {\bibinfo  {journal} {Phys. Rev. B}\ }\textbf {\bibinfo {volume} {40}},\ \bibinfo {pages} {12013} (\bibinfo {year} {1989})}\BibitemShut {NoStop}%
\bibitem [{\citenamefont {Willett}\ \emph {et~al.}(1990)\citenamefont {Willett}, \citenamefont {Paalanen}, \citenamefont {Ruel}, \citenamefont {West}, \citenamefont {Pfeiffer},\ and\ \citenamefont {Bishop}}]{Willett1990}%
  \BibitemOpen
  \bibfield  {author} {\bibinfo {author} {\bibfnamefont {R.~L.}\ \bibnamefont {Willett}}, \bibinfo {author} {\bibfnamefont {M.~A.}\ \bibnamefont {Paalanen}}, \bibinfo {author} {\bibfnamefont {R.~R.}\ \bibnamefont {Ruel}}, \bibinfo {author} {\bibfnamefont {K.~W.}\ \bibnamefont {West}}, \bibinfo {author} {\bibfnamefont {L.~N.}\ \bibnamefont {Pfeiffer}},\ and\ \bibinfo {author} {\bibfnamefont {D.~J.}\ \bibnamefont {Bishop}},\ }\bibfield  {title} {\bibinfo {title} {{Anomalous sound propagation at \ensuremath{\nu}=1/2 in a 2D electron gas: Observation of a spontaneously broken translational symmetry?}},\ }\href {https://doi.org/10.1103/PhysRevLett.65.112} {\bibfield  {journal} {\bibinfo  {journal} {Phys. Rev. Lett.}\ }\textbf {\bibinfo {volume} {65}},\ \bibinfo {pages} {112} (\bibinfo {year} {1990})}\BibitemShut {NoStop}%
\bibitem [{\citenamefont {Willett}\ \emph {et~al.}(1993{\natexlab{a}})\citenamefont {Willett}, \citenamefont {Ruel}, \citenamefont {Paalanen}, \citenamefont {West},\ and\ \citenamefont {Pfeiffer}}]{Willett1993}%
  \BibitemOpen
  \bibfield  {author} {\bibinfo {author} {\bibfnamefont {R.~L.}\ \bibnamefont {Willett}}, \bibinfo {author} {\bibfnamefont {R.~R.}\ \bibnamefont {Ruel}}, \bibinfo {author} {\bibfnamefont {M.~A.}\ \bibnamefont {Paalanen}}, \bibinfo {author} {\bibfnamefont {K.~W.}\ \bibnamefont {West}},\ and\ \bibinfo {author} {\bibfnamefont {L.~N.}\ \bibnamefont {Pfeiffer}},\ }\bibfield  {title} {\bibinfo {title} {{Enhanced finite-wave-vector conductivity at multiple even-denominator filling factors in two-dimensional electron systems}},\ }\href {https://doi.org/10.1103/PhysRevB.47.7344} {\bibfield  {journal} {\bibinfo  {journal} {Phys. Rev. B}\ }\textbf {\bibinfo {volume} {47}},\ \bibinfo {pages} {7344} (\bibinfo {year} {1993}{\natexlab{a}})}\BibitemShut {NoStop}%
\bibitem [{\citenamefont {Kang}\ \emph {et~al.}(1993)\citenamefont {Kang}, \citenamefont {Stormer}, \citenamefont {Pfeiffer}, \citenamefont {Baldwin},\ and\ \citenamefont {West}}]{Kang1993}%
  \BibitemOpen
  \bibfield  {author} {\bibinfo {author} {\bibfnamefont {W.}~\bibnamefont {Kang}}, \bibinfo {author} {\bibfnamefont {H.~L.}\ \bibnamefont {Stormer}}, \bibinfo {author} {\bibfnamefont {L.~N.}\ \bibnamefont {Pfeiffer}}, \bibinfo {author} {\bibfnamefont {K.~W.}\ \bibnamefont {Baldwin}},\ and\ \bibinfo {author} {\bibfnamefont {K.~W.}\ \bibnamefont {West}},\ }\bibfield  {title} {\bibinfo {title} {{How real are composite fermions?}},\ }\href {https://doi.org/10.1103/PhysRevLett.71.3850} {\bibfield  {journal} {\bibinfo  {journal} {Phys. Rev. Lett.}\ }\textbf {\bibinfo {volume} {71}},\ \bibinfo {pages} {3850} (\bibinfo {year} {1993})}\BibitemShut {NoStop}%
\bibitem [{\citenamefont {Willett}\ \emph {et~al.}(1993{\natexlab{b}})\citenamefont {Willett}, \citenamefont {Ruel}, \citenamefont {West},\ and\ \citenamefont {Pfeiffer}}]{Willett1993_2}%
  \BibitemOpen
  \bibfield  {author} {\bibinfo {author} {\bibfnamefont {R.~L.}\ \bibnamefont {Willett}}, \bibinfo {author} {\bibfnamefont {R.~R.}\ \bibnamefont {Ruel}}, \bibinfo {author} {\bibfnamefont {K.~W.}\ \bibnamefont {West}},\ and\ \bibinfo {author} {\bibfnamefont {L.~N.}\ \bibnamefont {Pfeiffer}},\ }\bibfield  {title} {\bibinfo {title} {{Experimental demonstration of a Fermi surface at one-half filling of the lowest Landau level}},\ }\href {https://doi.org/10.1103/PhysRevLett.71.3846} {\bibfield  {journal} {\bibinfo  {journal} {Phys. Rev. Lett.}\ }\textbf {\bibinfo {volume} {71}},\ \bibinfo {pages} {3846} (\bibinfo {year} {1993}{\natexlab{b}})}\BibitemShut {NoStop}%
\bibitem [{\citenamefont {Du}\ \emph {et~al.}(1993)\citenamefont {Du}, \citenamefont {Stormer}, \citenamefont {Tsui}, \citenamefont {Pfeiffer},\ and\ \citenamefont {West}}]{Du1993}%
  \BibitemOpen
  \bibfield  {author} {\bibinfo {author} {\bibfnamefont {R.~R.}\ \bibnamefont {Du}}, \bibinfo {author} {\bibfnamefont {H.~L.}\ \bibnamefont {Stormer}}, \bibinfo {author} {\bibfnamefont {D.~C.}\ \bibnamefont {Tsui}}, \bibinfo {author} {\bibfnamefont {L.~N.}\ \bibnamefont {Pfeiffer}},\ and\ \bibinfo {author} {\bibfnamefont {K.~W.}\ \bibnamefont {West}},\ }\bibfield  {title} {\bibinfo {title} {{Experimental evidence for new particles in the fractional quantum {H}all effect}},\ }\href {https://doi.org/10.1103/PhysRevLett.70.2944} {\bibfield  {journal} {\bibinfo  {journal} {Phys. Rev. Lett.}\ }\textbf {\bibinfo {volume} {70}},\ \bibinfo {pages} {2944} (\bibinfo {year} {1993})}\BibitemShut {NoStop}%
\bibitem [{\citenamefont {Willett}(1997)}]{Willett1997}%
  \BibitemOpen
  \bibfield  {author} {\bibinfo {author} {\bibfnamefont {R.~L.}\ \bibnamefont {Willett}},\ }\bibfield  {title} {\bibinfo {title} {{Experimental evidence for composite fermions}},\ }\href {https://doi.org/10.1080/00018739700101528} {\bibfield  {journal} {\bibinfo  {journal} {Advances in Physics}\ }\textbf {\bibinfo {volume} {46}},\ \bibinfo {pages} {447} (\bibinfo {year} {1997})},\ \Eprint {https://arxiv.org/abs/https://doi.org/10.1080/00018739700101528} {https://doi.org/10.1080/00018739700101528} \BibitemShut {NoStop}%
\bibitem [{\citenamefont {Yoshioka}\ \emph {et~al.}(1983)\citenamefont {Yoshioka}, \citenamefont {Halperin},\ and\ \citenamefont {Lee}}]{YHL1983}%
  \BibitemOpen
  \bibfield  {author} {\bibinfo {author} {\bibfnamefont {D.}~\bibnamefont {Yoshioka}}, \bibinfo {author} {\bibfnamefont {B.~I.}\ \bibnamefont {Halperin}},\ and\ \bibinfo {author} {\bibfnamefont {P.~A.}\ \bibnamefont {Lee}},\ }\bibfield  {title} {\bibinfo {title} {{Ground State of Two-Dimensional Electrons in Strong Magnetic Fields and $\frac{1}{3}$ Quantized {H}all Effect}},\ }\href {https://doi.org/10.1103/PhysRevLett.50.1219} {\bibfield  {journal} {\bibinfo  {journal} {Phys. Rev. Lett.}\ }\textbf {\bibinfo {volume} {50}},\ \bibinfo {pages} {1219} (\bibinfo {year} {1983})}\BibitemShut {NoStop}%
\bibitem [{\citenamefont {Laughlin}(1983{\natexlab{b}})}]{Laughlin1983b}%
  \BibitemOpen
  \bibfield  {author} {\bibinfo {author} {\bibfnamefont {R.~B.}\ \bibnamefont {Laughlin}},\ }\bibfield  {title} {\bibinfo {title} {{Quantized motion of three two-dimensional electrons in a strong magnetic field}},\ }\href {https://doi.org/10.1103/PhysRevB.27.3383} {\bibfield  {journal} {\bibinfo  {journal} {Phys. Rev. B}\ }\textbf {\bibinfo {volume} {27}},\ \bibinfo {pages} {3383} (\bibinfo {year} {1983}{\natexlab{b}})}\BibitemShut {NoStop}%
\bibitem [{\citenamefont {Haldane}(1983)}]{Haldane1983}%
  \BibitemOpen
  \bibfield  {author} {\bibinfo {author} {\bibfnamefont {F.~D.~M.}\ \bibnamefont {Haldane}},\ }\bibfield  {title} {\bibinfo {title} {{Fractional Quantization of the {H}all Effect: A Hierarchy of Incompressible Quantum Fluid States}},\ }\href {https://doi.org/10.1103/PhysRevLett.51.605} {\bibfield  {journal} {\bibinfo  {journal} {Phys. Rev. Lett.}\ }\textbf {\bibinfo {volume} {51}},\ \bibinfo {pages} {605} (\bibinfo {year} {1983})}\BibitemShut {NoStop}%
\bibitem [{\citenamefont {Haldane}\ and\ \citenamefont {Rezayi}(1985)}]{Haldane1985}%
  \BibitemOpen
  \bibfield  {author} {\bibinfo {author} {\bibfnamefont {F.~D.~M.}\ \bibnamefont {Haldane}}\ and\ \bibinfo {author} {\bibfnamefont {E.~H.}\ \bibnamefont {Rezayi}},\ }\bibfield  {title} {\bibinfo {title} {{Finite-Size Studies of the Incompressible State of the Fractionally Quantized {H}all Effect and its Excitations}},\ }\href {https://doi.org/10.1103/PhysRevLett.54.237} {\bibfield  {journal} {\bibinfo  {journal} {Phys. Rev. Lett.}\ }\textbf {\bibinfo {volume} {54}},\ \bibinfo {pages} {237} (\bibinfo {year} {1985})}\BibitemShut {NoStop}%
\bibitem [{\citenamefont {Fano}\ \emph {et~al.}(1986)\citenamefont {Fano}, \citenamefont {Ortolani},\ and\ \citenamefont {Colombo}}]{Fano1986}%
  \BibitemOpen
  \bibfield  {author} {\bibinfo {author} {\bibfnamefont {G.}~\bibnamefont {Fano}}, \bibinfo {author} {\bibfnamefont {F.}~\bibnamefont {Ortolani}},\ and\ \bibinfo {author} {\bibfnamefont {E.}~\bibnamefont {Colombo}},\ }\bibfield  {title} {\bibinfo {title} {{Configuration-interaction calculations on the fractional quantum {H}all effect}},\ }\href {https://doi.org/10.1103/PhysRevB.34.2670} {\bibfield  {journal} {\bibinfo  {journal} {Phys. Rev. B}\ }\textbf {\bibinfo {volume} {34}},\ \bibinfo {pages} {2670} (\bibinfo {year} {1986})}\BibitemShut {NoStop}%
\bibitem [{\citenamefont {d'Ambrumenil}\ and\ \citenamefont {Morf}(1989)}]{d'Ambrunmenil1989}%
  \BibitemOpen
  \bibfield  {author} {\bibinfo {author} {\bibfnamefont {N.}~\bibnamefont {d'Ambrumenil}}\ and\ \bibinfo {author} {\bibfnamefont {R.}~\bibnamefont {Morf}},\ }\bibfield  {title} {\bibinfo {title} {{Hierarchical classification of fractional quantum {H}all states}},\ }\href {https://doi.org/10.1103/PhysRevB.40.6108} {\bibfield  {journal} {\bibinfo  {journal} {Phys. Rev. B}\ }\textbf {\bibinfo {volume} {40}},\ \bibinfo {pages} {6108} (\bibinfo {year} {1989})}\BibitemShut {NoStop}%
\bibitem [{\citenamefont {Morf}\ \emph {et~al.}(2002)\citenamefont {Morf}, \citenamefont {d'Ambrumenil},\ and\ \citenamefont {Das~Sarma}}]{Morf2002}%
  \BibitemOpen
  \bibfield  {author} {\bibinfo {author} {\bibfnamefont {R.~H.}\ \bibnamefont {Morf}}, \bibinfo {author} {\bibfnamefont {N.}~\bibnamefont {d'Ambrumenil}},\ and\ \bibinfo {author} {\bibfnamefont {S.}~\bibnamefont {Das~Sarma}},\ }\bibfield  {title} {\bibinfo {title} {{Excitation gaps in fractional quantum {H}all states: An exact diagonalization study}},\ }\href {https://doi.org/10.1103/PhysRevB.66.075408} {\bibfield  {journal} {\bibinfo  {journal} {Phys. Rev. B}\ }\textbf {\bibinfo {volume} {66}},\ \bibinfo {pages} {075408} (\bibinfo {year} {2002})}\BibitemShut {NoStop}%
\bibitem [{\citenamefont {Shibata}\ and\ \citenamefont {Yoshioka}(2001)}]{Naokazu2001}%
  \BibitemOpen
  \bibfield  {author} {\bibinfo {author} {\bibfnamefont {N.}~\bibnamefont {Shibata}}\ and\ \bibinfo {author} {\bibfnamefont {D.}~\bibnamefont {Yoshioka}},\ }\bibfield  {title} {\bibinfo {title} {{Ground-State Phase Diagram of 2D Electrons in a High {L}andau Level: A Density-Matrix Renormalization Group Study}},\ }\href {https://doi.org/10.1103/PhysRevLett.86.5755} {\bibfield  {journal} {\bibinfo  {journal} {Phys. Rev. Lett.}\ }\textbf {\bibinfo {volume} {86}},\ \bibinfo {pages} {5755} (\bibinfo {year} {2001})}\BibitemShut {NoStop}%
\bibitem [{\citenamefont {Feiguin}\ \emph {et~al.}(2008)\citenamefont {Feiguin}, \citenamefont {Rezayi}, \citenamefont {Nayak},\ and\ \citenamefont {Das~Sarma}}]{Feiguin2008}%
  \BibitemOpen
  \bibfield  {author} {\bibinfo {author} {\bibfnamefont {A.~E.}\ \bibnamefont {Feiguin}}, \bibinfo {author} {\bibfnamefont {E.}~\bibnamefont {Rezayi}}, \bibinfo {author} {\bibfnamefont {C.}~\bibnamefont {Nayak}},\ and\ \bibinfo {author} {\bibfnamefont {S.}~\bibnamefont {Das~Sarma}},\ }\bibfield  {title} {\bibinfo {title} {{Density Matrix Renormalization Group Study of Incompressible Fractional Quantum {H}all States}},\ }\href {https://doi.org/10.1103/PhysRevLett.100.166803} {\bibfield  {journal} {\bibinfo  {journal} {Phys. Rev. Lett.}\ }\textbf {\bibinfo {volume} {100}},\ \bibinfo {pages} {166803} (\bibinfo {year} {2008})}\BibitemShut {NoStop}%
\bibitem [{\citenamefont {Kovrizhin}(2010)}]{Kovrizhin2010}%
  \BibitemOpen
  \bibfield  {author} {\bibinfo {author} {\bibfnamefont {D.~L.}\ \bibnamefont {Kovrizhin}},\ }\bibfield  {title} {\bibinfo {title} {{Density matrix renormalization group for bosonic quantum {H}all effect}},\ }\href {https://doi.org/10.1103/PhysRevB.81.125130} {\bibfield  {journal} {\bibinfo  {journal} {Phys. Rev. B}\ }\textbf {\bibinfo {volume} {81}},\ \bibinfo {pages} {125130} (\bibinfo {year} {2010})}\BibitemShut {NoStop}%
\bibitem [{\citenamefont {Zaletel}\ and\ \citenamefont {Mong}(2012)}]{Zaletel2012}%
  \BibitemOpen
  \bibfield  {author} {\bibinfo {author} {\bibfnamefont {M.~P.}\ \bibnamefont {Zaletel}}\ and\ \bibinfo {author} {\bibfnamefont {R.~S.~K.}\ \bibnamefont {Mong}},\ }\bibfield  {title} {\bibinfo {title} {{Exact matrix product states for quantum {H}all wave functions}},\ }\href {https://doi.org/10.1103/PhysRevB.86.245305} {\bibfield  {journal} {\bibinfo  {journal} {Phys. Rev. B}\ }\textbf {\bibinfo {volume} {86}},\ \bibinfo {pages} {245305} (\bibinfo {year} {2012})}\BibitemShut {NoStop}%
\bibitem [{\citenamefont {Hu}\ \emph {et~al.}(2012)\citenamefont {Hu}, \citenamefont {Papić}, \citenamefont {Johri}, \citenamefont {Bhatt},\ and\ \citenamefont {Schmitteckert}}]{Hu2012}%
  \BibitemOpen
  \bibfield  {author} {\bibinfo {author} {\bibfnamefont {Z.-X.}\ \bibnamefont {Hu}}, \bibinfo {author} {\bibfnamefont {Z.}~\bibnamefont {Papić}}, \bibinfo {author} {\bibfnamefont {S.}~\bibnamefont {Johri}}, \bibinfo {author} {\bibfnamefont {R.}~\bibnamefont {Bhatt}},\ and\ \bibinfo {author} {\bibfnamefont {P.}~\bibnamefont {Schmitteckert}},\ }\bibfield  {title} {\bibinfo {title} {{Comparison of the density-matrix renormalization group method applied to fractional quantum {H}all systems in different geometries}},\ }\href {https://doi.org/https://doi.org/10.1016/j.physleta.2012.05.031} {\bibfield  {journal} {\bibinfo  {journal} {Physics Letters A}\ }\textbf {\bibinfo {volume} {376}},\ \bibinfo {pages} {2157} (\bibinfo {year} {2012})}\BibitemShut {NoStop}%
\bibitem [{\citenamefont {Estienne}\ \emph {et~al.}(2013)\citenamefont {Estienne}, \citenamefont {Papi\ifmmode~\acute{c}\else \'{c}\fi{}}, \citenamefont {Regnault},\ and\ \citenamefont {Bernevig}}]{Estienne2013}%
  \BibitemOpen
  \bibfield  {author} {\bibinfo {author} {\bibfnamefont {B.}~\bibnamefont {Estienne}}, \bibinfo {author} {\bibfnamefont {Z.}~\bibnamefont {Papi\ifmmode~\acute{c}\else \'{c}\fi{}}}, \bibinfo {author} {\bibfnamefont {N.}~\bibnamefont {Regnault}},\ and\ \bibinfo {author} {\bibfnamefont {B.~A.}\ \bibnamefont {Bernevig}},\ }\bibfield  {title} {\bibinfo {title} {{Matrix product states for trial quantum {H}all states}},\ }\href {https://doi.org/10.1103/PhysRevB.87.161112} {\bibfield  {journal} {\bibinfo  {journal} {Phys. Rev. B}\ }\textbf {\bibinfo {volume} {87}},\ \bibinfo {pages} {161112(R)} (\bibinfo {year} {2013})}\BibitemShut {NoStop}%
\bibitem [{\citenamefont {Zaletel}\ \emph {et~al.}(2013)\citenamefont {Zaletel}, \citenamefont {Mong},\ and\ \citenamefont {Pollmann}}]{Zaletel2013}%
  \BibitemOpen
  \bibfield  {author} {\bibinfo {author} {\bibfnamefont {M.~P.}\ \bibnamefont {Zaletel}}, \bibinfo {author} {\bibfnamefont {R.~S.~K.}\ \bibnamefont {Mong}},\ and\ \bibinfo {author} {\bibfnamefont {F.}~\bibnamefont {Pollmann}},\ }\bibfield  {title} {\bibinfo {title} {{Topological Characterization of Fractional Quantum {H}all Ground States from Microscopic {H}amiltonians}},\ }\href {https://doi.org/10.1103/PhysRevLett.110.236801} {\bibfield  {journal} {\bibinfo  {journal} {Phys. Rev. Lett.}\ }\textbf {\bibinfo {volume} {110}},\ \bibinfo {pages} {236801} (\bibinfo {year} {2013})}\BibitemShut {NoStop}%
\bibitem [{\citenamefont {Fukuyama}\ \emph {et~al.}(1979)\citenamefont {Fukuyama}, \citenamefont {Platzman},\ and\ \citenamefont {Anderson}}]{FPA1979}%
  \BibitemOpen
  \bibfield  {author} {\bibinfo {author} {\bibfnamefont {H.}~\bibnamefont {Fukuyama}}, \bibinfo {author} {\bibfnamefont {P.~M.}\ \bibnamefont {Platzman}},\ and\ \bibinfo {author} {\bibfnamefont {P.~W.}\ \bibnamefont {Anderson}},\ }\bibfield  {title} {\bibinfo {title} {{Two-dimensional electron gas in a strong magnetic field}},\ }\href {https://doi.org/10.1103/PhysRevB.19.5211} {\bibfield  {journal} {\bibinfo  {journal} {Phys. Rev. B}\ }\textbf {\bibinfo {volume} {19}},\ \bibinfo {pages} {5211} (\bibinfo {year} {1979})}\BibitemShut {NoStop}%
\bibitem [{\citenamefont {Yoshioka}\ and\ \citenamefont {Fukuyama}(1979)}]{Yoshioka1979}%
  \BibitemOpen
  \bibfield  {author} {\bibinfo {author} {\bibfnamefont {D.}~\bibnamefont {Yoshioka}}\ and\ \bibinfo {author} {\bibfnamefont {H.}~\bibnamefont {Fukuyama}},\ }\bibfield  {title} {\bibinfo {title} {{Charge Density Wave State of Two-Dimensional Electrons in Strong Magnetic Fields}},\ }\href {https://doi.org/10.1143/JPSJ.47.394} {\bibfield  {journal} {\bibinfo  {journal} {Journal of the Physical Society of Japan}\ }\textbf {\bibinfo {volume} {47}},\ \bibinfo {pages} {394} (\bibinfo {year} {1979})},\ \Eprint {https://arxiv.org/abs/https://doi.org/10.1143/JPSJ.47.394} {https://doi.org/10.1143/JPSJ.47.394} \BibitemShut {NoStop}%
\bibitem [{\citenamefont {Yoshioka}\ and\ \citenamefont {Lee}(1983)}]{Yoshioka1983}%
  \BibitemOpen
  \bibfield  {author} {\bibinfo {author} {\bibfnamefont {D.}~\bibnamefont {Yoshioka}}\ and\ \bibinfo {author} {\bibfnamefont {P.~A.}\ \bibnamefont {Lee}},\ }\bibfield  {title} {\bibinfo {title} {{Ground-state energy of a two-dimensional charge-density-wave state in a strong magnetic field}},\ }\href {https://doi.org/10.1103/PhysRevB.27.4986} {\bibfield  {journal} {\bibinfo  {journal} {Phys. Rev. B}\ }\textbf {\bibinfo {volume} {27}},\ \bibinfo {pages} {4986} (\bibinfo {year} {1983})}\BibitemShut {NoStop}%
\bibitem [{\citenamefont {Tao}\ and\ \citenamefont {Thouless}(1983)}]{TaoThouless1983}%
  \BibitemOpen
  \bibfield  {author} {\bibinfo {author} {\bibfnamefont {R.}~\bibnamefont {Tao}}\ and\ \bibinfo {author} {\bibfnamefont {D.~J.}\ \bibnamefont {Thouless}},\ }\bibfield  {title} {\bibinfo {title} {{Fractional quantization of {H}all conductance}},\ }\href {https://doi.org/10.1103/PhysRevB.28.1142} {\bibfield  {journal} {\bibinfo  {journal} {Phys. Rev. B}\ }\textbf {\bibinfo {volume} {28}},\ \bibinfo {pages} {1142} (\bibinfo {year} {1983})}\BibitemShut {NoStop}%
\bibitem [{\citenamefont {Tao}(1984)}]{Tao1984}%
  \BibitemOpen
  \bibfield  {author} {\bibinfo {author} {\bibfnamefont {R.}~\bibnamefont {Tao}},\ }\bibfield  {title} {\bibinfo {title} {{Fractional quantization of Hall conductance. II}},\ }\href {https://doi.org/10.1103/PhysRevB.29.636} {\bibfield  {journal} {\bibinfo  {journal} {Phys. Rev. B}\ }\textbf {\bibinfo {volume} {29}},\ \bibinfo {pages} {636} (\bibinfo {year} {1984})}\BibitemShut {NoStop}%
\bibitem [{\citenamefont {Haussmann}(1996)}]{Haussmann1996}%
  \BibitemOpen
  \bibfield  {author} {\bibinfo {author} {\bibfnamefont {R.}~\bibnamefont {Haussmann}},\ }\bibfield  {title} {\bibinfo {title} {{Electronic spectral function for a two-dimensional electron system in the fractional quantum {H}all regime}},\ }\href {https://doi.org/10.1103/PhysRevB.53.7357} {\bibfield  {journal} {\bibinfo  {journal} {Phys. Rev. B}\ }\textbf {\bibinfo {volume} {53}},\ \bibinfo {pages} {7357} (\bibinfo {year} {1996})}\BibitemShut {NoStop}%
\bibitem [{\citenamefont {Ashoori}\ \emph {et~al.}(1990)\citenamefont {Ashoori}, \citenamefont {Lebens}, \citenamefont {Bigelow},\ and\ \citenamefont {Silsbee}}]{Ashoori1990}%
  \BibitemOpen
  \bibfield  {author} {\bibinfo {author} {\bibfnamefont {R.~C.}\ \bibnamefont {Ashoori}}, \bibinfo {author} {\bibfnamefont {J.~A.}\ \bibnamefont {Lebens}}, \bibinfo {author} {\bibfnamefont {N.~P.}\ \bibnamefont {Bigelow}},\ and\ \bibinfo {author} {\bibfnamefont {R.~H.}\ \bibnamefont {Silsbee}},\ }\bibfield  {title} {\bibinfo {title} {{Equilibrium tunneling from the two-dimensional electron gas in GaAs: Evidence for a magnetic-field-induced energy gap}},\ }\href {https://doi.org/10.1103/PhysRevLett.64.681} {\bibfield  {journal} {\bibinfo  {journal} {Phys. Rev. Lett.}\ }\textbf {\bibinfo {volume} {64}},\ \bibinfo {pages} {681} (\bibinfo {year} {1990})}\BibitemShut {NoStop}%
\bibitem [{\citenamefont {Ashoori}\ \emph {et~al.}(1993)\citenamefont {Ashoori}, \citenamefont {Lebens}, \citenamefont {Bigelow},\ and\ \citenamefont {Silsbee}}]{Ashoori1993}%
  \BibitemOpen
  \bibfield  {author} {\bibinfo {author} {\bibfnamefont {R.~C.}\ \bibnamefont {Ashoori}}, \bibinfo {author} {\bibfnamefont {J.~A.}\ \bibnamefont {Lebens}}, \bibinfo {author} {\bibfnamefont {N.~P.}\ \bibnamefont {Bigelow}},\ and\ \bibinfo {author} {\bibfnamefont {R.~H.}\ \bibnamefont {Silsbee}},\ }\bibfield  {title} {\bibinfo {title} {{Energy gaps of the two-dimensional electron gas explored with equilibrium tunneling spectroscopy}},\ }\href {https://doi.org/10.1103/PhysRevB.48.4616} {\bibfield  {journal} {\bibinfo  {journal} {Phys. Rev. B}\ }\textbf {\bibinfo {volume} {48}},\ \bibinfo {pages} {4616} (\bibinfo {year} {1993})}\BibitemShut {NoStop}%
\bibitem [{\citenamefont {Eisenstein}\ \emph {et~al.}(1992{\natexlab{a}})\citenamefont {Eisenstein}, \citenamefont {Pfeiffer},\ and\ \citenamefont {West}}]{Eisenstein1992}%
  \BibitemOpen
  \bibfield  {author} {\bibinfo {author} {\bibfnamefont {J.~P.}\ \bibnamefont {Eisenstein}}, \bibinfo {author} {\bibfnamefont {L.~N.}\ \bibnamefont {Pfeiffer}},\ and\ \bibinfo {author} {\bibfnamefont {K.~W.}\ \bibnamefont {West}},\ }\bibfield  {title} {\bibinfo {title} {{Coulomb barrier to tunneling between parallel two-dimensional electron systems}},\ }\href {https://doi.org/10.1103/PhysRevLett.69.3804} {\bibfield  {journal} {\bibinfo  {journal} {Phys. Rev. Lett.}\ }\textbf {\bibinfo {volume} {69}},\ \bibinfo {pages} {3804} (\bibinfo {year} {1992}{\natexlab{a}})}\BibitemShut {NoStop}%
\bibitem [{\citenamefont {Brown}\ \emph {et~al.}(1994)\citenamefont {Brown}, \citenamefont {Turner}, \citenamefont {Nicholls}, \citenamefont {Linfield}, \citenamefont {Pepper}, \citenamefont {Ritchie},\ and\ \citenamefont {Jones}}]{Brown1994}%
  \BibitemOpen
  \bibfield  {author} {\bibinfo {author} {\bibfnamefont {K.~M.}\ \bibnamefont {Brown}}, \bibinfo {author} {\bibfnamefont {N.}~\bibnamefont {Turner}}, \bibinfo {author} {\bibfnamefont {J.~T.}\ \bibnamefont {Nicholls}}, \bibinfo {author} {\bibfnamefont {E.~H.}\ \bibnamefont {Linfield}}, \bibinfo {author} {\bibfnamefont {M.}~\bibnamefont {Pepper}}, \bibinfo {author} {\bibfnamefont {D.~A.}\ \bibnamefont {Ritchie}},\ and\ \bibinfo {author} {\bibfnamefont {G.~A.~C.}\ \bibnamefont {Jones}},\ }\bibfield  {title} {\bibinfo {title} {{Tunneling between two-dimensional electron gases in a strong magnetic field}},\ }\href {https://doi.org/10.1103/PhysRevB.50.15465} {\bibfield  {journal} {\bibinfo  {journal} {Phys. Rev. B}\ }\textbf {\bibinfo {volume} {50}},\ \bibinfo {pages} {15465} (\bibinfo {year} {1994})}\BibitemShut {NoStop}%
\bibitem [{\citenamefont {Eisenstein}\ \emph {et~al.}(1995)\citenamefont {Eisenstein}, \citenamefont {Pfeiffer},\ and\ \citenamefont {West}}]{Eisenstein1995}%
  \BibitemOpen
  \bibfield  {author} {\bibinfo {author} {\bibfnamefont {J.~P.}\ \bibnamefont {Eisenstein}}, \bibinfo {author} {\bibfnamefont {L.~N.}\ \bibnamefont {Pfeiffer}},\ and\ \bibinfo {author} {\bibfnamefont {K.~W.}\ \bibnamefont {West}},\ }\bibfield  {title} {\bibinfo {title} {{Evidence for an Interlayer Exciton in Tunneling between Two-Dimensional Electron Systems}},\ }\href {https://doi.org/10.1103/PhysRevLett.74.1419} {\bibfield  {journal} {\bibinfo  {journal} {Phys. Rev. Lett.}\ }\textbf {\bibinfo {volume} {74}},\ \bibinfo {pages} {1419} (\bibinfo {year} {1995})}\BibitemShut {NoStop}%
\bibitem [{\citenamefont {Eisenstein}\ \emph {et~al.}(2016)\citenamefont {Eisenstein}, \citenamefont {Khaire}, \citenamefont {Nandi}, \citenamefont {Finck}, \citenamefont {Pfeiffer},\ and\ \citenamefont {West}}]{Eisenstein2016}%
  \BibitemOpen
  \bibfield  {author} {\bibinfo {author} {\bibfnamefont {J.~P.}\ \bibnamefont {Eisenstein}}, \bibinfo {author} {\bibfnamefont {T.}~\bibnamefont {Khaire}}, \bibinfo {author} {\bibfnamefont {D.}~\bibnamefont {Nandi}}, \bibinfo {author} {\bibfnamefont {A.~D.~K.}\ \bibnamefont {Finck}}, \bibinfo {author} {\bibfnamefont {L.~N.}\ \bibnamefont {Pfeiffer}},\ and\ \bibinfo {author} {\bibfnamefont {K.~W.}\ \bibnamefont {West}},\ }\bibfield  {title} {\bibinfo {title} {{Spin and the Coulomb gap in the half-filled lowest {L}andau level}},\ }\href {https://doi.org/10.1103/PhysRevB.94.125409} {\bibfield  {journal} {\bibinfo  {journal} {Phys. Rev. B}\ }\textbf {\bibinfo {volume} {94}},\ \bibinfo {pages} {125409} (\bibinfo {year} {2016})}\BibitemShut {NoStop}%
\bibitem [{\citenamefont {Chang}\ \emph {et~al.}(1983)\citenamefont {Chang}, \citenamefont {Paalanen}, \citenamefont {Tsui}, \citenamefont {St\"ormer},\ and\ \citenamefont {Hwang}}]{Chang1983}%
  \BibitemOpen
  \bibfield  {author} {\bibinfo {author} {\bibfnamefont {A.~M.}\ \bibnamefont {Chang}}, \bibinfo {author} {\bibfnamefont {M.~A.}\ \bibnamefont {Paalanen}}, \bibinfo {author} {\bibfnamefont {D.~C.}\ \bibnamefont {Tsui}}, \bibinfo {author} {\bibfnamefont {H.~L.}\ \bibnamefont {St\"ormer}},\ and\ \bibinfo {author} {\bibfnamefont {J.~C.~M.}\ \bibnamefont {Hwang}},\ }\bibfield  {title} {\bibinfo {title} {{Fractional quantum {H}all effect at low temperatures}},\ }\href {https://doi.org/10.1103/PhysRevB.28.6133} {\bibfield  {journal} {\bibinfo  {journal} {Phys. Rev. B}\ }\textbf {\bibinfo {volume} {28}},\ \bibinfo {pages} {6133} (\bibinfo {year} {1983})}\BibitemShut {NoStop}%
\bibitem [{\citenamefont {Hedin}(1965)}]{Hedin1965}%
  \BibitemOpen
  \bibfield  {author} {\bibinfo {author} {\bibfnamefont {L.}~\bibnamefont {Hedin}},\ }\bibfield  {title} {\bibinfo {title} {{New Method for Calculating the One-Particle Green's Function with Application to the Electron-Gas Problem}},\ }\href {https://doi.org/10.1103/PhysRev.139.A796} {\bibfield  {journal} {\bibinfo  {journal} {Phys. Rev.}\ }\textbf {\bibinfo {volume} {139}},\ \bibinfo {pages} {A796} (\bibinfo {year} {1965})}\BibitemShut {NoStop}%
\bibitem [{\citenamefont {{Van Houcke}}\ \emph {et~al.}(2010)\citenamefont {{Van Houcke}}, \citenamefont {Kozik}, \citenamefont {Prokof'ev},\ and\ \citenamefont {Svistunov}}]{VanHoucke2010}%
  \BibitemOpen
  \bibfield  {author} {\bibinfo {author} {\bibfnamefont {K.}~\bibnamefont {{Van Houcke}}}, \bibinfo {author} {\bibfnamefont {E.}~\bibnamefont {Kozik}}, \bibinfo {author} {\bibfnamefont {N.}~\bibnamefont {Prokof'ev}},\ and\ \bibinfo {author} {\bibfnamefont {B.}~\bibnamefont {Svistunov}},\ }\bibfield  {title} {\bibinfo {title} {{Diagrammatic Monte Carlo}},\ }\href {https://doi.org/10.1016/j.phpro.2010.09.034} {\bibfield  {journal} {\bibinfo  {journal} {Phys. Procedia}\ }\textbf {\bibinfo {volume} {6}},\ \bibinfo {pages} {95} (\bibinfo {year} {2010})}\BibitemShut {NoStop}%
\bibitem [{\citenamefont {Kozik}\ \emph {et~al.}(2010)\citenamefont {Kozik}, \citenamefont {{Van Houcke}}, \citenamefont {Gull}, \citenamefont {Pollet}, \citenamefont {Prokof'ev}, \citenamefont {Svistunov},\ and\ \citenamefont {Troyer}}]{Kozik2010}%
  \BibitemOpen
  \bibfield  {author} {\bibinfo {author} {\bibfnamefont {E.}~\bibnamefont {Kozik}}, \bibinfo {author} {\bibfnamefont {K.}~\bibnamefont {{Van Houcke}}}, \bibinfo {author} {\bibfnamefont {E.}~\bibnamefont {Gull}}, \bibinfo {author} {\bibfnamefont {L.}~\bibnamefont {Pollet}}, \bibinfo {author} {\bibfnamefont {N.}~\bibnamefont {Prokof'ev}}, \bibinfo {author} {\bibfnamefont {B.}~\bibnamefont {Svistunov}},\ and\ \bibinfo {author} {\bibfnamefont {M.}~\bibnamefont {Troyer}},\ }\bibfield  {title} {\bibinfo {title} {{Diagrammatic Monte Carlo for correlated fermions}},\ }\href {https://doi.org/10.1209/0295-5075/90/10004} {\bibfield  {journal} {\bibinfo  {journal} {Europhys. Lett.}\ }\textbf {\bibinfo {volume} {90}},\ \bibinfo {pages} {10004} (\bibinfo {year} {2010})}\BibitemShut {NoStop}%
\bibitem [{\citenamefont {Van~Houcke}\ \emph {et~al.}(2019)\citenamefont {Van~Houcke}, \citenamefont {Werner}, \citenamefont {Ohgoe}, \citenamefont {Prokof'ev},\ and\ \citenamefont {Svistunov}}]{VanHoucke2019}%
  \BibitemOpen
  \bibfield  {author} {\bibinfo {author} {\bibfnamefont {K.}~\bibnamefont {Van~Houcke}}, \bibinfo {author} {\bibfnamefont {F.}~\bibnamefont {Werner}}, \bibinfo {author} {\bibfnamefont {T.}~\bibnamefont {Ohgoe}}, \bibinfo {author} {\bibfnamefont {N.~V.}\ \bibnamefont {Prokof'ev}},\ and\ \bibinfo {author} {\bibfnamefont {B.~V.}\ \bibnamefont {Svistunov}},\ }\bibfield  {title} {\bibinfo {title} {{Diagrammatic Monte Carlo algorithm for the resonant Fermi gas}},\ }\href {https://doi.org/10.1103/PhysRevB.99.035140} {\bibfield  {journal} {\bibinfo  {journal} {Phys. Rev. B}\ }\textbf {\bibinfo {volume} {99}},\ \bibinfo {pages} {035140} (\bibinfo {year} {2019})}\BibitemShut {NoStop}%
\bibitem [{\citenamefont {Chen}\ and\ \citenamefont {Haule}(2019)}]{Chen2019}%
  \BibitemOpen
  \bibfield  {author} {\bibinfo {author} {\bibfnamefont {K.}~\bibnamefont {Chen}}\ and\ \bibinfo {author} {\bibfnamefont {K.}~\bibnamefont {Haule}},\ }\bibfield  {title} {\bibinfo {title} {{A combined variational and diagrammatic quantum Monte Carlo approach to the many-electron problem}},\ }\href {https://doi.org/10.1038/s41467-019-11708-6} {\bibfield  {journal} {\bibinfo  {journal} {Nat. Commun.}\ }\textbf {\bibinfo {volume} {10}},\ \bibinfo {pages} {3725} (\bibinfo {year} {2019})}\BibitemShut {NoStop}%
\bibitem [{\citenamefont {Kozik}(2023)}]{Kozik2023combinatorial}%
  \BibitemOpen
  \bibfield  {author} {\bibinfo {author} {\bibfnamefont {E.}~\bibnamefont {Kozik}},\ }\href {https://arxiv.org/abs/2309.13774} {\bibinfo {title} {{Combinatorial summation of Feynman diagrams: Equation of state of the 2D SU(N) Hubbard model}}} (\bibinfo {year} {2023}),\ \Eprint {https://arxiv.org/abs/2309.13774} {arXiv:2309.13774 [cond-mat.str-el]} \BibitemShut {NoStop}%
\bibitem [{\citenamefont {Sachdev}\ and\ \citenamefont {Ye}(1993)}]{Sachdev1993}%
  \BibitemOpen
  \bibfield  {author} {\bibinfo {author} {\bibfnamefont {S.}~\bibnamefont {Sachdev}}\ and\ \bibinfo {author} {\bibfnamefont {J.}~\bibnamefont {Ye}},\ }\bibfield  {title} {\bibinfo {title} {{Gapless spin-fluid ground state in a random quantum {H}eisenberg magnet}},\ }\href {https://doi.org/10.1103/PhysRevLett.70.3339} {\bibfield  {journal} {\bibinfo  {journal} {Phys. Rev. Lett.}\ }\textbf {\bibinfo {volume} {70}},\ \bibinfo {pages} {3339} (\bibinfo {year} {1993})}\BibitemShut {NoStop}%
\bibitem [{\citenamefont {Kitaev}(2015)}]{Kitaev2015}%
  \BibitemOpen
  \bibfield  {author} {\bibinfo {author} {\bibfnamefont {A.}~\bibnamefont {Kitaev}},\ }\href@noop {} {\bibinfo {title} {{A simple model of quantum holography}}},\ \bibinfo {howpublished} {proceedings of the KITP Program: Entanglement in Strongly-Correlated Quantum Matter} (\bibinfo {year} {2015})\BibitemShut {NoStop}%
\bibitem [{\citenamefont {Chowdhury}\ \emph {et~al.}(2022)\citenamefont {Chowdhury}, \citenamefont {Georges}, \citenamefont {Parcollet},\ and\ \citenamefont {Sachdev}}]{SYKReview2022}%
  \BibitemOpen
  \bibfield  {author} {\bibinfo {author} {\bibfnamefont {D.}~\bibnamefont {Chowdhury}}, \bibinfo {author} {\bibfnamefont {A.}~\bibnamefont {Georges}}, \bibinfo {author} {\bibfnamefont {O.}~\bibnamefont {Parcollet}},\ and\ \bibinfo {author} {\bibfnamefont {S.}~\bibnamefont {Sachdev}},\ }\bibfield  {title} {\bibinfo {title} {{Sachdev-{Y}e-{K}itaev models and beyond: Window into non-{F}ermi liquids}},\ }\href {https://doi.org/10.1103/RevModPhys.94.035004} {\bibfield  {journal} {\bibinfo  {journal} {Rev. Mod. Phys.}\ }\textbf {\bibinfo {volume} {94}},\ \bibinfo {pages} {035004} (\bibinfo {year} {2022})}\BibitemShut {NoStop}%
\bibitem [{\citenamefont {Esterlis}\ and\ \citenamefont {Schmalian}(2019)}]{EsterlisSchmalian2019}%
  \BibitemOpen
  \bibfield  {author} {\bibinfo {author} {\bibfnamefont {I.}~\bibnamefont {Esterlis}}\ and\ \bibinfo {author} {\bibfnamefont {J.}~\bibnamefont {Schmalian}},\ }\bibfield  {title} {\bibinfo {title} {{Cooper pairing of incoherent electrons: An electron-phonon version of the {S}achdev-{Y}e-{K}itaev model}},\ }\href {https://doi.org/10.1103/PhysRevB.100.115132} {\bibfield  {journal} {\bibinfo  {journal} {Phys. Rev. B}\ }\textbf {\bibinfo {volume} {100}},\ \bibinfo {pages} {115132} (\bibinfo {year} {2019})}\BibitemShut {NoStop}%
\bibitem [{\citenamefont {Wang}(2020)}]{Wang2020}%
  \BibitemOpen
  \bibfield  {author} {\bibinfo {author} {\bibfnamefont {Y.}~\bibnamefont {Wang}},\ }\bibfield  {title} {\bibinfo {title} {{Solvable Strong-Coupling Quantum-Dot Model with a Non-Fermi-Liquid Pairing Transition}},\ }\href {https://doi.org/10.1103/PhysRevLett.124.017002} {\bibfield  {journal} {\bibinfo  {journal} {Phys. Rev. Lett.}\ }\textbf {\bibinfo {volume} {124}},\ \bibinfo {pages} {017002} (\bibinfo {year} {2020})}\BibitemShut {NoStop}%
\bibitem [{\citenamefont {Hauck}\ \emph {et~al.}(2020)\citenamefont {Hauck}, \citenamefont {Klug}, \citenamefont {Esterlis},\ and\ \citenamefont {Schmalian}}]{Hauck2020}%
  \BibitemOpen
  \bibfield  {author} {\bibinfo {author} {\bibfnamefont {D.}~\bibnamefont {Hauck}}, \bibinfo {author} {\bibfnamefont {M.~J.}\ \bibnamefont {Klug}}, \bibinfo {author} {\bibfnamefont {I.}~\bibnamefont {Esterlis}},\ and\ \bibinfo {author} {\bibfnamefont {J.}~\bibnamefont {Schmalian}},\ }\bibfield  {title} {\bibinfo {title} {{Eliashberg equations for an electron–phonon version of the Sachdev–Ye–Kitaev model: Pair breaking in non-Fermi liquid superconductors}},\ }\href {https://doi.org/https://doi.org/10.1016/j.aop.2020.168120} {\bibfield  {journal} {\bibinfo  {journal} {Annals of Physics}\ }\textbf {\bibinfo {volume} {417}},\ \bibinfo {pages} {168120} (\bibinfo {year} {2020})},\ \bibinfo {note} {eliashberg theory at 60: Strong-coupling superconductivity and beyond}\BibitemShut {NoStop}%
\bibitem [{\citenamefont {Sachdev}(2015)}]{Sachdev2015}%
  \BibitemOpen
  \bibfield  {author} {\bibinfo {author} {\bibfnamefont {S.}~\bibnamefont {Sachdev}},\ }\bibfield  {title} {\bibinfo {title} {{Bekenstein-Hawking Entropy and Strange Metals}},\ }\href {https://doi.org/10.1103/PhysRevX.5.041025} {\bibfield  {journal} {\bibinfo  {journal} {Phys. Rev. X}\ }\textbf {\bibinfo {volume} {5}},\ \bibinfo {pages} {041025} (\bibinfo {year} {2015})}\BibitemShut {NoStop}%
\bibitem [{\citenamefont {Maldacena}\ and\ \citenamefont {Stanford}(2016)}]{Maldacena2016}%
  \BibitemOpen
  \bibfield  {author} {\bibinfo {author} {\bibfnamefont {J.}~\bibnamefont {Maldacena}}\ and\ \bibinfo {author} {\bibfnamefont {D.}~\bibnamefont {Stanford}},\ }\bibfield  {title} {\bibinfo {title} {{Remarks on the Sachdev-Ye-Kitaev model}},\ }\href {https://doi.org/10.1103/PhysRevD.94.106002} {\bibfield  {journal} {\bibinfo  {journal} {Phys. Rev. D}\ }\textbf {\bibinfo {volume} {94}},\ \bibinfo {pages} {106002} (\bibinfo {year} {2016})}\BibitemShut {NoStop}%
\bibitem [{\citenamefont {Jensen}(2016)}]{Jensen2016}%
  \BibitemOpen
  \bibfield  {author} {\bibinfo {author} {\bibfnamefont {K.}~\bibnamefont {Jensen}},\ }\bibfield  {title} {\bibinfo {title} {{Chaos in ${\mathrm{AdS}}_{2}$ Holography}},\ }\href {https://doi.org/10.1103/PhysRevLett.117.111601} {\bibfield  {journal} {\bibinfo  {journal} {Phys. Rev. Lett.}\ }\textbf {\bibinfo {volume} {117}},\ \bibinfo {pages} {111601} (\bibinfo {year} {2016})}\BibitemShut {NoStop}%
\bibitem [{\citenamefont {Song}\ \emph {et~al.}(2017)\citenamefont {Song}, \citenamefont {Jian},\ and\ \citenamefont {Balents}}]{Song2017}%
  \BibitemOpen
  \bibfield  {author} {\bibinfo {author} {\bibfnamefont {X.-Y.}\ \bibnamefont {Song}}, \bibinfo {author} {\bibfnamefont {C.-M.}\ \bibnamefont {Jian}},\ and\ \bibinfo {author} {\bibfnamefont {L.}~\bibnamefont {Balents}},\ }\bibfield  {title} {\bibinfo {title} {{Strongly Correlated Metal Built from {S}achdev-{Y}e-{K}itaev Models}},\ }\href {https://doi.org/10.1103/PhysRevLett.119.216601} {\bibfield  {journal} {\bibinfo  {journal} {Phys. Rev. Lett.}\ }\textbf {\bibinfo {volume} {119}},\ \bibinfo {pages} {216601} (\bibinfo {year} {2017})}\BibitemShut {NoStop}%
\bibitem [{\citenamefont {Chowdhury}\ \emph {et~al.}(2018)\citenamefont {Chowdhury}, \citenamefont {Werman}, \citenamefont {Berg},\ and\ \citenamefont {Senthil}}]{Chowdhury2018}%
  \BibitemOpen
  \bibfield  {author} {\bibinfo {author} {\bibfnamefont {D.}~\bibnamefont {Chowdhury}}, \bibinfo {author} {\bibfnamefont {Y.}~\bibnamefont {Werman}}, \bibinfo {author} {\bibfnamefont {E.}~\bibnamefont {Berg}},\ and\ \bibinfo {author} {\bibfnamefont {T.}~\bibnamefont {Senthil}},\ }\bibfield  {title} {\bibinfo {title} {{Translationally Invariant Non-Fermi-Liquid Metals with Critical Fermi Surfaces: Solvable Models}},\ }\href {https://doi.org/10.1103/PhysRevX.8.031024} {\bibfield  {journal} {\bibinfo  {journal} {Phys. Rev. X}\ }\textbf {\bibinfo {volume} {8}},\ \bibinfo {pages} {031024} (\bibinfo {year} {2018})}\BibitemShut {NoStop}%
\bibitem [{\citenamefont {Patel}\ \emph {et~al.}(2018)\citenamefont {Patel}, \citenamefont {McGreevy}, \citenamefont {Arovas},\ and\ \citenamefont {Sachdev}}]{Patel2018}%
  \BibitemOpen
  \bibfield  {author} {\bibinfo {author} {\bibfnamefont {A.~A.}\ \bibnamefont {Patel}}, \bibinfo {author} {\bibfnamefont {J.}~\bibnamefont {McGreevy}}, \bibinfo {author} {\bibfnamefont {D.~P.}\ \bibnamefont {Arovas}},\ and\ \bibinfo {author} {\bibfnamefont {S.}~\bibnamefont {Sachdev}},\ }\bibfield  {title} {\bibinfo {title} {{Magnetotransport in a Model of a Disordered Strange Metal}},\ }\href {https://doi.org/10.1103/PhysRevX.8.021049} {\bibfield  {journal} {\bibinfo  {journal} {Phys. Rev. X}\ }\textbf {\bibinfo {volume} {8}},\ \bibinfo {pages} {021049} (\bibinfo {year} {2018})}\BibitemShut {NoStop}%
\bibitem [{\citenamefont {Cha}\ \emph {et~al.}(2020)\citenamefont {Cha}, \citenamefont {Wentzell}, \citenamefont {Parcollet}, \citenamefont {Georges},\ and\ \citenamefont {Kim}}]{Cha2020}%
  \BibitemOpen
  \bibfield  {author} {\bibinfo {author} {\bibfnamefont {P.}~\bibnamefont {Cha}}, \bibinfo {author} {\bibfnamefont {N.}~\bibnamefont {Wentzell}}, \bibinfo {author} {\bibfnamefont {O.}~\bibnamefont {Parcollet}}, \bibinfo {author} {\bibfnamefont {A.}~\bibnamefont {Georges}},\ and\ \bibinfo {author} {\bibfnamefont {E.-A.}\ \bibnamefont {Kim}},\ }\bibfield  {title} {\bibinfo {title} {{Linear resistivity and {S}achdev-{Y}e-{K}itaev ({SYK}) spin liquid behavior in a quantum critical metal with spin-1/2 fermions}},\ }\href {https://doi.org/10.1073/pnas.2003179117} {\bibfield  {journal} {\bibinfo  {journal} {Proceedings of the National Academy of Sciences}\ }\textbf {\bibinfo {volume} {117}},\ \bibinfo {pages} {18341} (\bibinfo {year} {2020})},\ \Eprint {https://arxiv.org/abs/https://www.pnas.org/doi/pdf/10.1073/pnas.2003179117} {https://www.pnas.org/doi/pdf/10.1073/pnas.2003179117} \BibitemShut {NoStop}%
\bibitem [{\citenamefont {Martin}\ \emph {et~al.}(1990)\citenamefont {Martin}, \citenamefont {Fiory}, \citenamefont {Fleming}, \citenamefont {Schneemeyer},\ and\ \citenamefont {Waszczak}}]{Martin1990}%
  \BibitemOpen
  \bibfield  {author} {\bibinfo {author} {\bibfnamefont {S.}~\bibnamefont {Martin}}, \bibinfo {author} {\bibfnamefont {A.~T.}\ \bibnamefont {Fiory}}, \bibinfo {author} {\bibfnamefont {R.~M.}\ \bibnamefont {Fleming}}, \bibinfo {author} {\bibfnamefont {L.~F.}\ \bibnamefont {Schneemeyer}},\ and\ \bibinfo {author} {\bibfnamefont {J.~V.}\ \bibnamefont {Waszczak}},\ }\bibfield  {title} {\bibinfo {title} {{Normal-state transport properties of ${\mathrm{Bi}}_{2+\mathit{x}}$${\mathrm{Sr}}_{2\mathrm{\ensuremath{-}}\mathit{y}}$${\mathrm{CuO}}_{6+\mathrm{\ensuremath{\delta}}}$ crystals}},\ }\href {https://doi.org/10.1103/PhysRevB.41.846} {\bibfield  {journal} {\bibinfo  {journal} {Phys. Rev. B}\ }\textbf {\bibinfo {volume} {41}},\ \bibinfo {pages} {846} (\bibinfo {year} {1990})}\BibitemShut {NoStop}%
\bibitem [{\citenamefont {Takagi}\ \emph {et~al.}(1992)\citenamefont {Takagi}, \citenamefont {Batlogg}, \citenamefont {Kao}, \citenamefont {Kwo}, \citenamefont {Cava}, \citenamefont {Krajewski},\ and\ \citenamefont {Peck}}]{Takagi1992}%
  \BibitemOpen
  \bibfield  {author} {\bibinfo {author} {\bibfnamefont {H.}~\bibnamefont {Takagi}}, \bibinfo {author} {\bibfnamefont {B.}~\bibnamefont {Batlogg}}, \bibinfo {author} {\bibfnamefont {H.~L.}\ \bibnamefont {Kao}}, \bibinfo {author} {\bibfnamefont {J.}~\bibnamefont {Kwo}}, \bibinfo {author} {\bibfnamefont {R.~J.}\ \bibnamefont {Cava}}, \bibinfo {author} {\bibfnamefont {J.~J.}\ \bibnamefont {Krajewski}},\ and\ \bibinfo {author} {\bibfnamefont {W.~F.}\ \bibnamefont {Peck}},\ }\bibfield  {title} {\bibinfo {title} {{Systematic evolution of temperature-dependent resistivity in ${\mathrm{La}}_{2\mathrm{\ensuremath{-}}\mathit{x}}$${\mathrm{Sr}}_{\mathit{x}}$${\mathrm{CuO}}_{4}$}},\ }\href {https://doi.org/10.1103/PhysRevLett.69.2975} {\bibfield  {journal} {\bibinfo  {journal} {Phys. Rev. Lett.}\ }\textbf {\bibinfo {volume} {69}},\ \bibinfo {pages} {2975} (\bibinfo {year} {1992})}\BibitemShut {NoStop}%
\bibitem [{\citenamefont {Valla}\ \emph {et~al.}(2000)\citenamefont {Valla}, \citenamefont {Fedorov}, \citenamefont {Johnson}, \citenamefont {Li}, \citenamefont {Gu},\ and\ \citenamefont {Koshizuka}}]{Valla2000}%
  \BibitemOpen
  \bibfield  {author} {\bibinfo {author} {\bibfnamefont {T.}~\bibnamefont {Valla}}, \bibinfo {author} {\bibfnamefont {A.~V.}\ \bibnamefont {Fedorov}}, \bibinfo {author} {\bibfnamefont {P.~D.}\ \bibnamefont {Johnson}}, \bibinfo {author} {\bibfnamefont {Q.}~\bibnamefont {Li}}, \bibinfo {author} {\bibfnamefont {G.~D.}\ \bibnamefont {Gu}},\ and\ \bibinfo {author} {\bibfnamefont {N.}~\bibnamefont {Koshizuka}},\ }\bibfield  {title} {\bibinfo {title} {{Temperature Dependent Scattering Rates at the Fermi Surface of Optimally Doped ${\mathrm{Bi}}_{2}{\mathrm{Sr}}_{2}{\mathrm{CaCu}}_{2}{O}_{8+\mathit{\ensuremath{\delta}}}$}},\ }\href {https://doi.org/10.1103/PhysRevLett.85.828} {\bibfield  {journal} {\bibinfo  {journal} {Phys. Rev. Lett.}\ }\textbf {\bibinfo {volume} {85}},\ \bibinfo {pages} {828} (\bibinfo {year} {2000})}\BibitemShut {NoStop}%
\bibitem [{\citenamefont {Chen}\ \emph {et~al.}(2006)\citenamefont {Chen}, \citenamefont {Sambandamurthy}, \citenamefont {Wang}, \citenamefont {Lewis}, \citenamefont {Engel}, \citenamefont {Tsui}, \citenamefont {Ye}, \citenamefont {Pfeiffer},\ and\ \citenamefont {West}}]{Chen2006}%
  \BibitemOpen
  \bibfield  {author} {\bibinfo {author} {\bibfnamefont {Y.~P.}\ \bibnamefont {Chen}}, \bibinfo {author} {\bibfnamefont {G.}~\bibnamefont {Sambandamurthy}}, \bibinfo {author} {\bibfnamefont {Z.~H.}\ \bibnamefont {Wang}}, \bibinfo {author} {\bibfnamefont {R.~M.}\ \bibnamefont {Lewis}}, \bibinfo {author} {\bibfnamefont {L.~W.}\ \bibnamefont {Engel}}, \bibinfo {author} {\bibfnamefont {D.~C.}\ \bibnamefont {Tsui}}, \bibinfo {author} {\bibfnamefont {P.~D.}\ \bibnamefont {Ye}}, \bibinfo {author} {\bibfnamefont {L.~N.}\ \bibnamefont {Pfeiffer}},\ and\ \bibinfo {author} {\bibfnamefont {K.~W.}\ \bibnamefont {West}},\ }\bibfield  {title} {\bibinfo {title} {{Melting of a 2D quantum electron solid in high magnetic field}},\ }\href {https://doi.org/10.1038/nphys322} {\bibfield  {journal} {\bibinfo  {journal} {Nature Physics}\ }\textbf {\bibinfo {volume} {2}},\ \bibinfo {pages} {452} (\bibinfo {year} {2006})}\BibitemShut {NoStop}%
\bibitem [{\citenamefont {Azeyanagi}\ \emph {et~al.}(2018)\citenamefont {Azeyanagi}, \citenamefont {Ferrari},\ and\ \citenamefont {Massolo}}]{SYKPhaseDiagram2018}%
  \BibitemOpen
  \bibfield  {author} {\bibinfo {author} {\bibfnamefont {T.}~\bibnamefont {Azeyanagi}}, \bibinfo {author} {\bibfnamefont {F.}~\bibnamefont {Ferrari}},\ and\ \bibinfo {author} {\bibfnamefont {F.~I.~S.}\ \bibnamefont {Massolo}},\ }\bibfield  {title} {\bibinfo {title} {{Phase Diagram of Planar Matrix Quantum Mechanics, Tensor, and {S}achdev-{Y}e-{K}itaev Models}},\ }\href {https://doi.org/10.1103/PhysRevLett.120.061602} {\bibfield  {journal} {\bibinfo  {journal} {Phys. Rev. Lett.}\ }\textbf {\bibinfo {volume} {120}},\ \bibinfo {pages} {061602} (\bibinfo {year} {2018})}\BibitemShut {NoStop}%
\bibitem [{\citenamefont {Ferrari}\ and\ \citenamefont {Schaposnik~Massolo}(2019)}]{PhasesofMelonicQM2019}%
  \BibitemOpen
  \bibfield  {author} {\bibinfo {author} {\bibfnamefont {F.}~\bibnamefont {Ferrari}}\ and\ \bibinfo {author} {\bibfnamefont {F.~I.}\ \bibnamefont {Schaposnik~Massolo}},\ }\bibfield  {title} {\bibinfo {title} {{Phases of melonic quantum mechanics}},\ }\href {https://doi.org/10.1103/PhysRevD.100.026007} {\bibfield  {journal} {\bibinfo  {journal} {Phys. Rev. D}\ }\textbf {\bibinfo {volume} {100}},\ \bibinfo {pages} {026007} (\bibinfo {year} {2019})}\BibitemShut {NoStop}%
\bibitem [{\citenamefont {Louw}\ and\ \citenamefont {Kehrein}(2023)}]{Louw2023}%
  \BibitemOpen
  \bibfield  {author} {\bibinfo {author} {\bibfnamefont {J.~C.}\ \bibnamefont {Louw}}\ and\ \bibinfo {author} {\bibfnamefont {S.}~\bibnamefont {Kehrein}},\ }\bibfield  {title} {\bibinfo {title} {{Shared universality of charged black holes and the complex large-$q$ Sachdev-Ye-Kitaev model}},\ }\href {https://doi.org/10.1103/PhysRevB.107.075132} {\bibfield  {journal} {\bibinfo  {journal} {Phys. Rev. B}\ }\textbf {\bibinfo {volume} {107}},\ \bibinfo {pages} {075132} (\bibinfo {year} {2023})}\BibitemShut {NoStop}%
\bibitem [{\citenamefont {Willett}\ \emph {et~al.}(1988)\citenamefont {Willett}, \citenamefont {Stormer}, \citenamefont {Tsui}, \citenamefont {Pfeiffer}, \citenamefont {West},\ and\ \citenamefont {Baldwin}}]{Willett1988}%
  \BibitemOpen
  \bibfield  {author} {\bibinfo {author} {\bibfnamefont {R.~L.}\ \bibnamefont {Willett}}, \bibinfo {author} {\bibfnamefont {H.~L.}\ \bibnamefont {Stormer}}, \bibinfo {author} {\bibfnamefont {D.~C.}\ \bibnamefont {Tsui}}, \bibinfo {author} {\bibfnamefont {L.~N.}\ \bibnamefont {Pfeiffer}}, \bibinfo {author} {\bibfnamefont {K.~W.}\ \bibnamefont {West}},\ and\ \bibinfo {author} {\bibfnamefont {K.~W.}\ \bibnamefont {Baldwin}},\ }\bibfield  {title} {\bibinfo {title} {{Termination of the series of fractional quantum {H}all states at small filling factors}},\ }\href {https://doi.org/10.1103/PhysRevB.38.7881} {\bibfield  {journal} {\bibinfo  {journal} {Phys. Rev. B}\ }\textbf {\bibinfo {volume} {38}},\ \bibinfo {pages} {7881} (\bibinfo {year} {1988})}\BibitemShut {NoStop}%
\bibitem [{\citenamefont {Jiang}\ \emph {et~al.}(1990)\citenamefont {Jiang}, \citenamefont {Willett}, \citenamefont {Stormer}, \citenamefont {Tsui}, \citenamefont {Pfeiffer},\ and\ \citenamefont {West}}]{Jiang1990}%
  \BibitemOpen
  \bibfield  {author} {\bibinfo {author} {\bibfnamefont {H.~W.}\ \bibnamefont {Jiang}}, \bibinfo {author} {\bibfnamefont {R.~L.}\ \bibnamefont {Willett}}, \bibinfo {author} {\bibfnamefont {H.~L.}\ \bibnamefont {Stormer}}, \bibinfo {author} {\bibfnamefont {D.~C.}\ \bibnamefont {Tsui}}, \bibinfo {author} {\bibfnamefont {L.~N.}\ \bibnamefont {Pfeiffer}},\ and\ \bibinfo {author} {\bibfnamefont {K.~W.}\ \bibnamefont {West}},\ }\bibfield  {title} {\bibinfo {title} {{Quantum liquid versus electron solid around \ensuremath{\nu}=1/5 {L}andau-level filling}},\ }\href {https://doi.org/10.1103/PhysRevLett.65.633} {\bibfield  {journal} {\bibinfo  {journal} {Phys. Rev. Lett.}\ }\textbf {\bibinfo {volume} {65}},\ \bibinfo {pages} {633} (\bibinfo {year} {1990})}\BibitemShut {NoStop}%
\bibitem [{\citenamefont {Goldman}\ \emph {et~al.}(1990)\citenamefont {Goldman}, \citenamefont {Santos}, \citenamefont {Shayegan},\ and\ \citenamefont {Cunningham}}]{Goldman1990}%
  \BibitemOpen
  \bibfield  {author} {\bibinfo {author} {\bibfnamefont {V.~J.}\ \bibnamefont {Goldman}}, \bibinfo {author} {\bibfnamefont {M.}~\bibnamefont {Santos}}, \bibinfo {author} {\bibfnamefont {M.}~\bibnamefont {Shayegan}},\ and\ \bibinfo {author} {\bibfnamefont {J.~E.}\ \bibnamefont {Cunningham}},\ }\bibfield  {title} {\bibinfo {title} {{Evidence for two-dimentional quantum {W}igner crystal}},\ }\href {https://doi.org/10.1103/PhysRevLett.65.2189} {\bibfield  {journal} {\bibinfo  {journal} {Phys. Rev. Lett.}\ }\textbf {\bibinfo {volume} {65}},\ \bibinfo {pages} {2189} (\bibinfo {year} {1990})}\BibitemShut {NoStop}%
\bibitem [{\citenamefont {Jiang}\ \emph {et~al.}(1991)\citenamefont {Jiang}, \citenamefont {Stormer}, \citenamefont {Tsui}, \citenamefont {Pfeiffer},\ and\ \citenamefont {West}}]{Jiang1991}%
  \BibitemOpen
  \bibfield  {author} {\bibinfo {author} {\bibfnamefont {H.~W.}\ \bibnamefont {Jiang}}, \bibinfo {author} {\bibfnamefont {H.~L.}\ \bibnamefont {Stormer}}, \bibinfo {author} {\bibfnamefont {D.~C.}\ \bibnamefont {Tsui}}, \bibinfo {author} {\bibfnamefont {L.~N.}\ \bibnamefont {Pfeiffer}},\ and\ \bibinfo {author} {\bibfnamefont {K.~W.}\ \bibnamefont {West}},\ }\bibfield  {title} {\bibinfo {title} {{Magnetotransport studies of the insulating phase around \ensuremath{\nu}=1/5 {L}andau-level filling}},\ }\href {https://doi.org/10.1103/PhysRevB.44.8107} {\bibfield  {journal} {\bibinfo  {journal} {Phys. Rev. B}\ }\textbf {\bibinfo {volume} {44}},\ \bibinfo {pages} {8107} (\bibinfo {year} {1991})}\BibitemShut {NoStop}%
\bibitem [{\citenamefont {Williams}\ \emph {et~al.}(1991)\citenamefont {Williams}, \citenamefont {Wright}, \citenamefont {Clark}, \citenamefont {Andrei}, \citenamefont {Deville}, \citenamefont {Glattli}, \citenamefont {Probst}, \citenamefont {Etienne}, \citenamefont {Dorin}, \citenamefont {Foxon},\ and\ \citenamefont {Harris}}]{Williams1991}%
  \BibitemOpen
  \bibfield  {author} {\bibinfo {author} {\bibfnamefont {F.~I.~B.}\ \bibnamefont {Williams}}, \bibinfo {author} {\bibfnamefont {P.~A.}\ \bibnamefont {Wright}}, \bibinfo {author} {\bibfnamefont {R.~G.}\ \bibnamefont {Clark}}, \bibinfo {author} {\bibfnamefont {E.~Y.}\ \bibnamefont {Andrei}}, \bibinfo {author} {\bibfnamefont {G.}~\bibnamefont {Deville}}, \bibinfo {author} {\bibfnamefont {D.~C.}\ \bibnamefont {Glattli}}, \bibinfo {author} {\bibfnamefont {O.}~\bibnamefont {Probst}}, \bibinfo {author} {\bibfnamefont {B.}~\bibnamefont {Etienne}}, \bibinfo {author} {\bibfnamefont {C.}~\bibnamefont {Dorin}}, \bibinfo {author} {\bibfnamefont {C.~T.}\ \bibnamefont {Foxon}},\ and\ \bibinfo {author} {\bibfnamefont {J.~J.}\ \bibnamefont {Harris}},\ }\bibfield  {title} {\bibinfo {title} {{Conduction threshold and pinning frequency of magnetically induced Wigner solid}},\ }\href {https://doi.org/10.1103/PhysRevLett.66.3285} {\bibfield  {journal} {\bibinfo  {journal} {Phys. Rev. Lett.}\ }\textbf {\bibinfo {volume} {66}},\
  \bibinfo {pages} {3285} (\bibinfo {year} {1991})}\BibitemShut {NoStop}%
\bibitem [{\citenamefont {Paalanen}\ \emph {et~al.}(1992)\citenamefont {Paalanen}, \citenamefont {Willett}, \citenamefont {Ruel}, \citenamefont {Littlewood}, \citenamefont {West},\ and\ \citenamefont {Pfeiffer}}]{Paalanen1992_2}%
  \BibitemOpen
  \bibfield  {author} {\bibinfo {author} {\bibfnamefont {M.~A.}\ \bibnamefont {Paalanen}}, \bibinfo {author} {\bibfnamefont {R.~L.}\ \bibnamefont {Willett}}, \bibinfo {author} {\bibfnamefont {R.~R.}\ \bibnamefont {Ruel}}, \bibinfo {author} {\bibfnamefont {P.~B.}\ \bibnamefont {Littlewood}}, \bibinfo {author} {\bibfnamefont {K.~W.}\ \bibnamefont {West}},\ and\ \bibinfo {author} {\bibfnamefont {L.~N.}\ \bibnamefont {Pfeiffer}},\ }\bibfield  {title} {\bibinfo {title} {{Electrical conductivity and {W}igner crystallization}},\ }\href {https://doi.org/10.1103/PhysRevB.45.13784} {\bibfield  {journal} {\bibinfo  {journal} {Phys. Rev. B}\ }\textbf {\bibinfo {volume} {45}},\ \bibinfo {pages} {13784} (\bibinfo {year} {1992})}\BibitemShut {NoStop}%
\bibitem [{\citenamefont {Kukushkin}\ \emph {et~al.}(1993)\citenamefont {Kukushkin}, \citenamefont {Pulsford}, \citenamefont {von Klitzing}, \citenamefont {Haug}, \citenamefont {Ploog},\ and\ \citenamefont {Timofeev}}]{Kukushkin1993}%
  \BibitemOpen
  \bibfield  {author} {\bibinfo {author} {\bibfnamefont {I.~V.}\ \bibnamefont {Kukushkin}}, \bibinfo {author} {\bibfnamefont {N.~J.}\ \bibnamefont {Pulsford}}, \bibinfo {author} {\bibfnamefont {K.}~\bibnamefont {von Klitzing}}, \bibinfo {author} {\bibfnamefont {R.~J.}\ \bibnamefont {Haug}}, \bibinfo {author} {\bibfnamefont {K.}~\bibnamefont {Ploog}},\ and\ \bibinfo {author} {\bibfnamefont {V.~B.}\ \bibnamefont {Timofeev}},\ }\bibfield  {title} {\bibinfo {title} {{Wigner Solid vs. Incompressible Laughlin Liquid: Phase Diagram Derived from Time-Resolved Photoluminescence}},\ }\href {https://doi.org/10.1209/0295-5075/23/3/009} {\bibfield  {journal} {\bibinfo  {journal} {Europhysics Letters}\ }\textbf {\bibinfo {volume} {23}},\ \bibinfo {pages} {211} (\bibinfo {year} {1993})}\BibitemShut {NoStop}%
\bibitem [{\citenamefont {Bergholtz}\ and\ \citenamefont {Karlhede}(2005)}]{Bergholtz2005}%
  \BibitemOpen
  \bibfield  {author} {\bibinfo {author} {\bibfnamefont {E.~J.}\ \bibnamefont {Bergholtz}}\ and\ \bibinfo {author} {\bibfnamefont {A.}~\bibnamefont {Karlhede}},\ }\bibfield  {title} {\bibinfo {title} {{Half-Filled Lowest Landau Level on a Thin Torus}},\ }\href {https://doi.org/10.1103/PhysRevLett.94.026802} {\bibfield  {journal} {\bibinfo  {journal} {Phys. Rev. Lett.}\ }\textbf {\bibinfo {volume} {94}},\ \bibinfo {pages} {026802} (\bibinfo {year} {2005})}\BibitemShut {NoStop}%
\bibitem [{\citenamefont {Bergholtz}\ and\ \citenamefont {Karlhede}(2006)}]{Bergholtz_2006}%
  \BibitemOpen
  \bibfield  {author} {\bibinfo {author} {\bibfnamefont {E.~J.}\ \bibnamefont {Bergholtz}}\ and\ \bibinfo {author} {\bibfnamefont {A.}~\bibnamefont {Karlhede}},\ }\bibfield  {title} {\bibinfo {title} {{‘{O}ne-dimensional’ theory of the quantum {H}all system}},\ }\href {https://doi.org/10.1088/1742-5468/2006/04/L04001} {\bibfield  {journal} {\bibinfo  {journal} {Journal of Statistical Mechanics: Theory and Experiment}\ }\textbf {\bibinfo {volume} {2006}},\ \bibinfo {pages} {L04001} (\bibinfo {year} {2006})}\BibitemShut {NoStop}%
\bibitem [{\citenamefont {Bergholtz}\ \emph {et~al.}(2007)\citenamefont {Bergholtz}, \citenamefont {Hansson}, \citenamefont {Hermanns},\ and\ \citenamefont {Karlhede}}]{Bergholtz2007}%
  \BibitemOpen
  \bibfield  {author} {\bibinfo {author} {\bibfnamefont {E.~J.}\ \bibnamefont {Bergholtz}}, \bibinfo {author} {\bibfnamefont {T.~H.}\ \bibnamefont {Hansson}}, \bibinfo {author} {\bibfnamefont {M.}~\bibnamefont {Hermanns}},\ and\ \bibinfo {author} {\bibfnamefont {A.}~\bibnamefont {Karlhede}},\ }\bibfield  {title} {\bibinfo {title} {{Microscopic Theory of the Quantum {H}all Hierarchy}},\ }\href {https://doi.org/10.1103/PhysRevLett.99.256803} {\bibfield  {journal} {\bibinfo  {journal} {Phys. Rev. Lett.}\ }\textbf {\bibinfo {volume} {99}},\ \bibinfo {pages} {256803} (\bibinfo {year} {2007})}\BibitemShut {NoStop}%
\bibitem [{\citenamefont {Bergholtz}\ and\ \citenamefont {Karlhede}(2008)}]{Bergholtz2008}%
  \BibitemOpen
  \bibfield  {author} {\bibinfo {author} {\bibfnamefont {E.~J.}\ \bibnamefont {Bergholtz}}\ and\ \bibinfo {author} {\bibfnamefont {A.}~\bibnamefont {Karlhede}},\ }\bibfield  {title} {\bibinfo {title} {{Quantum {H}all system in {T}ao-{T}houless limit}},\ }\href {https://doi.org/10.1103/PhysRevB.77.155308} {\bibfield  {journal} {\bibinfo  {journal} {Phys. Rev. B}\ }\textbf {\bibinfo {volume} {77}},\ \bibinfo {pages} {155308} (\bibinfo {year} {2008})}\BibitemShut {NoStop}%
\bibitem [{\citenamefont {Cao}\ \emph {et~al.}(2020)\citenamefont {Cao}, \citenamefont {Chowdhury}, \citenamefont {Rodan-Legrain}, \citenamefont {Rubies-Bigorda}, \citenamefont {Watanabe}, \citenamefont {Taniguchi}, \citenamefont {Senthil},\ and\ \citenamefont {Jarillo-Herrero}}]{Cao2020}%
  \BibitemOpen
  \bibfield  {author} {\bibinfo {author} {\bibfnamefont {Y.}~\bibnamefont {Cao}}, \bibinfo {author} {\bibfnamefont {D.}~\bibnamefont {Chowdhury}}, \bibinfo {author} {\bibfnamefont {D.}~\bibnamefont {Rodan-Legrain}}, \bibinfo {author} {\bibfnamefont {O.}~\bibnamefont {Rubies-Bigorda}}, \bibinfo {author} {\bibfnamefont {K.}~\bibnamefont {Watanabe}}, \bibinfo {author} {\bibfnamefont {T.}~\bibnamefont {Taniguchi}}, \bibinfo {author} {\bibfnamefont {T.}~\bibnamefont {Senthil}},\ and\ \bibinfo {author} {\bibfnamefont {P.}~\bibnamefont {Jarillo-Herrero}},\ }\bibfield  {title} {\bibinfo {title} {{Strange Metal in Magic-Angle Graphene with near Planckian Dissipation}},\ }\href {https://doi.org/10.1103/PhysRevLett.124.076801} {\bibfield  {journal} {\bibinfo  {journal} {Phys. Rev. Lett.}\ }\textbf {\bibinfo {volume} {124}},\ \bibinfo {pages} {076801} (\bibinfo {year} {2020})}\BibitemShut {NoStop}%
\bibitem [{\citenamefont {Lyu}\ \emph {et~al.}(2021)\citenamefont {Lyu}, \citenamefont {Tuchfeld}, \citenamefont {Verma}, \citenamefont {Tian}, \citenamefont {Watanabe}, \citenamefont {Taniguchi}, \citenamefont {Lau}, \citenamefont {Randeria},\ and\ \citenamefont {Bockrath}}]{Lyu2021}%
  \BibitemOpen
  \bibfield  {author} {\bibinfo {author} {\bibfnamefont {R.}~\bibnamefont {Lyu}}, \bibinfo {author} {\bibfnamefont {Z.}~\bibnamefont {Tuchfeld}}, \bibinfo {author} {\bibfnamefont {N.}~\bibnamefont {Verma}}, \bibinfo {author} {\bibfnamefont {H.}~\bibnamefont {Tian}}, \bibinfo {author} {\bibfnamefont {K.}~\bibnamefont {Watanabe}}, \bibinfo {author} {\bibfnamefont {T.}~\bibnamefont {Taniguchi}}, \bibinfo {author} {\bibfnamefont {C.~N.}\ \bibnamefont {Lau}}, \bibinfo {author} {\bibfnamefont {M.}~\bibnamefont {Randeria}},\ and\ \bibinfo {author} {\bibfnamefont {M.}~\bibnamefont {Bockrath}},\ }\bibfield  {title} {\bibinfo {title} {{Strange metal behavior of the Hall angle in twisted bilayer graphene}},\ }\href {https://doi.org/10.1103/PhysRevB.103.245424} {\bibfield  {journal} {\bibinfo  {journal} {Phys. Rev. B}\ }\textbf {\bibinfo {volume} {103}},\ \bibinfo {pages} {245424} (\bibinfo {year} {2021})}\BibitemShut {NoStop}%
\bibitem [{\citenamefont {Jaoui}\ \emph {et~al.}(2022)\citenamefont {Jaoui}, \citenamefont {Das}, \citenamefont {Di~Battista}, \citenamefont {D{\'i}ez-M{\'e}rida}, \citenamefont {Lu}, \citenamefont {Watanabe}, \citenamefont {Taniguchi}, \citenamefont {Ishizuka}, \citenamefont {Levitov},\ and\ \citenamefont {Efetov}}]{Jaoui2022}%
  \BibitemOpen
  \bibfield  {author} {\bibinfo {author} {\bibfnamefont {A.}~\bibnamefont {Jaoui}}, \bibinfo {author} {\bibfnamefont {I.}~\bibnamefont {Das}}, \bibinfo {author} {\bibfnamefont {G.}~\bibnamefont {Di~Battista}}, \bibinfo {author} {\bibfnamefont {J.}~\bibnamefont {D{\'i}ez-M{\'e}rida}}, \bibinfo {author} {\bibfnamefont {X.}~\bibnamefont {Lu}}, \bibinfo {author} {\bibfnamefont {K.}~\bibnamefont {Watanabe}}, \bibinfo {author} {\bibfnamefont {T.}~\bibnamefont {Taniguchi}}, \bibinfo {author} {\bibfnamefont {H.}~\bibnamefont {Ishizuka}}, \bibinfo {author} {\bibfnamefont {L.}~\bibnamefont {Levitov}},\ and\ \bibinfo {author} {\bibfnamefont {D.~K.}\ \bibnamefont {Efetov}},\ }\bibfield  {title} {\bibinfo {title} {{Quantum critical behaviour in magic-angle twisted bilayer graphene}},\ }\href {https://doi.org/10.1038/s41567-022-01556-5} {\bibfield  {journal} {\bibinfo  {journal} {Nature Physics}\ }\textbf {\bibinfo {volume} {18}},\ \bibinfo {pages} {633} (\bibinfo {year} {2022})}\BibitemShut {NoStop}%
\bibitem [{\citenamefont {Cao}\ \emph {et~al.}(2018)\citenamefont {Cao}, \citenamefont {Fatemi}, \citenamefont {Demir}, \citenamefont {Fang}, \citenamefont {Tomarken}, \citenamefont {Luo}, \citenamefont {Sanchez-Yamagishi}, \citenamefont {Watanabe}, \citenamefont {Taniguchi}, \citenamefont {Kaxiras}, \citenamefont {Ashoori},\ and\ \citenamefont {Jarillo-Herrero}}]{Cao2018}%
  \BibitemOpen
  \bibfield  {author} {\bibinfo {author} {\bibfnamefont {Y.}~\bibnamefont {Cao}}, \bibinfo {author} {\bibfnamefont {V.}~\bibnamefont {Fatemi}}, \bibinfo {author} {\bibfnamefont {A.}~\bibnamefont {Demir}}, \bibinfo {author} {\bibfnamefont {S.}~\bibnamefont {Fang}}, \bibinfo {author} {\bibfnamefont {S.~L.}\ \bibnamefont {Tomarken}}, \bibinfo {author} {\bibfnamefont {J.~Y.}\ \bibnamefont {Luo}}, \bibinfo {author} {\bibfnamefont {J.~D.}\ \bibnamefont {Sanchez-Yamagishi}}, \bibinfo {author} {\bibfnamefont {K.}~\bibnamefont {Watanabe}}, \bibinfo {author} {\bibfnamefont {T.}~\bibnamefont {Taniguchi}}, \bibinfo {author} {\bibfnamefont {E.}~\bibnamefont {Kaxiras}}, \bibinfo {author} {\bibfnamefont {R.~C.}\ \bibnamefont {Ashoori}},\ and\ \bibinfo {author} {\bibfnamefont {P.}~\bibnamefont {Jarillo-Herrero}},\ }\bibfield  {title} {\bibinfo {title} {{Correlated insulator behaviour at half-filling in magic-angle graphene superlattices}},\ }\href {https://doi.org/10.1038/nature26154} {\bibfield  {journal} {\bibinfo
  {journal} {Nature}\ }\textbf {\bibinfo {volume} {556}},\ \bibinfo {pages} {80} (\bibinfo {year} {2018})}\BibitemShut {NoStop}%
\bibitem [{\citenamefont {Zhang}\ \emph {et~al.}(2022)\citenamefont {Zhang}, \citenamefont {Lu},\ and\ \citenamefont {Liu}}]{Zhang2022}%
  \BibitemOpen
  \bibfield  {author} {\bibinfo {author} {\bibfnamefont {S.}~\bibnamefont {Zhang}}, \bibinfo {author} {\bibfnamefont {X.}~\bibnamefont {Lu}},\ and\ \bibinfo {author} {\bibfnamefont {J.}~\bibnamefont {Liu}},\ }\bibfield  {title} {\bibinfo {title} {{Correlated Insulators, Density Wave States, and Their Nonlinear Optical Response in Magic-Angle Twisted Bilayer Graphene}},\ }\href {https://doi.org/10.1103/PhysRevLett.128.247402} {\bibfield  {journal} {\bibinfo  {journal} {Phys. Rev. Lett.}\ }\textbf {\bibinfo {volume} {128}},\ \bibinfo {pages} {247402} (\bibinfo {year} {2022})}\BibitemShut {NoStop}%
\bibitem [{\citenamefont {Chen}\ \emph {et~al.}(2018)\citenamefont {Chen}, \citenamefont {Ilan}, \citenamefont {de~Juan}, \citenamefont {Pikulin},\ and\ \citenamefont {Franz}}]{Chen2018}%
  \BibitemOpen
  \bibfield  {author} {\bibinfo {author} {\bibfnamefont {A.}~\bibnamefont {Chen}}, \bibinfo {author} {\bibfnamefont {R.}~\bibnamefont {Ilan}}, \bibinfo {author} {\bibfnamefont {F.}~\bibnamefont {de~Juan}}, \bibinfo {author} {\bibfnamefont {D.~I.}\ \bibnamefont {Pikulin}},\ and\ \bibinfo {author} {\bibfnamefont {M.}~\bibnamefont {Franz}},\ }\bibfield  {title} {\bibinfo {title} {{Quantum Holography in a Graphene Flake with an Irregular Boundary}},\ }\href {https://doi.org/10.1103/PhysRevLett.121.036403} {\bibfield  {journal} {\bibinfo  {journal} {Phys. Rev. Lett.}\ }\textbf {\bibinfo {volume} {121}},\ \bibinfo {pages} {036403} (\bibinfo {year} {2018})}\BibitemShut {NoStop}%
\bibitem [{\citenamefont {Brzezi\ifmmode~\acute{n}\else \'{n}\fi{}ska}\ \emph {et~al.}(2023)\citenamefont {Brzezi\ifmmode~\acute{n}\else \'{n}\fi{}ska}, \citenamefont {Guan}, \citenamefont {Yazyev}, \citenamefont {Sachdev},\ and\ \citenamefont {Kruchkov}}]{Brzezinska2023}%
  \BibitemOpen
  \bibfield  {author} {\bibinfo {author} {\bibfnamefont {M.}~\bibnamefont {Brzezi\ifmmode~\acute{n}\else \'{n}\fi{}ska}}, \bibinfo {author} {\bibfnamefont {Y.}~\bibnamefont {Guan}}, \bibinfo {author} {\bibfnamefont {O.~V.}\ \bibnamefont {Yazyev}}, \bibinfo {author} {\bibfnamefont {S.}~\bibnamefont {Sachdev}},\ and\ \bibinfo {author} {\bibfnamefont {A.}~\bibnamefont {Kruchkov}},\ }\bibfield  {title} {\bibinfo {title} {{Engineering {SYK} Interactions in Disordered Graphene Flakes under Realistic Experimental Conditions}},\ }\href {https://doi.org/10.1103/PhysRevLett.131.036503} {\bibfield  {journal} {\bibinfo  {journal} {Phys. Rev. Lett.}\ }\textbf {\bibinfo {volume} {131}},\ \bibinfo {pages} {036503} (\bibinfo {year} {2023})}\BibitemShut {NoStop}%
\bibitem [{\citenamefont {Seidel}\ \emph {et~al.}(2005)\citenamefont {Seidel}, \citenamefont {Fu}, \citenamefont {Lee}, \citenamefont {Leinaas},\ and\ \citenamefont {Moore}}]{Seidel2005}%
  \BibitemOpen
  \bibfield  {author} {\bibinfo {author} {\bibfnamefont {A.}~\bibnamefont {Seidel}}, \bibinfo {author} {\bibfnamefont {H.}~\bibnamefont {Fu}}, \bibinfo {author} {\bibfnamefont {D.-H.}\ \bibnamefont {Lee}}, \bibinfo {author} {\bibfnamefont {J.~M.}\ \bibnamefont {Leinaas}},\ and\ \bibinfo {author} {\bibfnamefont {J.}~\bibnamefont {Moore}},\ }\bibfield  {title} {\bibinfo {title} {{Incompressible Quantum Liquids and New Conservation Laws}},\ }\href {https://doi.org/10.1103/PhysRevLett.95.266405} {\bibfield  {journal} {\bibinfo  {journal} {Phys. Rev. Lett.}\ }\textbf {\bibinfo {volume} {95}},\ \bibinfo {pages} {266405} (\bibinfo {year} {2005})}\BibitemShut {NoStop}%
\bibitem [{\citenamefont {Tao}\ \emph {et~al.}(1989)\citenamefont {Tao}, \citenamefont {Widom}, \citenamefont {Tao},\ and\ \citenamefont {Sim}}]{Tao1989}%
  \BibitemOpen
  \bibfield  {author} {\bibinfo {author} {\bibfnamefont {R.}~\bibnamefont {Tao}}, \bibinfo {author} {\bibfnamefont {A.}~\bibnamefont {Widom}}, \bibinfo {author} {\bibfnamefont {Z.~C.}\ \bibnamefont {Tao}},\ and\ \bibinfo {author} {\bibfnamefont {H.-K.}\ \bibnamefont {Sim}},\ }\bibfield  {title} {\bibinfo {title} {{THERMODYNAMIC STABILITY OF THE TWO-DIMENSIONAL JELLIUM MODEL IN A STRONG MAGNETIC FIELD}},\ }\href {https://doi.org/10.1142/S0217979289000129} {\bibfield  {journal} {\bibinfo  {journal} {International Journal of Modern Physics B}\ }\textbf {\bibinfo {volume} {03}},\ \bibinfo {pages} {129} (\bibinfo {year} {1989})},\ \Eprint {https://arxiv.org/abs/https://doi.org/10.1142/S0217979289000129} {https://doi.org/10.1142/S0217979289000129} \BibitemShut {NoStop}%
\bibitem [{\citenamefont {Kravchenko}\ \emph {et~al.}(1990)\citenamefont {Kravchenko}, \citenamefont {Rinberg}, \citenamefont {Semenchinsky},\ and\ \citenamefont {Pudalov}}]{Kravchenko1990}%
  \BibitemOpen
  \bibfield  {author} {\bibinfo {author} {\bibfnamefont {S.~V.}\ \bibnamefont {Kravchenko}}, \bibinfo {author} {\bibfnamefont {D.~A.}\ \bibnamefont {Rinberg}}, \bibinfo {author} {\bibfnamefont {S.~G.}\ \bibnamefont {Semenchinsky}},\ and\ \bibinfo {author} {\bibfnamefont {V.~M.}\ \bibnamefont {Pudalov}},\ }\bibfield  {title} {\bibinfo {title} {{Evidence for the influence of electron-electron interaction on the chemical potential of the two-dimensional electron gas}},\ }\href {https://doi.org/10.1103/PhysRevB.42.3741} {\bibfield  {journal} {\bibinfo  {journal} {Phys. Rev. B}\ }\textbf {\bibinfo {volume} {42}},\ \bibinfo {pages} {3741} (\bibinfo {year} {1990})}\BibitemShut {NoStop}%
\bibitem [{\citenamefont {Eisenstein}\ \emph {et~al.}(1992{\natexlab{b}})\citenamefont {Eisenstein}, \citenamefont {Pfeiffer},\ and\ \citenamefont {West}}]{Eisenstein1992a}%
  \BibitemOpen
  \bibfield  {author} {\bibinfo {author} {\bibfnamefont {J.~P.}\ \bibnamefont {Eisenstein}}, \bibinfo {author} {\bibfnamefont {L.~N.}\ \bibnamefont {Pfeiffer}},\ and\ \bibinfo {author} {\bibfnamefont {K.~W.}\ \bibnamefont {West}},\ }\bibfield  {title} {\bibinfo {title} {{Negative compressibility of interacting two-dimensional electron and quasiparticle gases}},\ }\href {https://doi.org/10.1103/PhysRevLett.68.674} {\bibfield  {journal} {\bibinfo  {journal} {Phys. Rev. Lett.}\ }\textbf {\bibinfo {volume} {68}},\ \bibinfo {pages} {674} (\bibinfo {year} {1992}{\natexlab{b}})}\BibitemShut {NoStop}%
\bibitem [{\citenamefont {Eisenstein}\ \emph {et~al.}(1994)\citenamefont {Eisenstein}, \citenamefont {Pfeiffer},\ and\ \citenamefont {West}}]{Eisenstein1994}%
  \BibitemOpen
  \bibfield  {author} {\bibinfo {author} {\bibfnamefont {J.~P.}\ \bibnamefont {Eisenstein}}, \bibinfo {author} {\bibfnamefont {L.~N.}\ \bibnamefont {Pfeiffer}},\ and\ \bibinfo {author} {\bibfnamefont {K.~W.}\ \bibnamefont {West}},\ }\bibfield  {title} {\bibinfo {title} {{Compressibility of the two-dimensional electron gas: Measurements of the zero-field exchange energy and fractional quantum {H}all gap}},\ }\href {https://doi.org/10.1103/PhysRevB.50.1760} {\bibfield  {journal} {\bibinfo  {journal} {Phys. Rev. B}\ }\textbf {\bibinfo {volume} {50}},\ \bibinfo {pages} {1760} (\bibinfo {year} {1994})}\BibitemShut {NoStop}%
\bibitem [{\citenamefont {Prange}\ and\ \citenamefont {Girvin}(1989)}]{PrangeGirvin1990}%
  \BibitemOpen
  \bibinfo {editor} {\bibfnamefont {R.}~\bibnamefont {Prange}}\ and\ \bibinfo {editor} {\bibfnamefont {S.}~\bibnamefont {Girvin}},\ eds.,\ \href {https://books.google.co.uk/books?id=mxrSBwAAQBAJ} {\emph {\bibinfo {title} {{The Quantum Hall Effect}}}},\ Graduate Texts in Contemporary Physics\ (\bibinfo  {publisher} {Springer New York},\ \bibinfo {year} {1989})\BibitemShut {NoStop}%
\bibitem [{\citenamefont {Abrikosov}\ \emph {et~al.}(2012)\citenamefont {Abrikosov}, \citenamefont {Gorkov}, \citenamefont {Dzyaloshinski},\ and\ \citenamefont {Silverman}}]{AGD}%
  \BibitemOpen
  \bibfield  {author} {\bibinfo {author} {\bibfnamefont {A.}~\bibnamefont {Abrikosov}}, \bibinfo {author} {\bibfnamefont {L.}~\bibnamefont {Gorkov}}, \bibinfo {author} {\bibfnamefont {I.}~\bibnamefont {Dzyaloshinski}},\ and\ \bibinfo {author} {\bibfnamefont {R.}~\bibnamefont {Silverman}},\ }\href {https://books.google.ch/books?id=JYTCAgAAQBAJ} {\emph {\bibinfo {title} {{Methods of Quantum Field Theory in Statistical Physics}}}},\ Dover Books on Physics\ (\bibinfo  {publisher} {Dover Publications, New York},\ \bibinfo {year} {2012})\BibitemShut {NoStop}%
\bibitem [{\citenamefont {Dyson}(1952)}]{Dyson1952}%
  \BibitemOpen
  \bibfield  {author} {\bibinfo {author} {\bibfnamefont {F.~J.}\ \bibnamefont {Dyson}},\ }\bibfield  {title} {\bibinfo {title} {{Divergence of Perturbation Theory in Quantum Electrodynamics}},\ }\href {https://doi.org/10.1103/PhysRev.85.631} {\bibfield  {journal} {\bibinfo  {journal} {Phys. Rev.}\ }\textbf {\bibinfo {volume} {85}},\ \bibinfo {pages} {631} (\bibinfo {year} {1952})}\BibitemShut {NoStop}%
\bibitem [{\citenamefont {Wu}\ \emph {et~al.}(2017)\citenamefont {Wu}, \citenamefont {Ferrero}, \citenamefont {Georges},\ and\ \citenamefont {Kozik}}]{Wu2017}%
  \BibitemOpen
  \bibfield  {author} {\bibinfo {author} {\bibfnamefont {W.}~\bibnamefont {Wu}}, \bibinfo {author} {\bibfnamefont {M.}~\bibnamefont {Ferrero}}, \bibinfo {author} {\bibfnamefont {A.}~\bibnamefont {Georges}},\ and\ \bibinfo {author} {\bibfnamefont {E.}~\bibnamefont {Kozik}},\ }\bibfield  {title} {\bibinfo {title} {{Controlling Feynman diagrammatic expansions: Physical nature of the pseudogap in the two-dimensional Hubbard model}},\ }\href {https://doi.org/10.1103/PhysRevB.96.041105} {\bibfield  {journal} {\bibinfo  {journal} {Phys. Rev. B}\ }\textbf {\bibinfo {volume} {96}},\ \bibinfo {pages} {041105(R)} (\bibinfo {year} {2017})}\BibitemShut {NoStop}%
\bibitem [{\citenamefont {Rossi}\ \emph {et~al.}(2016)\citenamefont {Rossi}, \citenamefont {Werner}, \citenamefont {Prokof'ev},\ and\ \citenamefont {Svistunov}}]{Rossi2016}%
  \BibitemOpen
  \bibfield  {author} {\bibinfo {author} {\bibfnamefont {R.}~\bibnamefont {Rossi}}, \bibinfo {author} {\bibfnamefont {F.}~\bibnamefont {Werner}}, \bibinfo {author} {\bibfnamefont {N.}~\bibnamefont {Prokof'ev}},\ and\ \bibinfo {author} {\bibfnamefont {B.}~\bibnamefont {Svistunov}},\ }\bibfield  {title} {\bibinfo {title} {{Shifted-action expansion and applicability of dressed diagrammatic schemes}},\ }\href {https://doi.org/10.1103/PhysRevB.93.161102} {\bibfield  {journal} {\bibinfo  {journal} {Phys. Rev. B}\ }\textbf {\bibinfo {volume} {93}},\ \bibinfo {pages} {161102(R)} (\bibinfo {year} {2016})}\BibitemShut {NoStop}%
\bibitem [{\citenamefont {Kim}\ \emph {et~al.}(2021)\citenamefont {Kim}, \citenamefont {Prokof'ev}, \citenamefont {Svistunov},\ and\ \citenamefont {Kozik}}]{homotopic_action}%
  \BibitemOpen
  \bibfield  {author} {\bibinfo {author} {\bibfnamefont {A.~J.}\ \bibnamefont {Kim}}, \bibinfo {author} {\bibfnamefont {N.~V.}\ \bibnamefont {Prokof'ev}}, \bibinfo {author} {\bibfnamefont {B.~V.}\ \bibnamefont {Svistunov}},\ and\ \bibinfo {author} {\bibfnamefont {E.}~\bibnamefont {Kozik}},\ }\bibfield  {title} {\bibinfo {title} {{Homotopic Action: A Pathway to Convergent Diagrammatic Theories}},\ }\href {https://doi.org/10.1103/PhysRevLett.126.257001} {\bibfield  {journal} {\bibinfo  {journal} {Phys. Rev. Lett.}\ }\textbf {\bibinfo {volume} {126}},\ \bibinfo {pages} {257001} (\bibinfo {year} {2021})}\BibitemShut {NoStop}%
\bibitem [{\citenamefont {Prokof'ev}\ and\ \citenamefont {Svistunov}(1998)}]{Prokofev1998}%
  \BibitemOpen
  \bibfield  {author} {\bibinfo {author} {\bibfnamefont {N.~V.}\ \bibnamefont {Prokof'ev}}\ and\ \bibinfo {author} {\bibfnamefont {B.~V.}\ \bibnamefont {Svistunov}},\ }\bibfield  {title} {\bibinfo {title} {{Polaron Problem by Diagrammatic Quantum Monte Carlo}},\ }\href {https://doi.org/10.1103/PhysRevLett.81.2514} {\bibfield  {journal} {\bibinfo  {journal} {Phys. Rev. Lett.}\ }\textbf {\bibinfo {volume} {81}},\ \bibinfo {pages} {2514} (\bibinfo {year} {1998})}\BibitemShut {NoStop}%
\bibitem [{\citenamefont {{Van Houcke}}\ \emph {et~al.}(2012)\citenamefont {{Van Houcke}}, \citenamefont {Werner}, \citenamefont {Kozik}, \citenamefont {Prokof'ev}, \citenamefont {Svistunov}, \citenamefont {Ku}, \citenamefont {Sommer}, \citenamefont {Cheuk}, \citenamefont {Schirotzek},\ and\ \citenamefont {Zwierlein}}]{VanHoucke2012}%
  \BibitemOpen
  \bibfield  {author} {\bibinfo {author} {\bibfnamefont {K.}~\bibnamefont {{Van Houcke}}}, \bibinfo {author} {\bibfnamefont {F.}~\bibnamefont {Werner}}, \bibinfo {author} {\bibfnamefont {E.}~\bibnamefont {Kozik}}, \bibinfo {author} {\bibfnamefont {N.}~\bibnamefont {Prokof'ev}}, \bibinfo {author} {\bibfnamefont {B.}~\bibnamefont {Svistunov}}, \bibinfo {author} {\bibfnamefont {M.~J.~H.}\ \bibnamefont {Ku}}, \bibinfo {author} {\bibfnamefont {A.~T.}\ \bibnamefont {Sommer}}, \bibinfo {author} {\bibfnamefont {L.~W.}\ \bibnamefont {Cheuk}}, \bibinfo {author} {\bibfnamefont {A.}~\bibnamefont {Schirotzek}},\ and\ \bibinfo {author} {\bibfnamefont {M.~W.}\ \bibnamefont {Zwierlein}},\ }\bibfield  {title} {\bibinfo {title} {{Feynman diagrams versus Fermi-gas Feynman emulator}},\ }\href {https://doi.org/10.1038/nphys2273} {\bibfield  {journal} {\bibinfo  {journal} {Nat. Phys.}\ }\textbf {\bibinfo {volume} {8}},\ \bibinfo {pages} {366} (\bibinfo {year} {2012})}\BibitemShut {NoStop}%
\bibitem [{\citenamefont {Kaye}\ \emph {et~al.}(2022)\citenamefont {Kaye}, \citenamefont {Chen},\ and\ \citenamefont {Parcollet}}]{DLR2022}%
  \BibitemOpen
  \bibfield  {author} {\bibinfo {author} {\bibfnamefont {J.}~\bibnamefont {Kaye}}, \bibinfo {author} {\bibfnamefont {K.}~\bibnamefont {Chen}},\ and\ \bibinfo {author} {\bibfnamefont {O.}~\bibnamefont {Parcollet}},\ }\bibfield  {title} {\bibinfo {title} {{Discrete {L}ehmann representation of imaginary time {G}reen's functions}},\ }\href {https://doi.org/10.1103/PhysRevB.105.235115} {\bibfield  {journal} {\bibinfo  {journal} {Phys. Rev. B}\ }\textbf {\bibinfo {volume} {105}},\ \bibinfo {pages} {235115} (\bibinfo {year} {2022})}\BibitemShut {NoStop}%
\bibitem [{\citenamefont {Bruus}\ and\ \citenamefont {Flensberg}(2004)}]{Bruus2004}%
  \BibitemOpen
  \bibfield  {author} {\bibinfo {author} {\bibfnamefont {H.}~\bibnamefont {Bruus}}\ and\ \bibinfo {author} {\bibfnamefont {K.}~\bibnamefont {Flensberg}},\ }\href {https://books.google.co.uk/books?id=zeaMBAAAQBAJ} {\emph {\bibinfo {title} {{Many-Body Quantum Theory in Condensed Matter Physics: An Introduction}}}},\ Oxford Graduate Texts\ (\bibinfo  {publisher} {OUP Oxford},\ \bibinfo {year} {2004})\BibitemShut {NoStop}%
\bibitem [{\citenamefont {van Schilfgaarde}\ \emph {et~al.}(2006)\citenamefont {van Schilfgaarde}, \citenamefont {Kotani},\ and\ \citenamefont {Faleev}}]{Schilfgaarde2006}%
  \BibitemOpen
  \bibfield  {author} {\bibinfo {author} {\bibfnamefont {M.}~\bibnamefont {van Schilfgaarde}}, \bibinfo {author} {\bibfnamefont {T.}~\bibnamefont {Kotani}},\ and\ \bibinfo {author} {\bibfnamefont {S.}~\bibnamefont {Faleev}},\ }\bibfield  {title} {\bibinfo {title} {{Quasiparticle Self-Consistent {$GW$} Theory}},\ }\href {https://doi.org/10.1103/PhysRevLett.96.226402} {\bibfield  {journal} {\bibinfo  {journal} {Phys. Rev. Lett.}\ }\textbf {\bibinfo {volume} {96}},\ \bibinfo {pages} {226402} (\bibinfo {year} {2006})}\BibitemShut {NoStop}%
\bibitem [{\citenamefont {Kotani}\ \emph {et~al.}(2007)\citenamefont {Kotani}, \citenamefont {van Schilfgaarde}, \citenamefont {Faleev},\ and\ \citenamefont {Chantis}}]{Kotani2007}%
  \BibitemOpen
  \bibfield  {author} {\bibinfo {author} {\bibfnamefont {T.}~\bibnamefont {Kotani}}, \bibinfo {author} {\bibfnamefont {M.}~\bibnamefont {van Schilfgaarde}}, \bibinfo {author} {\bibfnamefont {S.~V.}\ \bibnamefont {Faleev}},\ and\ \bibinfo {author} {\bibfnamefont {A.}~\bibnamefont {Chantis}},\ }\bibfield  {title} {\bibinfo {title} {{Quasiparticle self-consistent GW method: a short summary}},\ }\href {https://doi.org/10.1088/0953-8984/19/36/365236} {\bibfield  {journal} {\bibinfo  {journal} {Journal of Physics: Condensed Matter}\ }\textbf {\bibinfo {volume} {19}},\ \bibinfo {pages} {365236} (\bibinfo {year} {2007})}\BibitemShut {NoStop}%
\bibitem [{\citenamefont {H\"user}\ \emph {et~al.}(2013)\citenamefont {H\"user}, \citenamefont {Olsen},\ and\ \citenamefont {Thygesen}}]{Huser2013}%
  \BibitemOpen
  \bibfield  {author} {\bibinfo {author} {\bibfnamefont {F.}~\bibnamefont {H\"user}}, \bibinfo {author} {\bibfnamefont {T.}~\bibnamefont {Olsen}},\ and\ \bibinfo {author} {\bibfnamefont {K.~S.}\ \bibnamefont {Thygesen}},\ }\bibfield  {title} {\bibinfo {title} {{Quasiparticle {GW} calculations for solids, molecules, and two-dimensional materials}},\ }\href {https://doi.org/10.1103/PhysRevB.87.235132} {\bibfield  {journal} {\bibinfo  {journal} {Phys. Rev. B}\ }\textbf {\bibinfo {volume} {87}},\ \bibinfo {pages} {235132} (\bibinfo {year} {2013})}\BibitemShut {NoStop}%
\bibitem [{\citenamefont {He}\ \emph {et~al.}(1993)\citenamefont {He}, \citenamefont {Platzman},\ and\ \citenamefont {Halperin}}]{HePlatzmanHalperin1993}%
  \BibitemOpen
  \bibfield  {author} {\bibinfo {author} {\bibfnamefont {S.}~\bibnamefont {He}}, \bibinfo {author} {\bibfnamefont {P.~M.}\ \bibnamefont {Platzman}},\ and\ \bibinfo {author} {\bibfnamefont {B.~I.}\ \bibnamefont {Halperin}},\ }\bibfield  {title} {\bibinfo {title} {{Tunneling into a two-dimensional electron system in a strong magnetic field}},\ }\href {https://doi.org/10.1103/PhysRevLett.71.777} {\bibfield  {journal} {\bibinfo  {journal} {Phys. Rev. Lett.}\ }\textbf {\bibinfo {volume} {71}},\ \bibinfo {pages} {777} (\bibinfo {year} {1993})}\BibitemShut {NoStop}%
\bibitem [{\citenamefont {Kim}\ and\ \citenamefont {Wen}(1994)}]{Kim1994}%
  \BibitemOpen
  \bibfield  {author} {\bibinfo {author} {\bibfnamefont {Y.~B.}\ \bibnamefont {Kim}}\ and\ \bibinfo {author} {\bibfnamefont {X.-G.}\ \bibnamefont {Wen}},\ }\bibfield  {title} {\bibinfo {title} {{Instantons and the spectral function of electrons in the half-filled {L}andau level}},\ }\href {https://doi.org/10.1103/PhysRevB.50.8078} {\bibfield  {journal} {\bibinfo  {journal} {Phys. Rev. B}\ }\textbf {\bibinfo {volume} {50}},\ \bibinfo {pages} {8078} (\bibinfo {year} {1994})}\BibitemShut {NoStop}%
\bibitem [{\citenamefont {Trugman}\ and\ \citenamefont {Kivelson}(1985)}]{Trugman1985}%
  \BibitemOpen
  \bibfield  {author} {\bibinfo {author} {\bibfnamefont {S.~A.}\ \bibnamefont {Trugman}}\ and\ \bibinfo {author} {\bibfnamefont {S.}~\bibnamefont {Kivelson}},\ }\bibfield  {title} {\bibinfo {title} {{Exact results for the fractional quantum Hall effect with general interactions}},\ }\href {https://doi.org/10.1103/PhysRevB.31.5280} {\bibfield  {journal} {\bibinfo  {journal} {Phys. Rev. B}\ }\textbf {\bibinfo {volume} {31}},\ \bibinfo {pages} {5280} (\bibinfo {year} {1985})}\BibitemShut {NoStop}%
\bibitem [{\citenamefont {Baym}\ and\ \citenamefont {Kadanoff}(1961)}]{Baym1961}%
  \BibitemOpen
  \bibfield  {author} {\bibinfo {author} {\bibfnamefont {G.}~\bibnamefont {Baym}}\ and\ \bibinfo {author} {\bibfnamefont {L.~P.}\ \bibnamefont {Kadanoff}},\ }\bibfield  {title} {\bibinfo {title} {{Conservation Laws and Correlation Functions}},\ }\href {https://doi.org/10.1103/PhysRev.124.287} {\bibfield  {journal} {\bibinfo  {journal} {Phys. Rev.}\ }\textbf {\bibinfo {volume} {124}},\ \bibinfo {pages} {287} (\bibinfo {year} {1961})}\BibitemShut {NoStop}%
\bibitem [{\citenamefont {Parcollet}\ and\ \citenamefont {Georges}(1999)}]{Parcollet1999}%
  \BibitemOpen
  \bibfield  {author} {\bibinfo {author} {\bibfnamefont {O.}~\bibnamefont {Parcollet}}\ and\ \bibinfo {author} {\bibfnamefont {A.}~\bibnamefont {Georges}},\ }\bibfield  {title} {\bibinfo {title} {{Non-Fermi-liquid regime of a doped Mott insulator}},\ }\href {https://doi.org/10.1103/PhysRevB.59.5341} {\bibfield  {journal} {\bibinfo  {journal} {Phys. Rev. B}\ }\textbf {\bibinfo {volume} {59}},\ \bibinfo {pages} {5341} (\bibinfo {year} {1999})}\BibitemShut {NoStop}%
\bibitem [{\citenamefont {van Loon}\ \emph {et~al.}(2020)\citenamefont {van Loon}, \citenamefont {Krien},\ and\ \citenamefont {Katanin}}]{vanLoon2020}%
  \BibitemOpen
  \bibfield  {author} {\bibinfo {author} {\bibfnamefont {E.~G. C.~P.}\ \bibnamefont {van Loon}}, \bibinfo {author} {\bibfnamefont {F.}~\bibnamefont {Krien}},\ and\ \bibinfo {author} {\bibfnamefont {A.~A.}\ \bibnamefont {Katanin}},\ }\bibfield  {title} {\bibinfo {title} {{Bethe-{S}alpeter Equation at the Critical End Point of the {M}ott Transition}},\ }\href {https://doi.org/10.1103/PhysRevLett.125.136402} {\bibfield  {journal} {\bibinfo  {journal} {Phys. Rev. Lett.}\ }\textbf {\bibinfo {volume} {125}},\ \bibinfo {pages} {136402} (\bibinfo {year} {2020})}\BibitemShut {NoStop}%
\bibitem [{\citenamefont {van Loon}(2022)}]{vanLoon2022}%
  \BibitemOpen
  \bibfield  {author} {\bibinfo {author} {\bibfnamefont {E.~G. C.~P.}\ \bibnamefont {van Loon}},\ }\bibfield  {title} {\bibinfo {title} {{Two-particle correlations and the metal-insulator transition: Iterated perturbation theory revisited}},\ }\href {https://doi.org/10.1103/PhysRevB.105.245104} {\bibfield  {journal} {\bibinfo  {journal} {Phys. Rev. B}\ }\textbf {\bibinfo {volume} {105}},\ \bibinfo {pages} {245104} (\bibinfo {year} {2022})}\BibitemShut {NoStop}%
\bibitem [{\citenamefont {Efros}\ and\ \citenamefont {Pikus}(1993)}]{Efros1993}%
  \BibitemOpen
  \bibfield  {author} {\bibinfo {author} {\bibfnamefont {A.~L.}\ \bibnamefont {Efros}}\ and\ \bibinfo {author} {\bibfnamefont {F.~G.}\ \bibnamefont {Pikus}},\ }\bibfield  {title} {\bibinfo {title} {{Classical approach to the gap in the tunneling density of states of a two-dimensional electron liquid in a strong magnetic field}},\ }\href {https://doi.org/10.1103/PhysRevB.48.14694} {\bibfield  {journal} {\bibinfo  {journal} {Phys. Rev. B}\ }\textbf {\bibinfo {volume} {48}},\ \bibinfo {pages} {14694} (\bibinfo {year} {1993})}\BibitemShut {NoStop}%
\bibitem [{\citenamefont {Johansson}\ and\ \citenamefont {Kinaret}(1993)}]{Johansson1993}%
  \BibitemOpen
  \bibfield  {author} {\bibinfo {author} {\bibfnamefont {P.}~\bibnamefont {Johansson}}\ and\ \bibinfo {author} {\bibfnamefont {J.~M.}\ \bibnamefont {Kinaret}},\ }\bibfield  {title} {\bibinfo {title} {{Magnetophonon shakeup in a {W}igner crystal: Applications to tunneling spectroscopy in the quantum {H}all regime}},\ }\href {https://doi.org/10.1103/PhysRevLett.71.1435} {\bibfield  {journal} {\bibinfo  {journal} {Phys. Rev. Lett.}\ }\textbf {\bibinfo {volume} {71}},\ \bibinfo {pages} {1435} (\bibinfo {year} {1993})}\BibitemShut {NoStop}%
\bibitem [{\citenamefont {Hatsugai}\ \emph {et~al.}(1993)\citenamefont {Hatsugai}, \citenamefont {Bares},\ and\ \citenamefont {Wen}}]{Hatsugai1993}%
  \BibitemOpen
  \bibfield  {author} {\bibinfo {author} {\bibfnamefont {Y.}~\bibnamefont {Hatsugai}}, \bibinfo {author} {\bibfnamefont {P.-A.}\ \bibnamefont {Bares}},\ and\ \bibinfo {author} {\bibfnamefont {X.~G.}\ \bibnamefont {Wen}},\ }\bibfield  {title} {\bibinfo {title} {{Electron spectral function of an interacting two dimensional electron gas in a strong magnetic field}},\ }\href {https://doi.org/10.1103/PhysRevLett.71.424} {\bibfield  {journal} {\bibinfo  {journal} {Phys. Rev. Lett.}\ }\textbf {\bibinfo {volume} {71}},\ \bibinfo {pages} {424} (\bibinfo {year} {1993})}\BibitemShut {NoStop}%
\bibitem [{\citenamefont {Halperin}(1994)}]{Halperin1994}%
  \BibitemOpen
  \bibfield  {author} {\bibinfo {author} {\bibfnamefont {B.~I.}\ \bibnamefont {Halperin}},\ }\bibfield  {title} {\bibinfo {title} {{Theories for v = 1/2 in single- and double-layer systems}},\ }\href {https://doi.org/https://doi.org/10.1016/0039-6028(94)90850-8} {\bibfield  {journal} {\bibinfo  {journal} {Surface Science}\ }\textbf {\bibinfo {volume} {305}},\ \bibinfo {pages} {1} (\bibinfo {year} {1994})}\BibitemShut {NoStop}%
\bibitem [{\citenamefont {Villegas~Rosales}\ \emph {et~al.}(2021)\citenamefont {Villegas~Rosales}, \citenamefont {Madathil}, \citenamefont {Chung}, \citenamefont {Pfeiffer}, \citenamefont {West}, \citenamefont {Baldwin},\ and\ \citenamefont {Shayegan}}]{VillegasRosales2021}%
  \BibitemOpen
  \bibfield  {author} {\bibinfo {author} {\bibfnamefont {K.~A.}\ \bibnamefont {Villegas~Rosales}}, \bibinfo {author} {\bibfnamefont {P.~T.}\ \bibnamefont {Madathil}}, \bibinfo {author} {\bibfnamefont {Y.~J.}\ \bibnamefont {Chung}}, \bibinfo {author} {\bibfnamefont {L.~N.}\ \bibnamefont {Pfeiffer}}, \bibinfo {author} {\bibfnamefont {K.~W.}\ \bibnamefont {West}}, \bibinfo {author} {\bibfnamefont {K.~W.}\ \bibnamefont {Baldwin}},\ and\ \bibinfo {author} {\bibfnamefont {M.}~\bibnamefont {Shayegan}},\ }\bibfield  {title} {\bibinfo {title} {{Fractional Quantum Hall Effect Energy Gaps: Role of Electron Layer Thickness}},\ }\href {https://doi.org/10.1103/PhysRevLett.127.056801} {\bibfield  {journal} {\bibinfo  {journal} {Phys. Rev. Lett.}\ }\textbf {\bibinfo {volume} {127}},\ \bibinfo {pages} {056801} (\bibinfo {year} {2021})}\BibitemShut {NoStop}%
\bibitem [{\citenamefont {Maryenko}\ \emph {et~al.}(2018)\citenamefont {Maryenko}, \citenamefont {McCollam}, \citenamefont {Falson}, \citenamefont {Kozuka}, \citenamefont {Bruin}, \citenamefont {Zeitler},\ and\ \citenamefont {Kawasaki}}]{Maryenko2018}%
  \BibitemOpen
  \bibfield  {author} {\bibinfo {author} {\bibfnamefont {D.}~\bibnamefont {Maryenko}}, \bibinfo {author} {\bibfnamefont {A.}~\bibnamefont {McCollam}}, \bibinfo {author} {\bibfnamefont {J.}~\bibnamefont {Falson}}, \bibinfo {author} {\bibfnamefont {Y.}~\bibnamefont {Kozuka}}, \bibinfo {author} {\bibfnamefont {J.}~\bibnamefont {Bruin}}, \bibinfo {author} {\bibfnamefont {U.}~\bibnamefont {Zeitler}},\ and\ \bibinfo {author} {\bibfnamefont {M.}~\bibnamefont {Kawasaki}},\ }\bibfield  {title} {\bibinfo {title} {{Composite fermion liquid to {W}igner solid transition in the lowest {L}andau level of zinc oxide}},\ }\href {https://doi.org/10.1038/s41467-018-06834-6} {\bibfield  {journal} {\bibinfo  {journal} {Nature Communications}\ }\textbf {\bibinfo {volume} {9}},\ \bibinfo {pages} {4356} (\bibinfo {year} {2018})}\BibitemShut {NoStop}%
\bibitem [{\citenamefont {Rohringer}\ \emph {et~al.}(2018)\citenamefont {Rohringer}, \citenamefont {Hafermann}, \citenamefont {Toschi}, \citenamefont {Katanin}, \citenamefont {Antipov}, \citenamefont {Katsnelson}, \citenamefont {Lichtenstein}, \citenamefont {Rubtsov},\ and\ \citenamefont {Held}}]{Rohringer2018}%
  \BibitemOpen
  \bibfield  {author} {\bibinfo {author} {\bibfnamefont {G.}~\bibnamefont {Rohringer}}, \bibinfo {author} {\bibfnamefont {H.}~\bibnamefont {Hafermann}}, \bibinfo {author} {\bibfnamefont {A.}~\bibnamefont {Toschi}}, \bibinfo {author} {\bibfnamefont {A.~A.}\ \bibnamefont {Katanin}}, \bibinfo {author} {\bibfnamefont {A.~E.}\ \bibnamefont {Antipov}}, \bibinfo {author} {\bibfnamefont {M.~I.}\ \bibnamefont {Katsnelson}}, \bibinfo {author} {\bibfnamefont {A.~I.}\ \bibnamefont {Lichtenstein}}, \bibinfo {author} {\bibfnamefont {A.~N.}\ \bibnamefont {Rubtsov}},\ and\ \bibinfo {author} {\bibfnamefont {K.}~\bibnamefont {Held}},\ }\bibfield  {title} {\bibinfo {title} {{Diagrammatic routes to nonlocal correlations beyond dynamical mean field theory}},\ }\href {https://doi.org/10.1103/RevModPhys.90.025003} {\bibfield  {journal} {\bibinfo  {journal} {Rev. Mod. Phys.}\ }\textbf {\bibinfo {volume} {90}},\ \bibinfo {pages} {025003} (\bibinfo {year} {2018})}\BibitemShut {NoStop}%
\bibitem [{\citenamefont {Giuliani}\ and\ \citenamefont {Quinn}(1985)}]{Giuliani1985}%
  \BibitemOpen
  \bibfield  {author} {\bibinfo {author} {\bibfnamefont {G.~F.}\ \bibnamefont {Giuliani}}\ and\ \bibinfo {author} {\bibfnamefont {J.~J.}\ \bibnamefont {Quinn}},\ }\bibfield  {title} {\bibinfo {title} {{Breakdown of the random-phase approximation in the fractional-quantum-{H}all-effect regime}},\ }\href {https://doi.org/10.1103/PhysRevB.31.3451} {\bibfield  {journal} {\bibinfo  {journal} {Phys. Rev. B}\ }\textbf {\bibinfo {volume} {31}},\ \bibinfo {pages} {3451} (\bibinfo {year} {1985})}\BibitemShut {NoStop}%
\bibitem [{\citenamefont {Thouless}(1985)}]{Thouless1985}%
  \BibitemOpen
  \bibfield  {author} {\bibinfo {author} {\bibfnamefont {D.~J.}\ \bibnamefont {Thouless}},\ }\bibfield  {title} {\bibinfo {title} {{Long-range order and the fractional quantum {H}all effect}},\ }\href {https://doi.org/10.1103/PhysRevB.31.8305} {\bibfield  {journal} {\bibinfo  {journal} {Phys. Rev. B}\ }\textbf {\bibinfo {volume} {31}},\ \bibinfo {pages} {8305} (\bibinfo {year} {1985})}\BibitemShut {NoStop}%
\end{thebibliography}%
\end{document}